\documentclass[11pt]{article}
\pdfoutput=1

\usepackage{graphicx,epsf, epsfig, amssymb}
\usepackage{bm}
\usepackage{longtable}
\usepackage{color}
\usepackage{amsfonts,amsmath,amssymb,mathrsfs}

\usepackage{graphicx}
\usepackage{amssymb}
\usepackage{amsmath}
\usepackage{enumerate}
\usepackage{amsfonts}
\usepackage[section]{placeins}
\usepackage[colorlinks=true,linktocpage=true,linkcolor=blue,citecolor=blue]{hyperref}

\newcommand{\be}{\begin{equation}}
\newcommand{\ee}{\end{equation}}
\newcommand{\bea}{\begin{eqnarray}}
\newcommand{\eea}{\end{eqnarray}}

\newcommand{\eqn}[1]{(\ref{#1})}

\newcommand{\dmoff}[1]{}
\newcommand{\mt}[1]{\textrm{\tiny #1}}

\newcommand{\lqcd}{\Lambda_\mt{QCD}}

\newcommand{\sac}{\, , \qquad}

\newcommand{\rbar}{\overline{r}}
\newcommand{\tw}[1]{\tilde{\omega}_{{#1}}}
\newcommand{\psione}{\Psi_{\mbox{\scriptsize{I}}}}
\newcommand{\psitwo}{\Psi_{\mbox{\scriptsize{II}}}}

\def\beq{\begin{eqnarray}}
\def\eeq{\end{eqnarray}}
\newcommand{\nn}{\nonumber}

\def\d{{\partial}}

\newcommand{\ie}{i.e.~}
\newcommand{\eg}{e.g.~}
\newcommand{\beqa}{\begin{eqnarray}}
\newcommand{\eeqa}{\end{eqnarray}}

\newcommand{\vcoff}[1]{}

\newcommand{\reoff}[1]{}

\newcommand{\lp}{\left(}
\newcommand{\rp}{\right)}

\newcommand{\Rmnum}[1]{\expandafter\@slowromancap\romannumeral #1@}

\numberwithin{equation}{section} 

\begin{document}
\bibliographystyle{hieeetr}

\pagestyle{plain}
\setcounter{page}{1}

\begin{titlepage}

\begin{center}
\vspace*{-1cm} \today \hfill ICCUB-13-231

\vskip 2.0cm

{\huge {\bf Holographic collisions in confining theories}}

\vskip 14mm

{\large  {\bf Vitor Cardoso,$^{1,2,3}$ Roberto Emparan,$^{4,5}$ David Mateos,$^{4,5}$ \\ Paolo Pani,$^{1,6}$ and Jorge V. Rocha$^{1}$}}

\vspace{0.5 cm}


${}^1$ {\it CENTRA, Departamento de F\'{\i}sica, Instituto Superior
  T\'ecnico, Universidade de Lisboa, Av.~Rovisco Pais 1, 1049
  Lisboa, Portugal.}

\medskip
  
${}^2$ {\it Perimeter Institute for Theoretical Physics,
Waterloo, Ontario N2J 2W9, Canada.}

\medskip

${}^3$ {\it Department of Physics and Astronomy, The University of Mississippi, University, MS 38677, USA.}

\medskip

${}^4$ {\it Instituci\'o Catalana de Recerca i Estudis Avan\c cats (ICREA), Passeig Llu\'\i s Companys 23, E-08010, Barcelona, Spain}

\medskip

${}^5$ {\it Departament de F\'\i sica Fonamental,  Institut de Ci\`encies del Cosmos, Universitat de Barcelona, Mart\'{\i}  i Franqu\`es 1, E-08028 Barcelona, Spain}

\medskip

${}^6$ {\it Institute for Theory $\&$ Computation, Harvard-Smithsonian
CfA, 60 Garden Street, Cambridge, MA, USA}

\vskip 0.5cm

{\tt  vitor.cardoso@ist.utl.pt, \, dmateos@icrea.cat, \, 
emparan@ub.edu \, paolo.pani@ist.utl.pt, \,
jorge.v.rocha@ist.utl.pt}

\vspace{5mm}

{\bf Abstract}
\end{center}
 \noindent
We study the gravitational dual of a high-energy collision in a confining gauge theory. We consider a linearized approach in which two point particles traveling in an AdS-soliton background suddenly collide to form an object at rest (presumably a black hole for large enough center-of-mass energies). The resulting radiation exhibits the features expected in a theory with a mass gap: late-time power law tails of the form $t^{-3/2}$, the failure of Huygens' principle and distortion of the wave pattern as it propagates.  The energy spectrum is exponentially suppressed for frequencies smaller than the gauge theory mass gap. Consequently, we observe no memory effect in the gravitational waveforms. At larger frequencies the spectrum has an upward-stairway structure, which corresponds to the excitation of the tower of massive states in the confining gauge theory. We discuss the importance of phenomenological cutoffs to regularize the divergent spectrum, and the aspects of the full non-linear collision that are expected to be captured by our approach.

\noindent


\end{titlepage}

\tableofcontents

\section{Introduction}
\label{sec:intro}

The study of collisions and their outcomes is one of the most important ways of obtaining information about a theory and of testing it experimentally. This is true both in particle physics, where collision experiments have been dominant for a century now, and in gravitational physics, with the expected imminent detection of the gravitational radiation from collisions of black holes and neutron stars. The advent of gauge/gravity dualities brings about a merging of these two fields~\cite{Cardoso:2012qm}: the collision of two high-energy particles in certain non-abelian gauge theories can be adequately described in terms of a dual gravitational collision in a higher-dimensional spacetime with negative curvature.

One phenomenon of current interest in this area is the collision at high energies of two objects (nuclei, nucleons, or partons) which, through the interactions of Quantum Chromodynamics (QCD), form a ball of quark-gluon plasma (QGP). Although the dual of QCD is not known, the analogous process in gauge theories with a gravity dual can be described via the collision of two objects of finite but small size that form a black hole in an asymptotically AdS spacetime.\footnote{See e.g.~\cite{CasalderreySolana:2011us} for a review of applications of the gauge/gravity duality to QCD.} 

The study of these collision processes is challenging because one must solve Einstein's equations in a dynamical setting, which generically must be done numerically. Several such studies have now been performed in cases in which the gauge theory is a Conformal Field Theory (CFT) \cite{Chesler:2010bi,Wu:2011yd,Casalderrey-Solana:2013aba,vanderSchee:2013pia}.\footnote{See \cite{Chesler:2008hg,Chesler:2009cy,Heller:2012km,Heller:2011ju,Heller:2012je,Bantilan:2012vu,vanderSchee:2012qj} for related numerical studies in AdS.} The goal of this paper is to give a first step towards extending this program to gravitational duals of confining gauge theories. For this purpose we will consider collisions in the so-called AdS-soliton \cite{Witten:1998zw,Horowitz:1998ha}.  

One motivation for this extension is that a CFT has a continuous spectrum, so the result of the collision cannot be directly interpreted in terms of e.g.~particle production. In contrast, we will see that, in a confining geometry, typical observables (e.g.~the emitted radiation) admit an immediate particle interpretation. Another motivation is to explore the effects of confinement on the produced QGP. In a real heavy ion collision at RHIC or LHC the temperature of the produced QGP is roughly $2T_c \lesssim T \lesssim 4T_c$, with $T_c \sim \lqcd$ the deconfinement temperature. This means that there is no hierarchical separation between the temperature of the fireball and the confinement scale, thus suggesting that the latter may play a role in the dynamics of the QGP. Similarly, in the range of temperatures above, the trace anomaly in QCD, which measures deviations from conformality, is still relatively sizable \cite{Panero:2009tv}, again suggesting a possible role of $\lqcd$.
\subsection{The Zero-Frequency Limit framework}
Our framework is very simple: we model the colliding objects as point particles moving along geodesics in a background spacetime, 
colliding instantaneously to form a single object at rest. The process amounts to specifying a conserved stress-energy tensor for point particles following these trajectories,
and the gravitational field that they create is treated as a linearized perturbation of the background. 
Treating the collision in this approximation is well motivated, and this model is sometimes known in the literature as `instantaneous collision framework' or `Zero Frequency Limit' (ZFL) approximation 
\cite{Weinberg:1964ew,Weinberg:1965nx,Rees:1974,Adler:1975dj,Smarr:1977fy,MadalenaThesis}. The ZFL has been applied in a variety of contexts, including electromagnetism where it can be used to compute the electromagnetic 
radiation given away in $\beta$-decay (see for instance Chapter 15 in~\cite{Jackson}).
Wheeler used the ZFL to estimate the emission of gravitational and electromagnetic radiation from impulsive events \cite{Wheeler:1963}.
In essence, we are reducing the full gravitational dynamics of the process to an effective theory of point particles interacting through a three-leg vertex, and then weakly coupling gravitational radiation to this system. This description is a reasonable one given what we know generically about black hole formation in similar collisions.

In asymptotically flat spacetimes, this simple approximation turns out to describe accurately all the main features of high-energy collisions of two equal-mass black holes~\cite{Berti:2010ce,Cardoso:2002ay,Sperhake:2008ga,East:2012mb}. 
\begin{figure*}[htb]
\begin{center}
\begin{tabular}{c}
\includegraphics[scale=0.33,angle=0,clip=true]{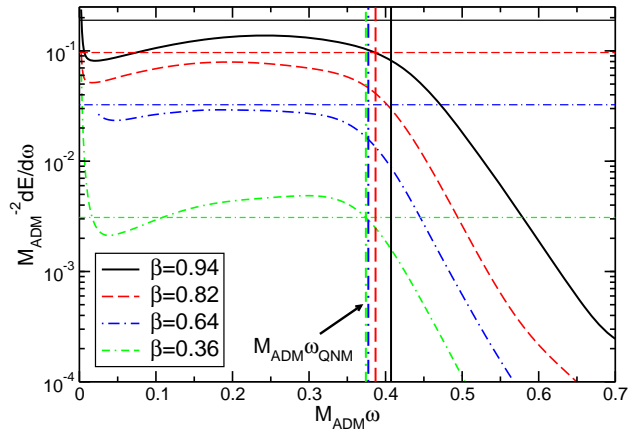}
\end{tabular}
\end{center}
\caption{\label{fig:ZFL_NR} 
  Energy spectrum for the dominant (quadrupolar,
  i.e. $l=2$) component of the gravitational radiation computed from numerical relativity (NR)
  simulations of the head-on collision of two equal-mass black holes (from
  \cite{Sperhake:2008ga}). The collision speed in the center-of-mass frame,
  $\beta=v/c$, is indicated in the legend. The energy spectrum is roughly flat
  (independent of frequency) up to the quasinormal mode (QNM) frequencies
  (marked by vertical lines), after which it decays exponentially. All
  quantities are normalized to the Arnowitt-Deser-Misner (ADM) mass of the
  system $M_{\rm ADM}$. The dashed horizontal lines are the ZFL prediction, obtained by a multipolar decomposition of \eqref{zfl_flat}.
  }
\end{figure*}
These results are summarized in Figure~\ref{fig:ZFL_NR}, and refer to head-on collisions of two equal-mass black holes, with center-of-mass energy parameterized by 
total energy $M_{ADM}$ and velocity $v$. The salient features of the ZFL analysis are:
\begin{enumerate}

\item A flat, frequency-independent energy spectrum. For the head-on collision of two equal-mass objects each with rest mass $M/2$, velocity $v$ in the center-of-mass (CM) frame and  Lorentz factor $\gamma$, the ZFL prediction \cite{Smarr:1977fy,MadalenaThesis} for the energy spectrum at an angle $\theta$ relative to the collision axis is
\be
\frac{d^2E}{d\omega d\Omega}=\frac{M^2\gamma^2v^4}{4\pi^2}\frac{\sin^4\theta}{\left(1-v^2\cos^2\theta\right)^2}\,.\label{zfl_flat}
\ee
The independence of this spectrum on frequency $\omega$ follows from simple arguments: since we are working in the linearized approximation to gravity, the right-hand side of \eqref{zfl_flat} must be proportional to $M^2$. This fixes all the leading-order dependence on $M$. The absence of any other dimensionful parameter in the problem then forbids, on dimensional grounds, any possible dependence on $\omega$. So the ZFL yields a flat spectrum, as shown in Fig.~\ref{fig:ZFL_NR} for different CM velocities $v$. The nonlinear results are in good quantitative agreement and do show an approximately flat energy spectrum.

\item The need for an appropriate, physical cutoff. Because the spectrum is flat, estimates for the total radiated energy or time-domain signals formally diverge. The dependence on $\omega$ that would cutoff the spectrum is lost when we neglect non-linear effects and thus eliminate all
the details of the interaction and the internal structure of the colliding and final objects.  We can nevertheless reintroduce the cutoff in frequency (or momentum, via the dispersion relations) in a phenomenological way, which in asymptotically flat spacetimes is essentially uniquely determined (up to numerical, order-one factors). Although dimensional arguments do not fully fix the cutoff ---besides the dimensionful scale $M$, there is a dimensionless parameter $\gamma$---
we can expect that it is the size $\sim M\gamma$ of the {\it final} black hole that sets the cutoff: black holes absorb very efficiently frequencies that are larger than its lowest quasinormal mode (QNM) frequency, so we may expect that any higher frequencies will not be radiated away, and therefore 
\beq\label{omcutoff}
\omega_\mathrm{cutoff}\sim \omega_{QNM}\sim 1/(M\gamma)\,.
\eeq 
This implies that the frequency cutoff decreases as the velocity of the colliding particles increases, \ie the quanta radiated are less energetic for larger CM energies --- a characteristic property of collisions that involve black holes. However, since the ZFL spectrum \eqref{zfl_flat} scales like $\gamma^2$, the total radiated energy scales like $\gamma$ and thus grows with $v$. 
Nonlinear simulations are in excellent agreement with this picture and show an exponential suppression of the spectrum for frequencies larger than 
the final black hole QNM frequency, as shown in Fig.~\ref{fig:ZFL_NR}.

\item A ``memory'' effect in the signal, which is a consequence of the identity
\be
\left(\bar{\dot{h}}\right)_{\omega=0}=\lim_{\omega\to 0}\int_{-\infty}^{+\infty}\dot{h}e^{-i\omega t}dt=h(t=+\infty)-h(t=-\infty)\,,\label{memory}
\ee
for the Fourier transform $\bar{\dot{h}}(\omega)$ of the time derivative of any metric perturbation $h(t)$ (we omitted unimportant overall factors in the definition of the transform). Thus, the low-frequency spectrum depends exclusively on the asymptotic state of the colliding particles, which can be readily computed from their Coulomb gravitational fields. Because the energy spectrum is related to $\bar{\dot{h}}(\omega)$ via
\be
\frac{dE}{d\Omega d\omega}\propto r^2\left(\bar{\dot{h}}\right)^2\,,
\ee
we immediately conclude that the energy spectrum at low-frequencies depends only on the asymptotic states \cite{Weinberg:1965nx,Adler:1975dj,Smarr:1977fy,MadalenaThesis,Berti:2010ce}. 

\item Finally, the angular distribution of radiation is nearly isotropic at large collision energies ($v\to 1$ in \eqref{zfl_flat}), when higher multipoles become increasingly more relevant. This is also in agreement with nonlinear simulations.

\end{enumerate}
The reason for the overall agreement of ZFL predictions with fully nonlinear simulations is not completely clear, one possibility being that nonlinearities are redshifted away. 
This simple model of particle collisions has provided useful benchmarking in the nonlinear simulations of asymptotically flat spacetimes \cite{Berti:2010ce,Cardoso:2002ay,Sperhake:2008ga,East:2012mb}; we expect similar benefits in asymptotically anti-de Sitter (AdS) spacetimes where full-blown nonlinear evolutions are specially hard to perform.
\subsection{The Zero-Frequency Limit in solitonic-AdS backgrounds}
\label{subsec:ZFLAdS}

\begin{figure*}[tb]
\begin{center}
\includegraphics[width=10cm,angle=0,clip=true]{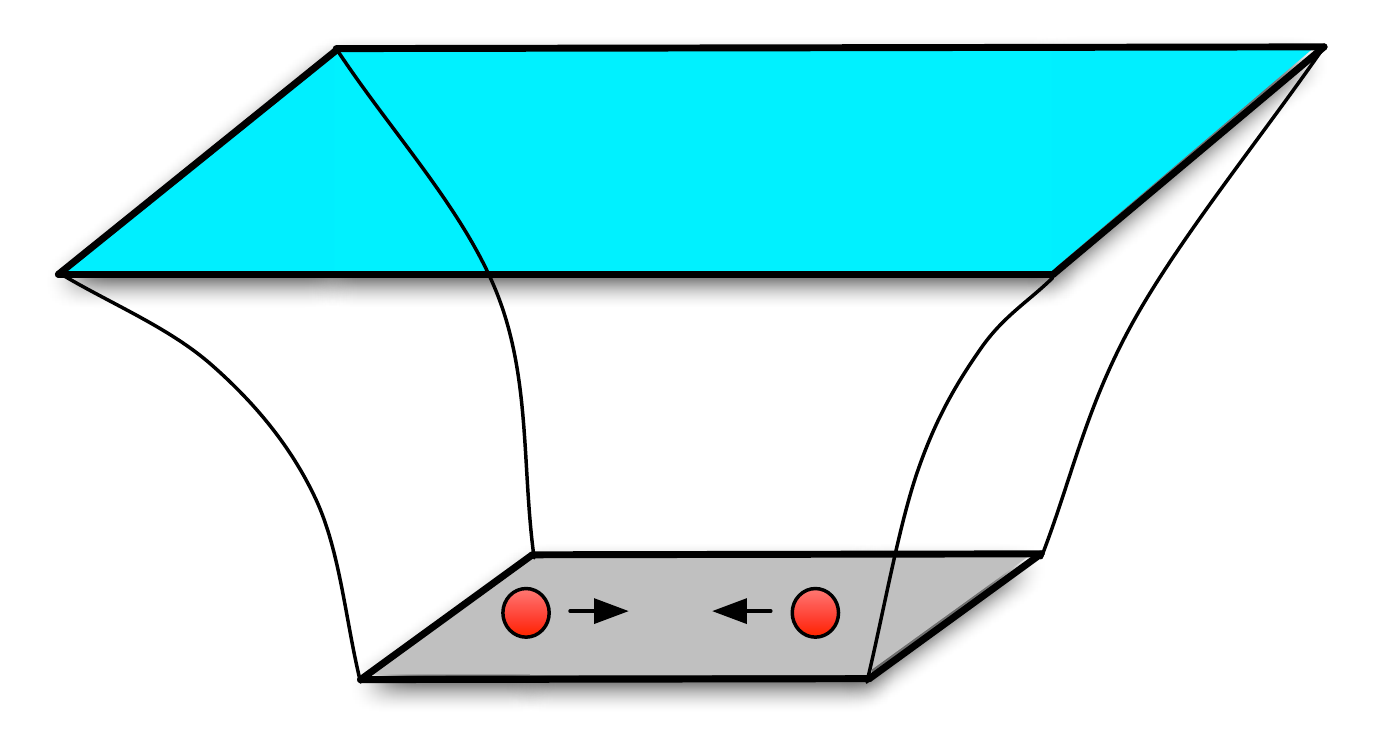}
\caption{The collision in the AdS-soliton. The vertical direction is the holographic radial direction. The horizontal directions are the gauge theory directions. The gauge theory can be thought of as living at the boundary of the space (top plane). The geometry ends smoothly at the plane at the bottom, where the extra circle of the AdS-soliton (not shown) shrinks to zero size. The two small-mass point-particles sit at this bottom and collide head-on with velocity $v$.
\label{fig:pictorial}}
\end{center}
\end{figure*}

A collision in the AdS-soliton background is depicted in Fig.~\ref{fig:pictorial}. 
These collisions differ from those in asymptotically flat spacetimes in several crucial respects. In particular, the process through which the initially-formed, highly-excited black hole radiates and relaxes to its equilibrium state is expected to be much more complex. 

\paragraph{Scales in the collision and horizon evolution.} In the Minkowski background, the entire collision and its evolution to a final state are characterized by the only scale in the problem,\footnote{This assumes that $\gamma$ is of order one, but when $\gamma$ is large the qualitative picture remains. In particular the fraction of energy radiated is relatively small even in ultrarelativistic cases.} namely the Schwarzschild radius $\sim M$. In particular, this scale controls both the properties of the linearized field of point particles and the properties of the final equilibrium black hole. 

In contrast, in the AdS-soliton there are two additional scales, namely the AdS curvature radius $L$ and the infrared length scale $r_0$ at which the geometry caps off smoothly. Actually, using coordinate reparametrizations they determine only one physical scale, which naturally may be taken as the gauge theory mass gap or confinement energy $\lqcd \sim r_0/L^2$. The presence of this scale besides the particle or black hole mass $M$ implies that the linearized field of a point particle in a confining background need not give a good estimate of the size and shape of the horizon of an equilibrium black hole localized in the infrared. In fact, it is a very poor approximation to it when the mass is large, $M\gg \lqcd$ \cite{Giddings:2002cd}. The black hole is dual to a plasma ball \cite{Aharony:2005bm} and is characterized by two very different scales: a thickness $\sim r_0$ in the holographic direction, and a much larger proper extent $\sim L(M/\lqcd)^{1/3}$ along the gauge-theory directions. Thus our approximation of the final state as a structureless point source is worse than in a Minkowski background.

Moreover, in a collision with $M\gg \lqcd$ we can expect that the initially formed horizon will be largely insensitive to the scale $r_0$. In this regime of energies the effects of the confinement scale can be neglected in this very early stage of the collision, but they will gradually appear in the relaxation to the final dual plasma ball.
Thus we expect a richer evolution from collision to relaxation in a confining AdS background than in a Minkowski background, with different stages being characterized by different horizon scales.

\paragraph{ZFL spectrum cutoff.} The linear approximation used in the ZFL neglects all the details of this process of horizon formation and relaxation, but some such information is nevertheless needed in order to specify the cutoff that renders finite the total radiated energy and time-domain signals. Here the differences between ZFL collisions in Minkowski and in AdS-soliton backgrounds can become significant in practice. In the former case, as we have seen, setting the frequency cutoff to be the final-state lowest QNM frequency is natural as this is the only scale in the problem. Moreover, in that background it does not matter whether we impose a frequency cutoff (the lowest QNM gives a characteristic time of horizon vibrations) or a momentum cutoff (implementing that a horizon only absorbs efficiently wavelengths shorter than its size). 

The correct choice of a cutoff in a confining background is  more convoluted. Attempting to find a cutoff on frequencies from the properties of the horizon formed in the collision is fraught with ambiguities: the horizon evolves through several complex stages and it is unclear to what extent these out-of-equilibrium horizons are well approximated by the properties of known stationary black holes. For instance,
in a high-energy collision with $M\gg \lqcd$ we might expect the initial horizon to be roughly similar to that of a large neutral black hole in global AdS, but even then it is not clear which kind of quasinormal modes would control the cutoff: `fast' modes with frequencies $\propto M^{1/4}$, proportional to the black hole temperature, which correspond to the highest gauge-theory energies that the horizon presumably probes; or `slow' modes, with much smaller frequencies $\sim 1/L$, which approximate hydrodynamic modes for very large black holes. Moreover, the dual plasma ball that the system relaxes to also has `slow' elastic modes associated to vibrations in the shape of the (dual) plasma ball. The complex time evolution of the horizon makes it unclear whether any of these QNMs can give a reliable frequency cutoff on the radiation.

The cutoff on the frequency of radiated AdS-soliton modes may  be obtained more plausibly, via their dispersion relation, from a momentum cutoff.
We would expect that the size of the horizon sets the largest wavelengths that it can efficiently absorb. For collision energies much larger than the confinement scale, at any stage the horizon will be larger (at least in some of its directions) than $1/\lqcd$, so it can absorb waves of momentum down to $k_\mathrm{cutoff}\ll \lqcd$. However, since the dispersion relation of the lowest AdS-soliton normal mode is $\omega^2\approx k^2+\lqcd^2$, the frequency is effectively cutoff by the mass gap.

While these estimates seem reasonable, the correct choice requires truly non-linear information about the formation and evolution of the horizon, which at present is unknown to us.
In order to deal with this uncertainty, we will not present our results in the time-domain, nor as an integrated total energy, but rather as frequency-domain signals, where the effect of the choice of different cutoffs is apparent and not obscured by integrating over the frequency spectrum. We will present some time-domain quantities only to illustrate their cutoff dependence.

It is important to note that in collisions at energies above the confinement scale, although it may be possible to neglect the confinement scale $r_0$ for certain aspects of the initial dynamics of the black hole formed, this scale is still crucial for the proper interpretation of the emitted radiation in terms of gauge theory particles, since it is responsible for the discreteness of the spectrum and the existence of a mass gap. These are universal, structure-independent features of the radiation produced in all such collisions that our model does capture.

\subsection{Plan}
The remainder of the paper is organized as follows. We begin by introducing the AdS-soliton geometry in Section~\ref{sec:AdSsoliton}, where we also collect its most relevant features for our study.  In Section~\ref{sec:scalar} we investigate, as a warm-up toy model, the scalar field radiation produced in the collision of two scalar-charged particles.  This is a simpler problem that shares many features of the gravitational problem --- except for the fact that in the former case the scalar charges in the initial particles simply add up to yield the final charge, whereas in the latter the kinetic energy of the colliding particles contributes to the final total charge (i.e.~mass).  Section~\ref{sec:grav} is concerned with the gravitational counterpart, the main objective of the paper.  In this section we study the gravitational radiation resulting from colliding two point particles in the AdS-soliton background and we present results for the far-region behavior of the stress-energy tensor of the dual confining field theory, obtained via the AdS/CFT correspondence. Several technical details of the derivation are relegated to the Appendices. Finally, we summarize our results in Section~\ref{sec:Conclusion}.

\section{The AdS-soliton background}
\label{sec:AdSsoliton}

As discussed in Section~\ref{sec:intro}, we will be interested in colliding point particles in a gravitational dual to a four-dimensional gauge theory in a confined phase.  The prototype for such a geometry is the six-dimensional AdS-soliton~\cite{Witten:1998zw,Horowitz:1998ha}.  We will take this spacetime as a background on top of which we then consider a linear analysis of perturbations induced by the point particles.

The AdS-soliton is a vacuum solution of Einstein's equations with a negative cosmological constant, which is given in terms of the AdS curvature radius by $-5/L^2$.  It asymptotes to AdS with one of the spatial coordinates periodically identified, thus forming an $S^1$.  This circle smoothly shrinks to zero size at a finite radial coordinate and consequently the geometry is regular everywhere, without possessing an event horizon.  The soliton metric is given by 
\begin{equation}
ds^2= g_{ab}dz^a dz^b = \frac{r^2}{L^2} \left[ -dt^2+dx_{(3)}^2 \right] + \frac{dr^2}{F(r)} + F(r)dy^2 \,,
\label{soliton}
\ee
where 
\be
F(r)=\frac{r^2}{L^2} \, f(r) \sac f(r) = 1-\frac{r_0^{5}}{r^{5}} \,,
\end{equation}
and $y$ has periodicity 
\be
\Delta y = \frac{4\pi L^2}{5r_0} \,.  
\ee 
We will group the coordinates as 
\be
x_i =  (x_1, x_2, x_3) \sac x^\mu=(t, x_i) \sac z^a = (t, x_i, r, y) \,.
\ee
The $S^1$ shrinks to zero size as $r\to r_0$ and the boundary lies at 
$r\to \infty$. The gauge theory lives on $\mbox{Mink}_{1,3} \times S^1$. As we will see, the Kaluza-Klein scale associated to the compact direction, 
\be
\lqcd = \frac{2\pi}{\Delta y} = \frac{5 r_0}{2 L^2} \,,
\ee
also sets the confining scale (hence our choice of notation), reflecting the well-known fact that these two scales cannot be decoupled within the gravity approximation.\footnote{See \cite{Mateos:2011bs} for a general discussion of this and other limitations in the applications of the the gauge/string duality to QCD.}

Of some interest for us is the eikonal limit of massless field propagation, which is generically associated to spacetime geodesics.  Let us focus on radial geodesics along the holographic direction $r$ and along the flat directions $x_i$.  These satisfy
\be
\pm dt = \frac{L}{r\sqrt{F}\sqrt{1-c_x^2+
\frac{r^2\epsilon}{L^2E_t^2}}} \, dr \,,
\ee
where $\epsilon=-1,0$ for timelike and null particles respectively, and $E_t$ is a conserved (dimensionless) energy parameter defined by 
\be
\frac{dt}{d\tau} = \frac{L^2E_t}{r^2} \,. 
\ee
The quantity
\be
c_x^2=\frac{E_{x_1}^2+E_{x_2}^2+E_{x_3}^2}{E_t^2} \,,
\label{c_x}
\ee
where 
\be
\frac{dx_i}{d\tau} = \frac{L^2E_{x_i}}{r^2} \,, 
\ee
defines the other conserved parameters and we set $E_y=0$.  Note that $dx_i/dt=E_{x_i}/E_t$ and therefore the projection of null geodesics onto flat (constant--$r$) slices follows straight lines.  Equation~\eqref{c_x} shows that $c_x$ represents the speed of light projected along the $x$--space.  In particular, light propagates along constant--$r$ slices with constant speed $c_x=1$, but a non vanishing component along the holographic direction implies $c_x<1$.

There are turning points at $r=r_0$ and at $r=r_i\equiv LE_t\sqrt{1-c_x^2}$.  In general the motion is bounded in the $r-$direction, and periodic.  For null or high-energy timelike particles, the (coordinate) time it takes for a roundtrip from $r=r_0$ to $r=\infty$ and back is
\be
{\cal P}^{\rm null}_{\rm roundtrip}=2\int_{r_0}^{\infty}\frac{L}{r\sqrt{F}\sqrt{1-c_x^2}}dr=\frac{2\sqrt{\pi}}{\sqrt{1-c_x^2}}\frac{ \Gamma\left(6/5\right)}{\Gamma\left(7/10\right)}\frac{L^2}{r_0}\sim \frac{2.507}{\sqrt{1-c_x^2}} \frac{L^2}{r_0}\,,\label{tnull}
\ee
with $\Gamma(x)$ a factorial-Gamma function.  This implies a characteristic frequency of 
\be
\omega_{\rm null\,geodesics}=\frac{2\pi}{{\cal P}_{\rm roundtrip}}=\sqrt{\pi}\sqrt{1-c_x^2}\frac{\Gamma\left(7/10\right)}{\Gamma\left(6/5\right)} \frac{r_0}{L^2} \sim 2.506 \sqrt{1-c_x^2}\frac{r_0}{L^2}\,.
\ee
This frequency is also the frequency of high-energy timelike particles.  On the opposite end we have low-energy particles, which oscillate between $r=r_i$ and $r=r_0$ with a period (for $c_x^2=0$, $r_i\sim r_0$, $E_t\sim r_0/L$)
\be
{\cal P}^{\rm timelike}_{\rm roundtrip}\sim\sqrt{\frac{2}{5}}\frac{\pi L^3E_t}{r_0^2}\sim\sqrt{\frac{2}{5}}\frac{\pi L^2}{r_0}  \,,
\ee
and a frequency
\be
\omega_{\rm timelike}\sim\frac{\sqrt{10}}{E_t}\frac{r_0^2}{L^3}\sim \sqrt{10} \, \frac{r_0}{L^2}\,.
\ee
Note that the oscillation period does not depend on the amplitude of the oscillation, in the small amplitude limit.

\section{A toy model: scalar interactions}
\label{sec:scalar}

Although our final goal is to study gravity, let us first take a look at a simplified problem, that of scalar fields in the background~\eqref{soliton}.  We will see later that this problem shares many common features with the more involved gravitational case.  We will study the stability of the spacetime against scalar field perturbations, compute the field of static scalar charges, and finally let these collide.

The action of our  generic setup is
\be
S_{\rm scalar}=
-\frac{1}{8\pi} \int d^{6}z \,  \sqrt{-g}  \, g^{ab}
\partial_a\Phi\,\partial_b\Phi\, 
- M \int ds \sqrt{-\dot Z^2} - 
Q \int ds \sqrt{-\dot Z^2} \, \Phi \,,
\label{eq:action}
\ee
where $\dot Z^2 = g_{ab}\dot{Z}^{a}\dot{Z}^{b}$, $\dot Z^a = dZ^a/ds$, and we have adopted the somewhat arbitrary $1/8\pi$ normalization for the scalar. This theory describes a point particle of mass $M$ and worldline $Z^a(s)$ minimally coupled  with strength (charge) $Q$ to a massless scalar field $\Phi$.  Note that the factor of $\sqrt{-\dot Z^2}$ in the last term is necessary to make this term invariant under worldline reparametrizations. One important consequence of this is that the coupling to the scalar field vanishes in the ultra-relativistic limit $\dot Z^2 \to 0$. We will see manifestations of this fact in our results below.

\subsection{Stability and normal modes}
\label{sec:scalar_modes}

We start by understanding the stability of the vacuum spacetime when the particle source is absent. The evolution of the scalar is then described by the massless Klein-Gordon (KG) equation 
\be
\Box\Phi= 
\frac{L^2}{r^2} \, \eta^{\mu\nu}\partial_\mu \partial_\nu \Phi +   
\frac{L^2}{r^2} \frac{1}{f} \, \partial^2_y \Phi + 
\frac{L^4}{r^4} \, \partial_r \left( \frac{r^6}{L^6} \, f \, \partial_r \Phi \right) =0 \,.
\ee
This equation separates under the ansatz
\be
\Phi (t, x_i, y, r)
=e^{-i \omega t + i k_i x_i + i n_y \lqcd y}\, \Psi(r)\,,
\ee
with $n_y = 0,1,2...$ due to periodicity $y\sim y+ 2\pi/\lqcd$, and yields
\be
- \partial_r \left( \frac{r^6}{L^6} \, f \, \partial_r \Psi \right) +
\frac{r^2}{L^2} \frac{1}{f} \, n_y^2 \lqcd^2 \, \Psi = 
\frac{r^2}{L^2} \, (\omega^2 - k^2)\, \Psi  
\,,
\label{SL2}
\ee
or more explicitly
\be
r \Big( r^5-r_0^5 \Big)^2 \, \Psi''+
\Big( r^5-r_0^5 \Big) \Big( 6r^5-r_0^5 \Big) \, \Psi' +
L^4r^2 \Big[ \Big( r^5-r_0^5 \Big) \Big( \omega^2-k^2 \Big) - 
n_y^2 \lqcd^2  r^5 \Big] \, \Psi = 0 \,,
\label{eqKGmodes}
\ee
with $k^2=k_1^2+k_2^2+k_3^2$.  Using a new variable $\rho=r/r_0$ the above can be written in manifestly dimensionless format,
\be
- \partial_\rho \Big( \rho^6 f \, \partial_\rho \Psi \Big) + \frac{25 n_y^2}{4 f} \, \Psi = \rho^2\tilde \omega^2 \, \Psi \,,
\label{SL}
\ee
or equivalently
\be
\rho (\rho^5-1)^2 \, \Psi''+(\rho^5-1)(6\rho^5-1) \, \Psi'+\rho^2\left[ (\rho^5-1)\, \tilde{\omega}^2-\left(\frac{5 n_y}{2}  \right)^2 \rho^5\right]\, \Psi=0\,,
\ee
where the dimensionless quantity 
\be
\tilde{\omega}^2 = \frac{L^4}{r_0^2}(\omega^2-k^2) = 
\frac{25}{4 \lqcd^2} ( \omega^2 - k^2)
\label{omegatilde}
\ee
is the invariant four-dimensional mass $\omega^2 - k^2$ measured in units of the confinement scale $\lqcd^2$.

Generically, the point $\rho=1$ is a regular singular point of the ODE, and close to this point the dominant asymptotic behavior is of the form $\Psi \sim (\rho-1)^{\pm n_y/2}$.  Specializing to $n_y=0$ modes, the equation simplifies to
\be
\rho (\rho^5-1) \, \Psi''+(6\rho^5-1) \, \Psi'+ \rho^2 \tilde{\omega}^2 \, \Psi=0\,,\label{waveeqscalarfields}
\ee
At $\rho=1$ the solutions behave as 
\be
\Psi \sim C^{r_0}_1\log(\rho-1)+C^{r_0}_2 \,. 
\ee
At infinity (for generic $n_y$) they behave as 
\be
\Psi \sim C^{\infty}_1+ \frac{C^{\infty}_2}{\rho^5} \,.  
\ee
We require as boundary conditions that the solution be regular in the infrared, i.e.~that $C^{r_0}_1=0$, and that it be normalizable near the boundary, i.e.~that $C^{\infty}_1=0$. 

\begin{figure}[tb]
\begin{center}
\includegraphics[width=10.5cm,angle=0,clip=true]{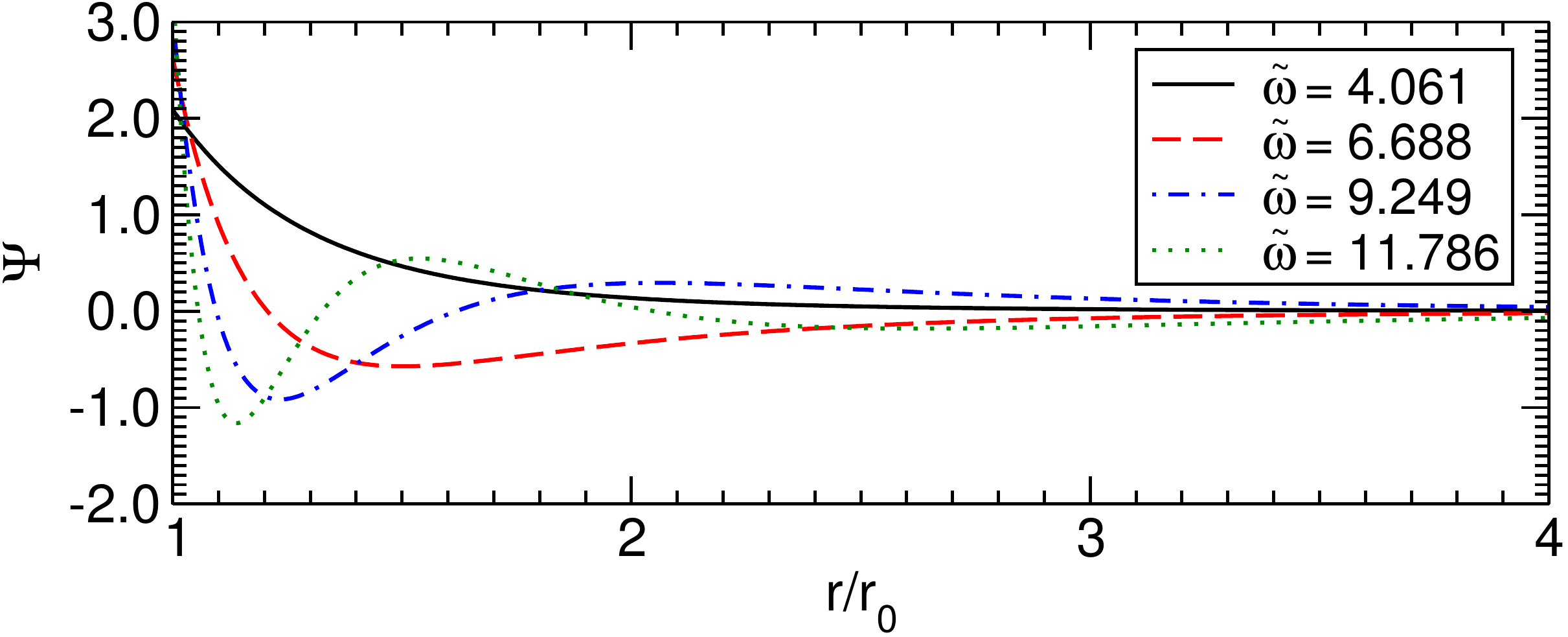}
\caption{Eigenfunctions [normalized according to the scalar product~\eqn{ortho}\label{fig:eigenscalar}] for the first four modes of the scalar field $\Psi$ listed in Table~\ref{tab:resonance}.}
\end{center}
\end{figure}

Equation \eqn{SL} together with these boundary conditions defines a Sturm-Liouville eigenvalue problem. The eigenvalues satisfy $\tilde \omega_n^2 > 0$ and physically they characterize the mass spectrum of scalar excitations in the gauge theory in the limit in which their possible mixing with higher-spin excitations is neglected, since we have ignored their possible mixing with e.g.~gravitational perturbations. We will denote the dimensionful eigenvalues in \eqn{SL2} as
\be
m_n^2 = (\omega^2 - k^2)_n > 0 .
\label{p}
\ee
In the case $n_y=0$, which will be our focus later, the corresponding eigenfunctions $\Psi_n$ satisfy
\be
\frac{L^4}{r^4} \partial_r \left( \frac{r^6}{L^6} \, f \, \partial_r \Psi_n \right) 
+ \frac{L^2}{r^2} \, m_n^2 \, \Psi_n = 0 \,,
\label{satisfy}
\ee
and are orthonormal with respect to the scalar product
\be
\int_{r_0}^\infty dr\, \frac{r^2}{L^2} \, \Psi_n(r) \, \Psi_m(r) = \delta_{mn} \,.
\label{ortho}
\ee 
By direct integration, with two independent codes, we find the modes in the first column of Table~\ref{tab:resonance}. The first eigenfunctions are shown in Fig.~\ref{fig:eigenscalar}. Note that the spectrum is discrete and gapped despite the fact that the radial direction is infinite, due to the fact that the AdS-soliton geometry acts like a `box'. As we will see, by expanding any function of the radial direction in the complete basis provided by these eigenfunctions, one may `Kaluza-Klein reduce' along the radial direction and reduce the problem to one dimension lower. 

The mass of the scalar mode with $n=0, n_y=0$ determines the mass gap of the gauge theory in the scalar channel as
\be
M_\mt{gap}^\mt{scalar} =  \frac{2 \sqrt{4.061}}{5}\, \lqcd \,.
\ee
We thus confirm that the mass gap is set by the Kaluza-Klein scale $\lqcd$, as anticipated above. In particular, this analysis implies that for 
$\omega<M_\mt{gap}^\mt{scalar}$ there can be no propagating mode of $\Phi$ in the geometry, as it would imply an imaginary wavenumber.  For very low frequencies the field is exponentially suppressed with distance; we will explicitly show this in Section~\ref{sec:scalar_collisions}.  These results are similar to those of waveguides in classical electromagnetism, as is the physical setting~\cite{Jackson}.

\begin{table}[tb]
\begin{center}
\begin{tabular}{cccccc}
\hline
\hline
&\multicolumn{2}{c}{$\tilde\omega_{\rm scalar}$}         &\multicolumn{3}{c}{$\tilde\omega_{\rm grav}$, $n_y=0$}\\ \hline
$n$       &   $n_y=0$            &$n_y=1$                    &vector I               & scalar  	& vector II \\
\hline 
0	&	4.061 	           & 5.700                   &	5.001               &	2.523		&	4.061\\
1		      &	6.688 	           & 8.198                   &	7.730               &	6.200		&	6.688\\
2		      &	9.249              & 10.694                  &	10.340 	            &	8.926		&	9.249\\
3		      &	11.786             & 13.192                  &	12.907              &	11.541	&	11.786\\
4		      &	14.312             & 15.692                  &	15.453              &	14.114	&	14.312\\
5	       	&	16.833	           & 18.194                  &	17.988              &	16.665	&	16.833 \\
6	      	&	19.349             & 20.697                  &	20.516              &	19.204	&	19.349\\
7	      	&	21.863             & 23.200                  &	23.039              &	21.735	&	21.863\\
8		      &	24.375             & 25.704                  &	25.558              &	24.261	&	24.375\\
38       	& 99.594             & 100.863                 & 100.826 		          &99.564	  &	99.594\\
39        & 102.100            & 103.369                 & 103.333		          &102.071	&	102.100 \\
40        & 104.606            & 105.875                 & 105.840		          &104.577	&	104.606\\
\hline
\hline
\end{tabular}
\caption{\label{tab:resonance} Resonances for Klein-Gordon modes (first two columns) and for vector and scalar gravitational modes (last three columns).  
Here $\tilde{\omega}^2 \equiv L^4(\omega^2-k^2)/r_0^2$.  Notice that for large overtone the spacing is roughly constant and equal to $\tw{n}-\tw{n-1}\sim 2.506$ for all fields.  
As we show in Sec.~\ref{sec:grav}, the second family of vector gravitational modes (vector II) is described by the same equation as in the Klein-Gordon case and therefore these two sets of modes exactly coincide. 
In the particle collisions we consider in Sec.~\ref{collision_grav}, only the gravitational scalar and vector-II modes are excited.
The spectrum of scalar gravitational modes agrees within numerical precision with the spectrum of scalar, $0^{++}$ glueballs found in \cite{appear} for the AdS$_6$ soliton.}
\end{center}
\end{table}

In Table~\ref{tab:resonance} we also present vector and 
scalar gravitational modes which are discussed in the next sections. Our results for scalar perturbations agree perfectly with those of Ref.~\cite{Haehl:2012tw} (see also~\cite{Constable:1999gb}).
For all families of modes and for large overtone number $n$, we find that the resonant frequencies are of the form $\tw{n}\sim \tw{0} + 2.506\, n$, with ${\tilde \omega}_{0}$ a field-dependent constant.  Notice that at large overtone, where the frequency is large and where the eikonal limit is valid, the geometrical optics regime of the wave equation should go over to the geodesic equation.  It is therefore pleasing to notice that the spacing of modes in this regime is
exactly (to within numerical precision) the same as the one predicted by a geodesic analysis.  This agreement can be made more formal by using the variable $\chi=1/\rho$, and the wavefunction 
\be
\Psi=\frac{\chi^2}{\sqrt{1-\chi^5}} \Theta\,.
\ee
The wave equation~\eqref{waveeqscalarfields} is now brought to the form
\be
\frac{d^2\Theta}{d\chi^2}+\tilde\omega^2 q \, \Theta=0\,,
\ee
with
\be
q=-\frac{-24+48\chi^5+ \chi^{10}}{4\chi^2{\tilde\omega^2}(1-\chi^5)^2}+\frac{1}{1-\chi^5}\,.\label{WKB}
\ee
A standard WKB analysis gives the asymptotic modes satisfying Dirichlet boundary conditions as~\cite{fedoryuk}
\be
\tw{n}=\pi n \left(\int_0^1\sqrt{q(t)}dt\right)^{-1}\,.
\ee
The quantity $q$ is to be evaluated at large $\tilde \omega$.  We then get a spacing of
\be
\pi \left(\int_0^1\sqrt{\frac{1}{1-t^5}}dt\right)^{-1}=\pi \left(\int_1^{\infty}\sqrt{\frac{1}{t^4(1-1/t^5)}}dt\right)^{-1}=\pi\left(\int_1^{\infty}\sqrt{\frac{1}{t^2F(t)}}dt\right)^{-1}\,.
\ee
A comparison with~\eqref{tnull} gives finally a spacing equal to $2\pi/{\cal P}_{\rm roundtrip}$, and the asymptotic relation is established.

\subsection{A static scalar charge in the AdS-soliton background}
\label{sec:static_scalar}

Let us consider now the solution of the KG equation with source at $x_i=X_i,\, r=b$ described by the action~\eqref{eq:action}.  Notice that, when $b>r_0$, this is in fact not a point particle, but rather a ring of matter that extends along the $y$-direction. Since this preserves the rotational symmetry along this direction,  only the $n_y=0$ mode will get sourced.  

The equation of motion is 
\be
{\square}\Phi =- 4\pi J 
= - \frac{4\pi Q L^3}{r^3}\delta(x_1-X_1)\delta(x_2-X_2)\delta(x_3-X_3)\delta(r-b)\,.
\ee
By Fourier expanding the field, 
\beq
\delta(x_i-X_i)&=&\frac{1}{2\pi}\int_{-\infty}^{+\infty}dk_i \, e^{ik_i(x_i-X_i)}\,,\\
\Phi(x_i,r)&=&\frac{1}{(2\pi)^3}\int_{-\infty}^{+\infty}dk_i \, e^{ik_i(x_i-X_i)}\,\Psi(k_i,r)\,, \label{Fourier}
\eeq
we get the following equation for the field of a static scalar charge,
\be
\frac{L^4}{r^4} \partial_r \left( \frac{r^6}{L^6} \, f \, \partial_r \Psi \right) 
- \frac{L^2}{r^2} \, k^2 \, \Psi =\frac{4\pi Q L^3}{r^3}\delta(r-b)\,,
\label{inho}
\ee
or equivalently 
\be
\frac{r(r^5-r_0^5)\Psi''+(6r^5-r_0^5)\Psi'-L^4r^2k^2\Psi}{L^2r^4}=\frac{4\pi Q L^3}{r^3}\delta(r-b)\,.
\label{eqscalarsource}
\ee
\noindent
There are several distinct but equivalent ways of solving this equation.  The standard procedure uses variation of parameters.  Let us introduce two linearly independent solutions, $\psione$ and $\psitwo$, of the homogeneous equations and their Wronskian ${\cal W}\equiv \psione\psitwo'-\psitwo\psione'$.
For our case
\be
{\cal W}(r)={\cal W}(r_1)\frac{r_1(r_1^5-r_0^5)}{r(r^5-r_0^5)}\,,
\label{wronskiano}
\ee
with $r_1$ an integration constant.
Now, all we have to do is define the homogeneous solutions $\psione,\,\psitwo$ such that 
\beq
\psione &\sim & \mbox{const.} \,,\quad r\to r_0\,, \label{eq:Psi1}\\
\psitwo &\sim& r^{-5} \,,\quad r\to \infty\,,\\
\psitwo &\sim& a_{\tilde\omega} \log(r-r_0)+c_{\tilde\omega}\,,\quad r\to r_0\,.\label{eq:aw}
\eeq

\noindent
The quantity $a_{\tilde \omega}$ is generically a function of $\tilde \omega^2$,  
and we provide a detailed characterization of it in Appendix~\ref{app:aomega}.  For the present, we are focusing on the case $\omega^2=0$ and thus in the static case $a_{\tilde \omega}=a_{\tilde \omega}(k)$. Note that $\psione$ is regular at $r=r_0$, that $\psitwo$ is normalizable, and that $\psitwo$ is singular at $r=r_0$ unless $a_{\tilde \omega}=0$. In this particular case $\psitwo$ is a regular and normalizable solution of the homogeneous equation. Recall that such solutions exist only for real and strictly positive values of $\omega^2 - k^2$. Since for 
$\omega=0$ this translates into strictly negative values of $k^2$, it follows that $a_{\tilde \omega}(k)$ has zeros at strictly imaginary values of $k$.
 
The solution can be written as
\be
\Psi= \left\{ 
\begin{array}{l}
A_k\psione(r)\,,\qquad r<b\,,\\
B_k\psitwo(r)\,,\qquad r>b\,.
\end{array}
\right.
\ee
Continuity at $r=b$ requires that $A_k\psione(b)=B_k\psitwo(b)$, and we can write
\be
\Psi= \left\{ 
\begin{array}{l}
C_k\psitwo(b)\psione(r)\,,\qquad r<b\,,\\
C_k\psione(b)\psitwo(r)\,,\qquad r>b\,.
\end{array}
\right.
\ee
Let us rewrite the KG equation as
\be
\frac{d}{dr}\left[r(r^5-r_0^5)\Psi'\right]-L^4r^2k^2\Psi
=4\pi L^5r Q\delta(r-b)\,.
\ee
Performing an integration on both sides from $b-\epsilon$ to $b+\epsilon$ we get
\be
C_k=\frac{4\pi Q L^5}{(b^5-r_0^5){\cal W}(b)}\,,
\ee
Finally,
\be
\Psi= \left\{ 
\begin{array}{l}
\frac{4\pi Q L^5}{(b^5-r_0^5){\cal W}(b)}\psitwo(b)\psione(r)\,,\qquad r<b\,,\\ \\
\frac{4\pi Q L^5}{(b^5-r_0^5){\cal W}(b)}\psione(b)\psitwo(r)\,,\qquad r>b\,.
\end{array}
\right.
\label{sol_scal_source}
\ee
This result is finite and continuous  everywhere, except in the limit $r\to b\to r_0$, since  in this case we have  (see Appendix \ref{app:aomega})
\be
{\cal W} \to \frac{a_{\tilde\omega} \psione(b)}{b-r_0} 
\ee
and therefore 
\be
\Psi(r) = \frac{4 \pi Q L^5}{5 a_{\tilde\omega}} \, \psitwo(r) \,.
\ee

\noindent
For $k=0$ the homogeneous equation can be solved exactly, with the result 
\be
\psitwo = \log\frac{r^5}{r^5-r_0^5} \sac \psione=1 \,. 
\ee
Thus $a_{\tilde\omega}(\tilde\omega=0)=-1$.  Furthermore, we find that at large $k$ the function $a_{\tilde\omega}$ increases exponentially (in absolute value), $a_{\tilde \omega \to-\infty} \sim -6.4\times 10^{-3}e^{-1.156\tilde \omega}$. 

A solution of the inhomogeneous equation \eqn{inho} can also be obtained by expanding $\Psi$ in the normal modes $\Psi_n$. Setting 
\be
\Psi (r) =  \sum_n c_n (k) \Psi_n (r) \,,
\label{exp}
\ee
substituting into \eqn{inho} and using \eqn{satisfy} we get
\be
\sum_n c_n (k) \, \Big(-m_n^2 -k^2 \Big)  
\,\frac{r^2}{L^2}\,  \Psi_n (r) =  
\frac{r^4}{L^4} \frac{4\pi Q L^3}{r^3}\delta(r-b)\,.
\ee
Multiplying both sides by $\Psi_m (r)$, integrating over $r$ and using the orthonormality conditions \eqn{ortho} we obtain
\be
c_n (k) = -\frac{b^4}{L^4} \, \frac{4\pi Q L^3}{b^3} \,
\frac{\Psi_n (b)}{m_n^2 + k^2} \,.
\label{coef}
\ee
We thus see that the coefficients in the expansion \eqn{exp} are proportional to the support of the corresponding wave function at the location of the particle. We also see that these coefficients have poles at (imaginary) values of $k$ determined by the mass spectrum of the normal modes. 
%

\subsubsection{An equivalent, matrix-valued Green function approach}
\label{sec:Green}

Here we discuss an equivalent approach, based on Green function techniques for coupled systems of ordinary differential equations (see e.g. Ref.~\cite{green_funct}).  This approach is advantageous because, as we shall discuss later, it can be directly extended to the gravitational case. 
Any second-order system of coupled ODEs can be written in a first-order form,
\be
\frac{d\mathbf{Y}}{dr}+\mathbf{V}\mathbf{Y}=\mathbf{S}\,,\label{system}
\ee
where $\mathbf{Y}$ and $\mathbf{S}$ are generically $n$ dimensional vectors and $\mathbf{V}$ is a $n\times n$ matrix.  We define the $n\times n$ matrix $\mathbf{X}$ whose $m$th column contains the $m$th solution of the homogeneous system $d\mathbf{x}/dr+\mathbf{V}\mathbf{x}=0$, i.e. $X_{ij}=x_i^{(j)}$, where the $j$ index denotes a solution of the homogeneous system and $i$ is the vector index.  The matrix $\mathbf{X}$ constructed in such a way is also a solution of the associated homogeneous system, in the sense that
\be
\frac{d\mathbf{X}}{d r}+\mathbf{V}\mathbf{X}=0 \label{homosyst}\,.
\ee
In order to solve~\eqref{system}, we impose the ansatz $\mathbf{Y}=\mathbf{X}\mathbf{\Xi}$, where $\mathbf{\Xi}$ is a vector to be determined.  Substituting this ansatz into the inhomogeneous system and using Eq.~\eqref{homosyst} we find
\be
 \frac{d\mathbf{\Xi}}{d r}=\mathbf{X}^{-1}\mathbf{S}\,,
\ee
and the solution to~\eqref{system} formally reads
\be
\mathbf{Y}=\mathbf{X}\int dr \mathbf{X}^{-1} \mathbf{S}\,.\label{gen_sol}
\ee

Let us now apply this method to Eq.~\eqref{eqscalarsource}.  In this case, $\mathbf{Y}\equiv (\Psi,\Psi')$ and 
\be
 \mathbf{V}= \left(\begin{array}{cc}
                    0 & -1 \\
                    -\frac{k^2 L^4 r}{r^5-r_0^5} & \frac{6 r^5-r_0^5}{r(r^5- r_0^5)}
                   \end{array}
\right)\,, \qquad 
 \mathbf{S}= \left(\begin{array}{c}
                    0\\
                    \frac{4 \pi Q L^5 \delta(r-b)}{r^5-r_0^5}
                   \end{array}
\right)\,.
\ee
Furthermore, using the same notation as in the previous section, the matrix of the homogeneous system reads 
\be
 \mathbf{X}= \left(\begin{array}{cc}
                    \psione &\psitwo \\
                   \psione' & \psitwo'
                   \end{array}\right)\,.
\ee
Evaluating the first component of Eq.~\eqref{gen_sol} we obtain
\be
 \Psi=4\pi Q L^5 \left[\psione(r)\int_r^\infty d\rbar\,\frac{\psitwo(\rbar)}{(\rbar^5-r_0^5){\cal W}(\rbar)}\delta(\rbar-b)+\psitwo(r)\int_{r_0}^r d\rbar\,\frac{\psione(\rbar)}{(\rbar^5-r_0^5){\cal W}(\rbar)}\delta(\rbar-b)\right]\,,
\ee
where the limits of integration were chosen in order to have the correct boundary conditions.
Finally, evaluating the expression above when $r<b$ and when $r>b$, we recover the same result as in Eq.~\eqref{sol_scal_source}.

\subsubsection{Yukawa-like potential at large distances}
\label{sec:Yukawa}
Let us now consider a point particle located at $r=b=r_0$ and $X_i=0$ and compute explicitly the large-$R$ behavior.  
From equation \eqref{eq:aw}, we get that the Wronskian between $\psione,\psitwo$ is ${\cal W}=a_{\tilde\omega}(k)/(r-r_0)$.
Using \eqref{sol_scal_source} we then find that for $r>r_0$,
\be
\Phi(x_i,r)=\frac{Q L^5}{10\pi^2 r_0^4}\int_{-\infty}^{+\infty}dk_i \, 
\frac{e^{i k_i x_i}}{a_{\tilde\omega}(k)} \, \psitwo(r)\,,
\ee
where the function $a_{\tilde\omega}$ depends only on $k=\sqrt{k_1^2+k_2^2+k_3^2}$.  At $r\sim r_0$, recall we have from Eq.~\eqref{eq:aw} that $\psitwo\sim c_{\tilde\omega}+ a_{\tilde\omega}\log(r-r_0)$ and so we get
\be
\Phi(x_i,r)=\frac{4\pi L^5}{5r_0^{4}}   Q \log(r-r_0)\delta(x_1)\delta(x_2)\delta(x_3)+\frac{Q L^5}{10\pi^2 r_0^4}\int_{-\infty}^{+\infty}dk_i\frac{c_{\tilde\omega}(k) e^{i k_i x_i}}{a_k(k)}\,.
\label{Psir0}
\ee
Thus, at leading order, the solution is localized in the radial direction. It is not our priority here, but it would be interesting to extract the coefficient $c_{\tilde\omega}$ defined in~\eqref{eq:aw}.  This coefficient would presumably dictate the small $R$, small $r_0$ dependence of the field away from the source location.

\begin{figure}[tb]
\begin{center}
\includegraphics[width=7.5cm,angle=0,clip=true]{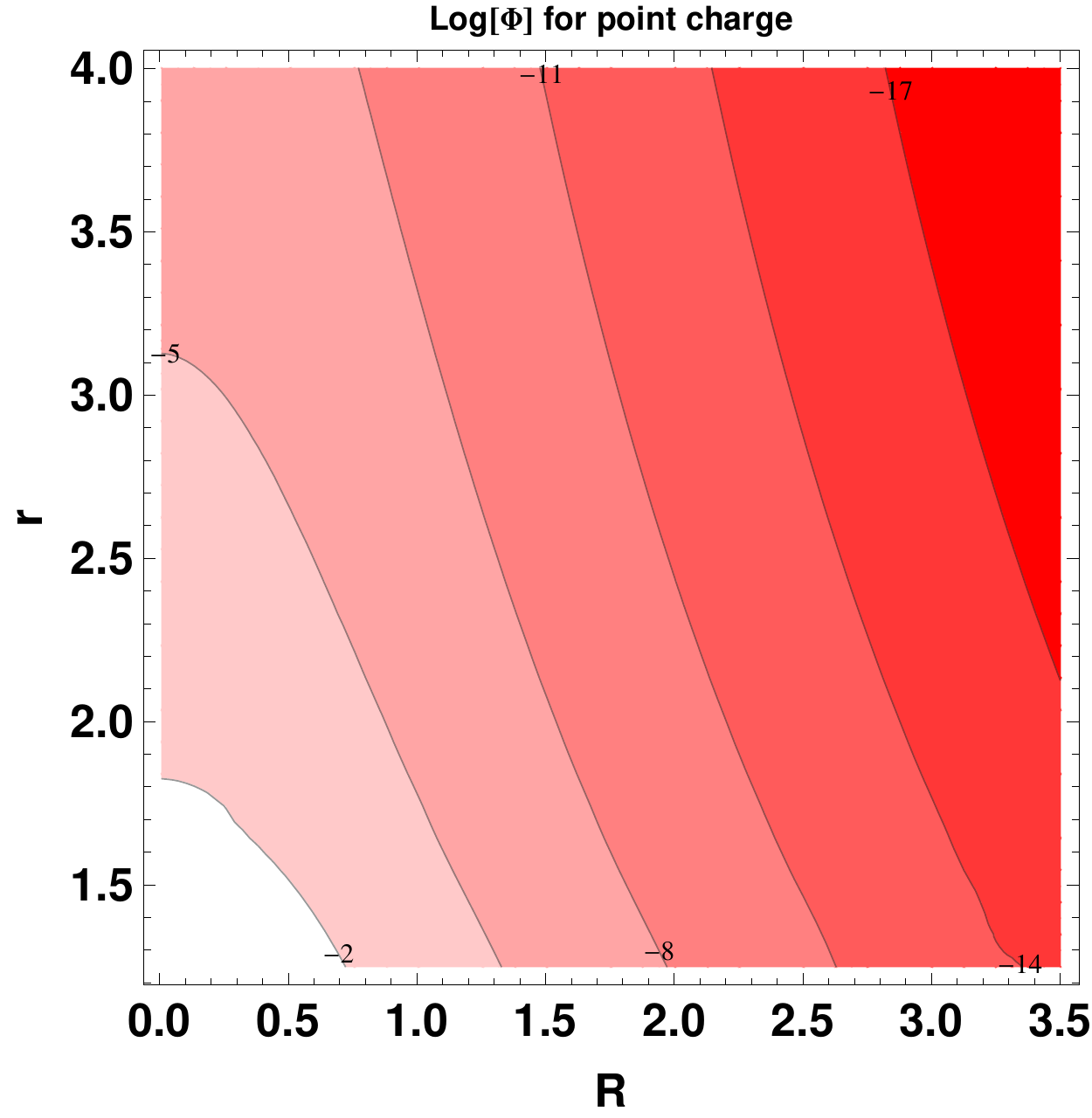}
\caption{Scalar field strength $\log\Phi(r,R)$ in the $R-r$ plane, with contour lines labeled by their respective magnitude. 
Notice how the field decays exponentially along the flat $R-$direction.
In this and in some subsequent plots we work with dimensionless units by setting $r_0=L=1$ and we have chosen, without loss of generality, $Q=1$.
\label{fig:scalarPPprofile}}
\end{center}
\end{figure}

\begin{figure*}[tb]
\begin{center}
\begin{tabular}{lr}
\includegraphics[width=7.4cm,angle=0,clip=true]{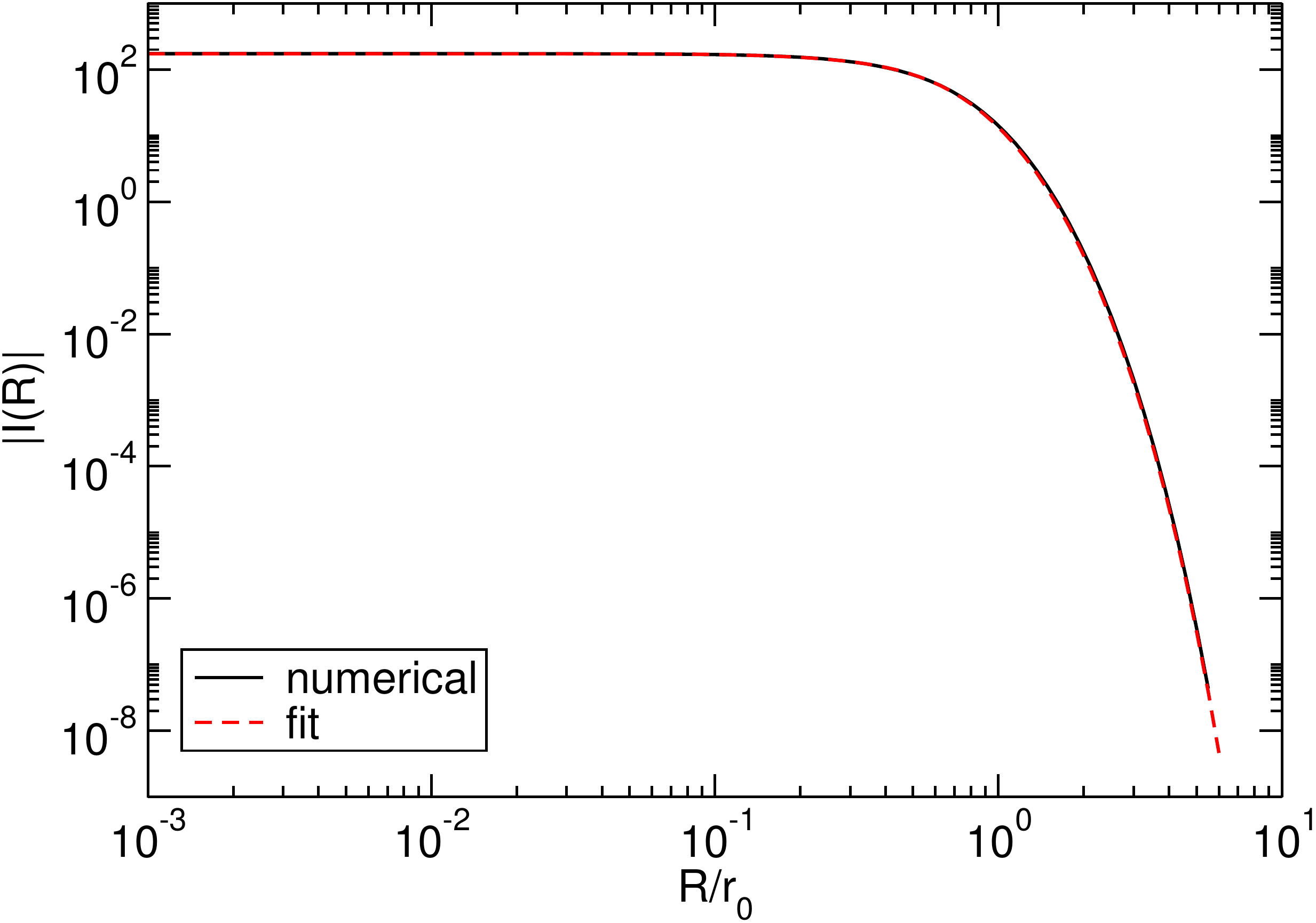}&
\includegraphics[width=7.4cm,angle=0,clip=true]{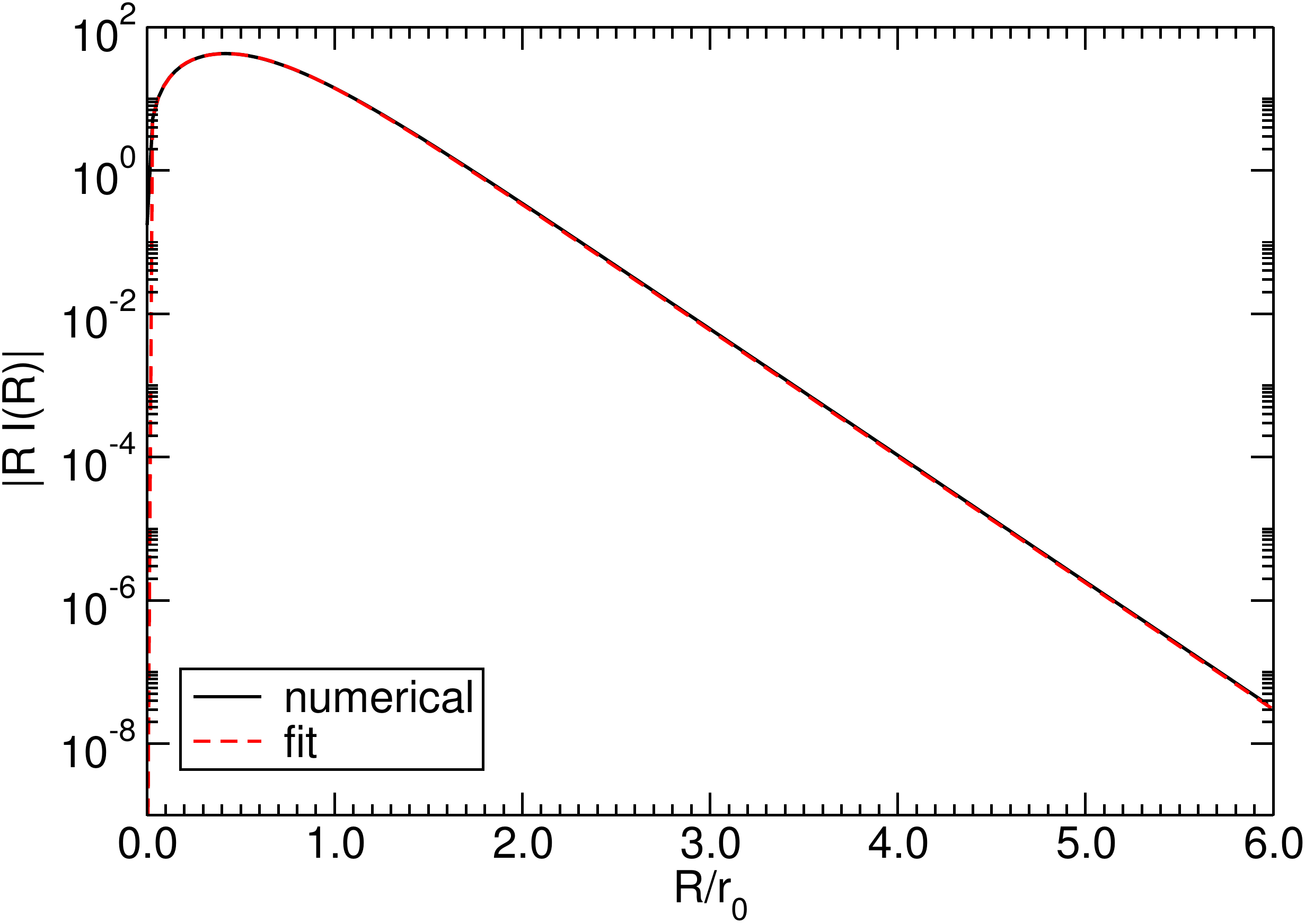}
\end{tabular}
\caption{Left: the integral $I(R)$~\eqref{integral} as a function of $R$ compared with a fit of the superposition of normal mode solutions~\eqref{scalar_superposition_normal} with $N=5$. Right: same for the quantity $R I(R)$. 
\label{fig:integral}}
\end{center}
\end{figure*}

Let us focus instead on the $r\to\infty$ regime, where $\psitwo\to 1/r^5$.  In this case, we get
\be
\Phi(x_i,r)=\frac{Q L^5}{10\pi^2 r_0^4\, r^5}I(x_1,x_2,x_3)\equiv \frac{Q L^5}{10\pi^2 r_0^4\, r^5}\int_{-\infty}^{+\infty}dk_i\frac{e^{i k_i x_i}}{a_{\tilde\omega}(k)}\,.
\label{pot_scalar_inf}
\ee
The integral above can be simplified,
\be
I(x_1,x_2,x_3)=I(R)=\frac{2\pi}{iR} \int_0^{+\infty}dk\, k\frac{e^{ikR}-e^{-ikR}}{a_{\tilde\omega}(k)}=\frac{2\pi}{iR} \int_{-\infty}^{+\infty}dk\, k\frac{e^{ikR}}{a_{\tilde\omega}(k)}\,,
\label{integral}
\ee
where $R=\sqrt{x_1^2+x_2^2+x_3^2}$ and we have used the fact that $a_{\tilde\omega}(k)$ is an even function of $k$.  Note that the values of $k$ corresponding to the normal modes of the system correspond to poles of the integral in the complex plane.  A contour plot of the scalar field strength $\Phi(r,R)$ in the $R-r$ plane is shown in Fig.~\ref{fig:scalarPPprofile}, whereas the the integral~\eqref{integral} is shown in Fig.~\ref{fig:integral} as a function of $R$. Note that the potential is finite as $R\to0$ and it decays exponentially for large values of $R$. 

Our numerical results are consistent with a Yukawa-like decay, $I\sim e^{-\mu R}/R$ (cf. right panel of Fig.~\ref{fig:integral}).  Indeed, the integral above can also be computed by deforming the integration contour in the complex plane and using the residue theorem.  We shall discuss this technique in detail in the following sections; here we just give the final result.  Using the fit~\eqref{fit_a_positive} to approximate the function $a_{\tilde\omega}(k)$ close to the poles, we obtain a sum of Yukawa-like potentials,\footnote{In this and in some subsequent expressions we work with dimensionless variables by setting $r_0=L=1$. We have also chosen $Q=1$ without loss of generality, since the scalar amplitudes for arbitrary $Q$ are simply proportional to $Q$.}
\be
I(R)\sim \sum_{n=0}^N \tilde c_n \frac{e^{-\mu_n R}}{R}\,,\label{scalar_superposition_normal}
\ee
where $\mu_n$ are the modes listed in the first column of Table~\ref{tab:resonance} and the coefficients $\tilde c_n$ read (cf. Section~\ref{sec:scalar_collisions} for details)
\be
\tilde c_n\approx\frac{2\pi}{3.9}{\cal P}(-1)^n\mu_n^{3.53}\,, \label{c_i}
\ee
where ${\cal P}=2\times 2.506$.  In Sec.~\ref{sec:vacuum_scalar} we shall confirm these results by obtaining the leading terms of Eq.~\eqref{scalar_superposition_normal} via a completely independent approach.  In Fig.~\ref{fig:integral} we compare the numerical results with the superposition of normal mode solutions~\eqref{scalar_superposition_normal} (cf.~also Section~\ref{sec:vacuum_scalar} below) truncated at $N=5$.

An asymptotic analysis of the integral $I$ in~\eqref{integral} at large $R$ confirms this Yukawa-like behavior: using a stationary phase approach, the relevant function to study is $f(k)=ka_{\tilde\omega}^{-1}$ at $k=0$~\cite{Wong}.  The asymptotic behavior at large $R$ is strongly dependent on the behavior of the even derivatives of $f(k)$ and in particular the existence of power-law behavior seems to be connected to non-zero even derivatives of this function.  As we show in Appendix~\ref{app:aomega}, the function $a_{\tilde \omega}$ is an even function of its argument for small enough argument.  Together with our numerical data, which is consistent with $f^{(2n)}=0$ at least for $n=0,2,4$, 
this kind of asymptotic analysis  also predicts what we find numerically, i.e.~a Yukawa-like suppression.

In fact, this behavior can be proven exactly by using the solution in the form \eqn{exp} with the coefficients \eqn{coef}. Through the Fourier transform \eqn{Fourier}, the poles in the coefficients produce precisely the Yukawa-like terms:
\be
\Phi (r,R) = \sum_n d_n \,  \Psi_n (r)  \, \frac{e^{- m_n R}}{R} \,,
\ee 
where  $d_n$ is proportional to the $k$-independent part of $c_n$. Note that this expression is valid for all values of $r, R$.

The Yukawa-like decay is not a peculiarity of having a point source.  Our results generalize to any compact distribution along the flat directions $x_i$, as long as the source is localized in the holographic direction.  In fact, the large distance behavior is controlled by the small-$\tilde \omega$ asymptotics, which is independent of how the source is distributed, as long as it is compact.  In subsection~\ref{sec:vacuum_scalar} below, we exhibit vacuum solutions not localized along the holographic direction and that also display a Yukawa-like decay.

\subsection{High-energy collisions of point particles}
\label{sec:scalar_collisions}

We now consider the collision of two scalar particles of equal mass moving towards each other at velocity $v$ along the flat directions, depicted in Fig.~\ref{fig:pictorial}.  For concreteness, we focus on two particles flying at each other along the $x_1$ axis and colliding at $t=0$.  Thus, we model the problem as
\be
{\square}\Phi= -4\pi \, J(t, x_i) \, \frac{L^3}{r^3} \, \delta (r-b) 
\label{eq:ZFl_scalar}
\ee
where
\bea
J(t, x_i) &=&- Q \, \gamma^{-1} \, \Theta(-t) 
\Big[\delta(x_1-vt)+\delta(x_1+ v t ) \Big]
\delta(x_2)\delta(x_3)
\nn \\ [1mm]
&&
-2 Q \, \epsilon_\mt{BH} \, \Theta(t)\,
\delta(x_1)\delta(x_2)\delta(x_3) \,, 
\label{eq:ZFl_scalar2}
\eea

\noindent
where $\gamma=1/\sqrt{1-v^2}$ is the relativistic boost factor, $\Theta(t)$ is the Heaviside function, and $\epsilon_\mt{BH}$ will be defined momentarily.

This is the Instantaneous Collision Framework or Zero Frequency Limit (ZFL) approximation described in the Introduction. It is well-motivated for the high-energy collision of two objects and is known to work well to describe classical processes in electromagnetism~\cite{Jackson,MadalenaThesis}.  Two objects flying at close to the speed of light barely feel each other's  field, and therefore the interaction takes place right at the moment of collision. Since we are eventually trying to describe the formation of a single black hole from the collision
of two objects, we let the final particle be at rest, as described by the second term on the right-hand side of \eqn{eq:ZFl_scalar}. 

Notice that already at this level a choice of the final state is crucial:  if the final state is charge-conserving then $\epsilon_\mt{BH}=1$.  If instead the final state is a black hole and black holes in this theory continue to have no hair, then the scalar charge of the final black hole is presumably zero, and 
$\epsilon_\mt{BH}=0$ in this case.  This technical detail yields a difference between a radiation output that scales as $\gamma^0$ if $\epsilon_\mt{BH}=1$ and a radiation that scales $\gamma^{-1}$ if $\epsilon_\mt{BH}=0$. The physical intuition behind this is that the radiation is the dislocation in the field created by the change in the source at $t=0$. For large $\gamma$ this change is of order unity if $\epsilon_\mt{BH}=1$ and of order $1/\gamma$ if $\epsilon_\mt{BH}=0$.

Let us proceed by Fourier analyzing the fields.  We expand any function ${\cal Z}$ as
\beq
{\cal Z}(t,x_i,r)&=&\frac{1}{(2\pi)^4}\int_{-\infty}^{+\infty}d\omega \int d^3k_i 
\, e^{-i\omega t} e^{ik_ix_i} \, {\cal Z}(\omega,k_i,r)\,, \label{fourier}\\ [2mm]
{\cal Z}(\omega,k_i,r)&=&\int_{-\infty}^{+\infty}dt\int d^3x_i \, e^{i\omega t} e^{-ik_i x_i} \, {\cal Z}(t,x_i,r)\,. 
\label{antifourier}
\eeq
In Fourier space,~\eqref{eq:ZFl_scalar} yields
\be
\frac{r(r^5-r_0^5)\Psi''+(6r^5-r_0^5)\Psi'+L^4r^2(\omega^2-k^2)\Psi}{L^2r^4}=\frac{r_0^4}{r^3 L^2} \, S(\omega,k_1) \, \delta(r-b) \,,
\label{finaleqscalarcollision}
\ee
where 
\be
S(\omega,k_1) = \frac{8\pi Q L^5}{i r_0^4} \left[
\frac{1}{\gamma^2} \, \frac{ \omega}
{(\omega-i \varepsilon)^2 - v^2k_1^2} - 
\frac{\epsilon_\mt{BH}}{\omega + i \varepsilon}
\right]  
\ee
and we have exhibited the appropriate $\varepsilon$-prescription, which we may not show explicitly in subsequent expressions.

With the same procedure as before, we define two homogeneous solutions $\psione, \psitwo$ by Eqs.~(\ref{eq:Psi1}--\ref{eq:aw}). The solution to the inhomogeneous problem can then be written as
\be
\Psi(\omega,k_i)= \left\{
\begin{array}{ll}
\frac{r_0^4\,S(\omega,k_1)}{(b^5-r_0^5){\cal W}(b)}\psitwo(b)\psione(r)\,,\qquad r<b\,,\\ \\
\frac{r_0^4\,S(\omega,k_1)}{(b^5-r_0^5){\cal W}(b)}\psione(b)\psitwo(r)\,,\qquad r>b\,.
\end{array}
\right.
\ee

\subsubsection{Reduction to four dimensions} 
We expand the five-dimensional field in the basis of normal modes as
\be
\Phi(t,x_i) = \sum_n \psi_n (t, x_i )\,  \Psi_n (r) \,.
\ee
Substituting into \eqn{eq:ZFl_scalar} and using \eqn{satisfy}, eq.~\eqn{eq:ZFl_scalar} becomes
\be
\sum_n \Big[ \left( \eta^{\mu \nu} \partial_\mu \partial_\nu - m_n^2 
\right) \psi_n (t, x_i ) \Big] \,  \Psi_n (r) = 
-4\pi \, J(t, x_i) \, \frac{L}{r} \, \delta (r-b) \,.
\ee
Multiplying by $(r^2/ L^2) \Psi_m (r)$, integrating over $r$ and using \eqn{ortho} we arrive at an infinite set of independent equations, one for each mode:
\be
\left( \eta^{\mu \nu} \partial_\mu \partial_\nu - m_n^2 
\right) \psi_n (t, x_i ) = -4\pi \, j_n (t, x_i) \,,
\ee
where
\be
 j_n (t, x_i) = \frac{b}{L} \, \Psi_n (b) \, J(t, x_i) \,.
 \ee
Thus the five-dimensional problem reduces to an infinite set of identical four-dimensional problems, each of them consisting of the determination of the massive scalar field generated by a source proportional to $J(t, x_i)$. Each of these problems is a classical bremsstrahlung problem in which two particles moving in opposite directions collide and come to a complete stop, thus emitting radiation into the massive scalar field that they couple to.

\subsubsection{Cutoffs \label{sec:cutoff}}

As we discussed in the introduction, in our approximation the spectrum needs to be cut off at high frequencies in order to yield sensible results for the total radiated energy and time-domain signals. While in the case of gravitational interactions one expects that non-linear effects, through black hole formation, dynamically introduce such a cutoff, in the case of scalar interactions the origin of the cutoff is less well defined; it would naturally depend on properties of the final object such as, possibly, its size scale. If this is larger than $1/\lqcd$, the argument in sec.~\ref{subsec:ZFLAdS} extends to this case and the frequency cutoff is set by $\lqcd$. At any rate, these details are of not of much interest to us as we are taking this calculation as a toy model for gravitational interactions and, if needed, we may simply borrow the cutoff from a gravitational estimate.

\subsubsection{High-energy collisions, conserved scalar charge
\label{sec:chargeconserving}}

In the large-velocity limit and assuming that the final state has conserved charge ($\epsilon_\mt{BH}=1$), the source function $S$ does not depend on $k_1$, and the process is spherically symmetric in the flat directions at leading order.  This is a manifestation of the fact that the source for the scalar field vanishes in the ultra relativistic limit, as mentioned above. Thus in this limit one is left with the spherically symmetric field sourced by the `sudden appearance' at $t=0$ of a point-like source at rest, namely by the second line of  \eqn{eq:ZFl_scalar}. 

For large holographic coordinate $r$ the field takes the form (see Eq.~\eqref{pot_scalar_inf})
\be
\Phi(\omega,R,r) \sim \frac{{4\pi}S(\omega)}{(2\pi)^35r^5}\int_{0}^{+\infty}dk\,k\frac{\sin kR}{R\, a_{\tilde \omega}} \;\longrightarrow \; \frac{{2}Q L^5}{5\pi r_0^4\,\omega R\,r^5}\int_{-\infty}^{+\infty}dk\,k\frac{e^{ik R}}{ a_{\tilde \omega}}\,. \label{intsphersymm}
\ee
Our final goal is to extract the stress-energy tensor of the dual gauge theory from the asymptotic behavior of the metric perturbations; we shall tackle that problem in Section~\ref{sec:grav}.  For now we are considering only a massless scalar field on a non-dynamical AdS-soliton geometry.  To complete this warm-up exercise we can determine the expectation value of the dual scalar operator, which is proportional to the coefficient of the $r^{-5}$ term~\cite{deHaro:2000xn}:
\be
 \langle{\cal O}(\omega,R)\rangle=\frac{{2}Q L^5}{\pi r_0^4 \omega R}\int_{-\infty}^{+\infty}dk\,k\frac{e^{ik R}}{ a_{\tilde \omega}}\,.
\ee

We will use contour integration to perform the integral, and for that we employ an extension of the standard approach to compute massive field propagators~\cite{bhabha,low}.  We start by estimating the residues of $1/a_{\tilde \omega}$.  Using the fit~\eqref{fit_a_positive}, we find close to the poles of this function that
\be
a_{\tilde \omega}\sim \frac{(-1)^{n} 2\pi}{{\cal P}}\times 3.9\,{\tw{n}}^{\;-2.53}\left({\tilde \omega}-\tw{n}\right)\equiv f_n \left({\tilde \omega}-\tw{n}\right) \,,\qquad {\tilde \omega}\to{\tilde \omega}_n\,, \label{anpoles}
\ee
where 
\be
\tw{n}\sim 5.25{\cal P}/(2\pi)+n{\cal P}/2 \sac {\cal P}=2\times 2.506 \,.
\ee
Let us focus on the $k$-integral, which has poles at
\be
k=\pm \sqrt{(\omega +i\epsilon)^2-L^4\tw{n}^2/r_0^2}\sim \pm\left(\sqrt{\omega^2-L^4\tw{n}^2/r_0^2}+i\epsilon' \omega \right)\,,
\ee
where $\epsilon'$ has the same sign as $\epsilon$. 
Because we intend to close the contour on the upper half plane, only poles in the upper half plane contribute.  These are located at
\bea
k_n=\left\{
\begin{array}{l}
\sqrt{\omega^2-L^4\tw{n}^2/r_0^2} \qquad\;\;\; \textrm{if} \;\; \omega>L^2\tw{n}/r_0\,,\\ [2mm]
-\sqrt{\omega^2-L^4\tw{n}^2/r_0^2} \qquad \textrm{if} \;\; \omega<-L^2\tw{n}/r_0\,, \\ [2mm]
i\sqrt{L^4\tw{n}^2/r_0^2-\omega^2} \qquad\;\; \textrm{otherwise}\,.
\end{array}
\right.
\eea
Finally, expressing
\be
{\tilde \omega}-\tw{n}\sim -\frac{L^4 k_n}{r_0^2\,\tw{n}}(k-k_n)\,,\label{polesak}
\ee
we obtain
\be
 \langle{\cal O}(\omega,R)\rangle= \frac{2iQ}{\pi r_0^2 }\sum_{n=0}^N\frac{(-1)^{n} \, {\cal P}}{3.9} \, \tw{n}^{3.53} \, \frac{e^{ik_n R}}{\omega R}\,. \label{Oscalar}
\ee
It is gratifying to recover the expected result that frequencies smaller than the effective mass are exponentially suppressed at large distances $R$.  On the other hand, for frequencies above the effective mass the field displays oscillatory behavior for large $R$.  The wavelength of these modes decreases as $\omega$ increases but each time the frequency crosses above a mode $\tilde{\omega}_n$ a new oscillatory term appears, with a wavelength that is shorter than the previous mode.

\begin{figure*}[tb]
\begin{center}
\includegraphics[width=10.5cm,angle=0,clip=true]{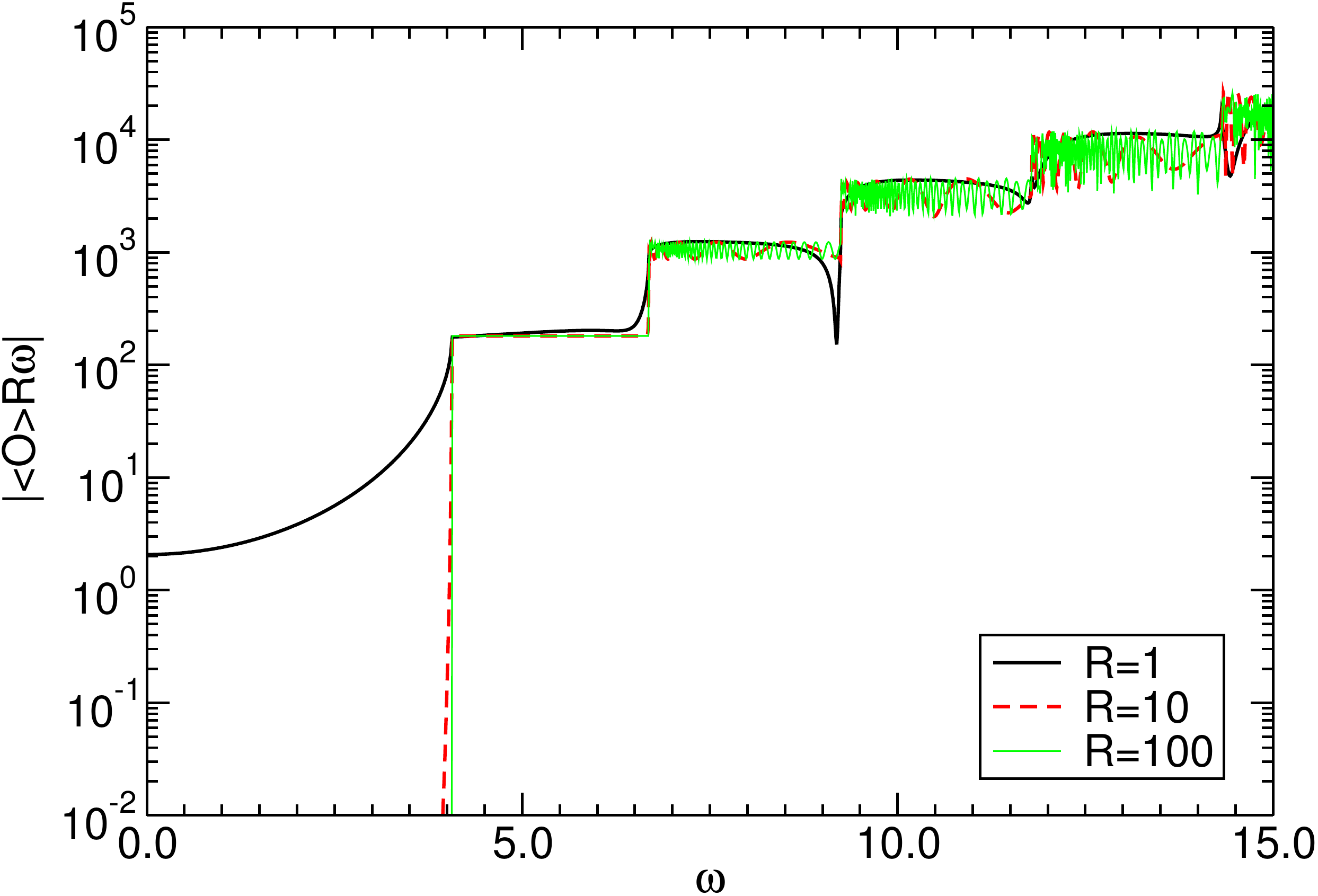}
\caption{The quantity $|\langle{\cal O}(\omega,R)\rangle| \omega R$ (modulo a constant factor proportional to $Q$) for $N=5$ in Eq.~\eqref{Oscalar}.  The effective mass term implies that the spectrum depends on the observation radius $R$ in the flat directions.  A common universal feature to all different observation points is the existence of plateaux and of a vanishingly small spectrum for frequencies smaller the the fundamental mode. 
\label{fig:spectrum}}
\end{center}
\end{figure*}

In Fig.~\ref{fig:spectrum} we show the quantity $|\langle{\cal O}(\omega,R)\rangle| \omega R$ (modulo a constant factor proportional to $Q$) as a function of the frequency. The different effective masses show up as plateaux in the spectrum, and the zero-frequency limit of the spectrum vanishes at large radii.  This is a universal solid prediction coming from this model.

\begin{figure*}[tb]
\begin{center}
\begin{tabular}{lr}
\includegraphics[width=7.4cm,angle=0,clip=true]{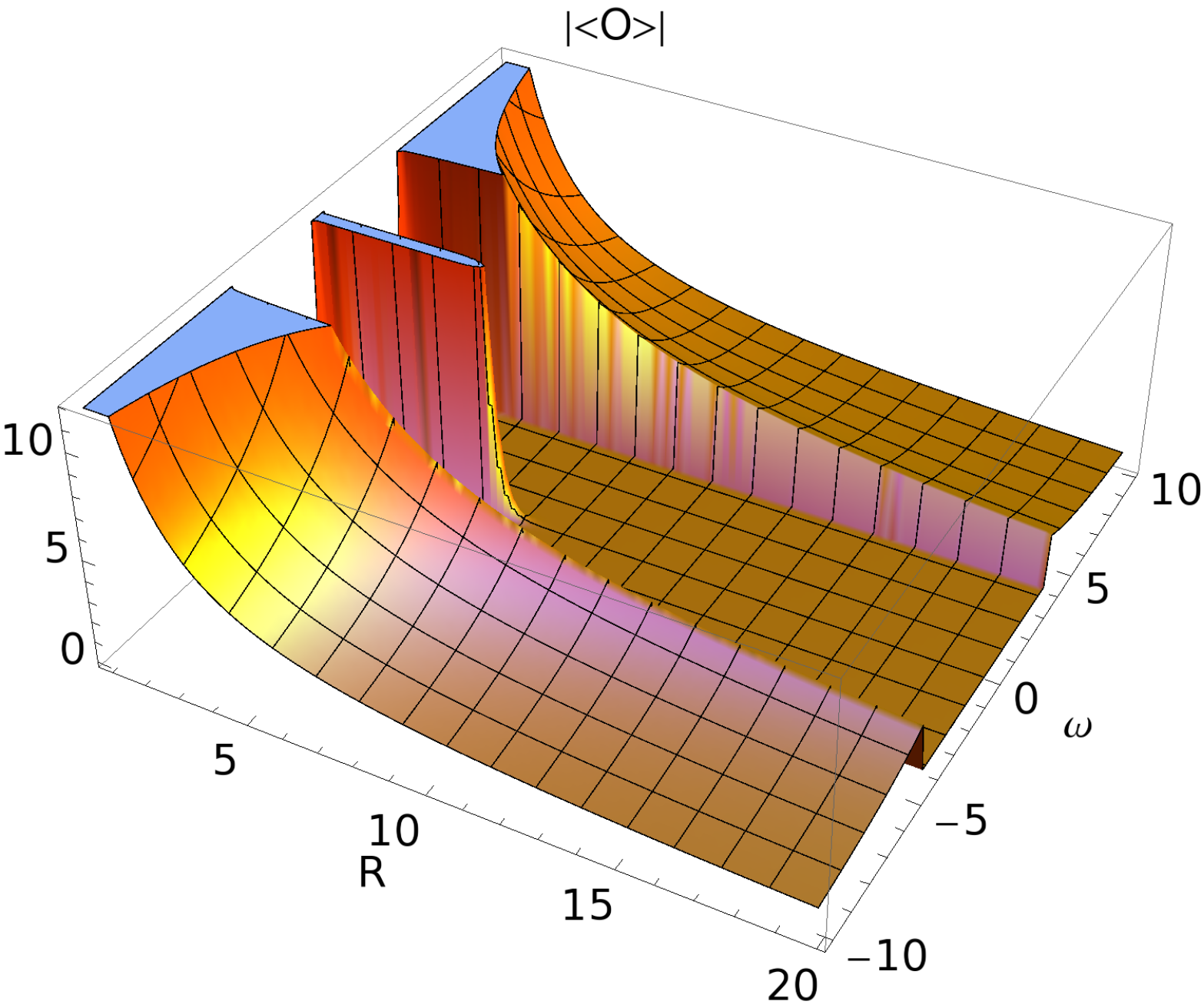} &
\includegraphics[width=7.4cm,angle=0,clip=true]{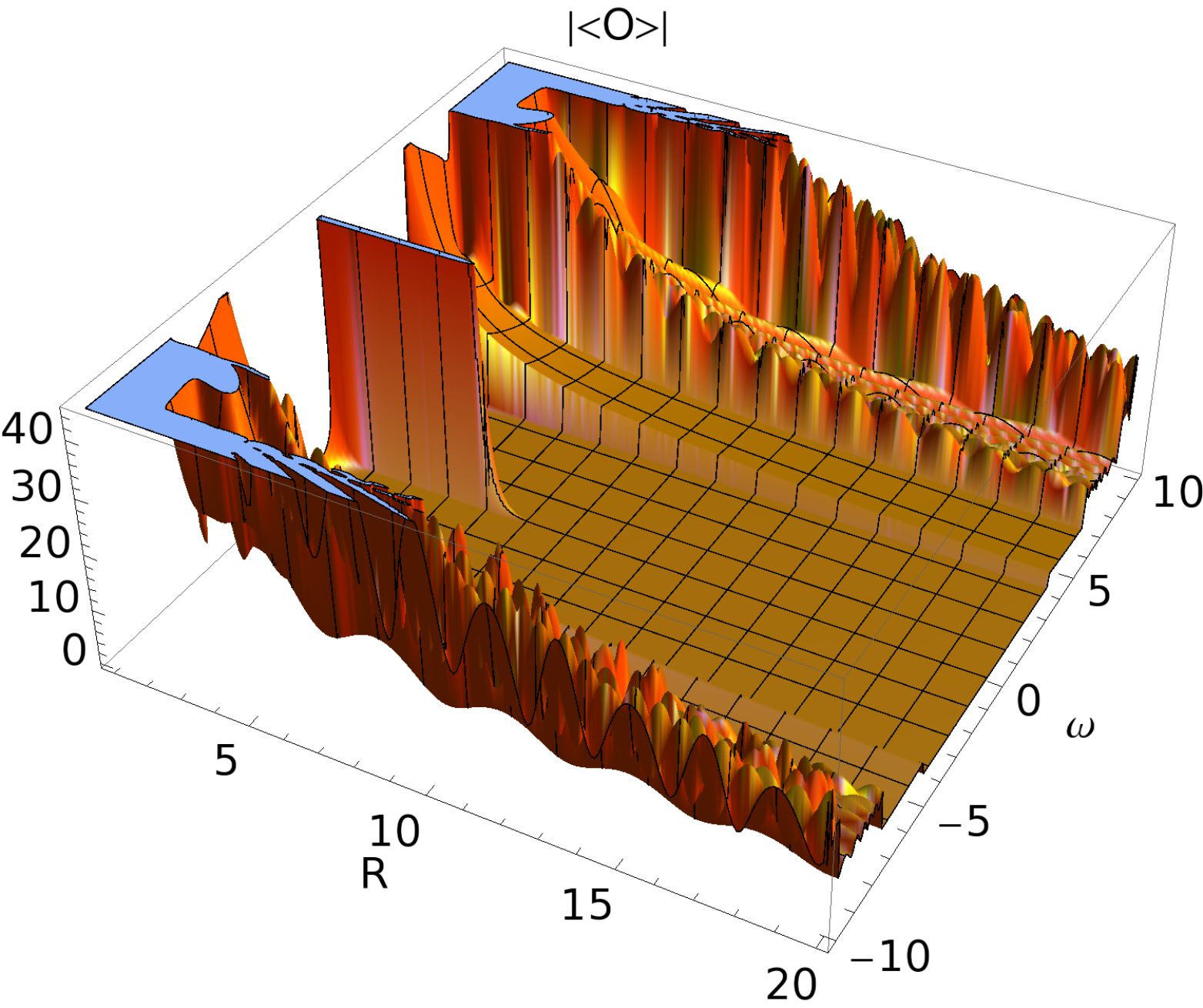}
\end{tabular}
\caption{The quantity $|\langle{\cal O}(\omega,R)\rangle|$, with the static part subtracted as in eq.~\eqn{subtracted}, as a function of $\omega$ and $R$ for $N=1$ (left panel) and $N=5$ (right panel), where $N$ is the number of massive modes included in Eq.~\eqref{Oscalar}.
\label{fig:scalar_3D}}
\end{center}
\end{figure*}

In Fig.~\ref{fig:scalar_3D} we show the full dependence of $|\langle{\cal O}(\omega,R)\rangle|$ on $\omega$ and $R$. As expected, the details of $\langle O \rangle$ depend on the number $N$ of modes included in Eq.~\eqref{Oscalar}. As $N$ increases, the waveform displays a complicated behavior due to the superposition of several massive modes. The various cutoffs for $\omega>\omega_n$ are also evident in the right panel of Fig.~\ref{fig:scalar_3D}.

\begin{figure*}[tb]
\begin{center}
\includegraphics[width=8.5cm,angle=0,clip=true]{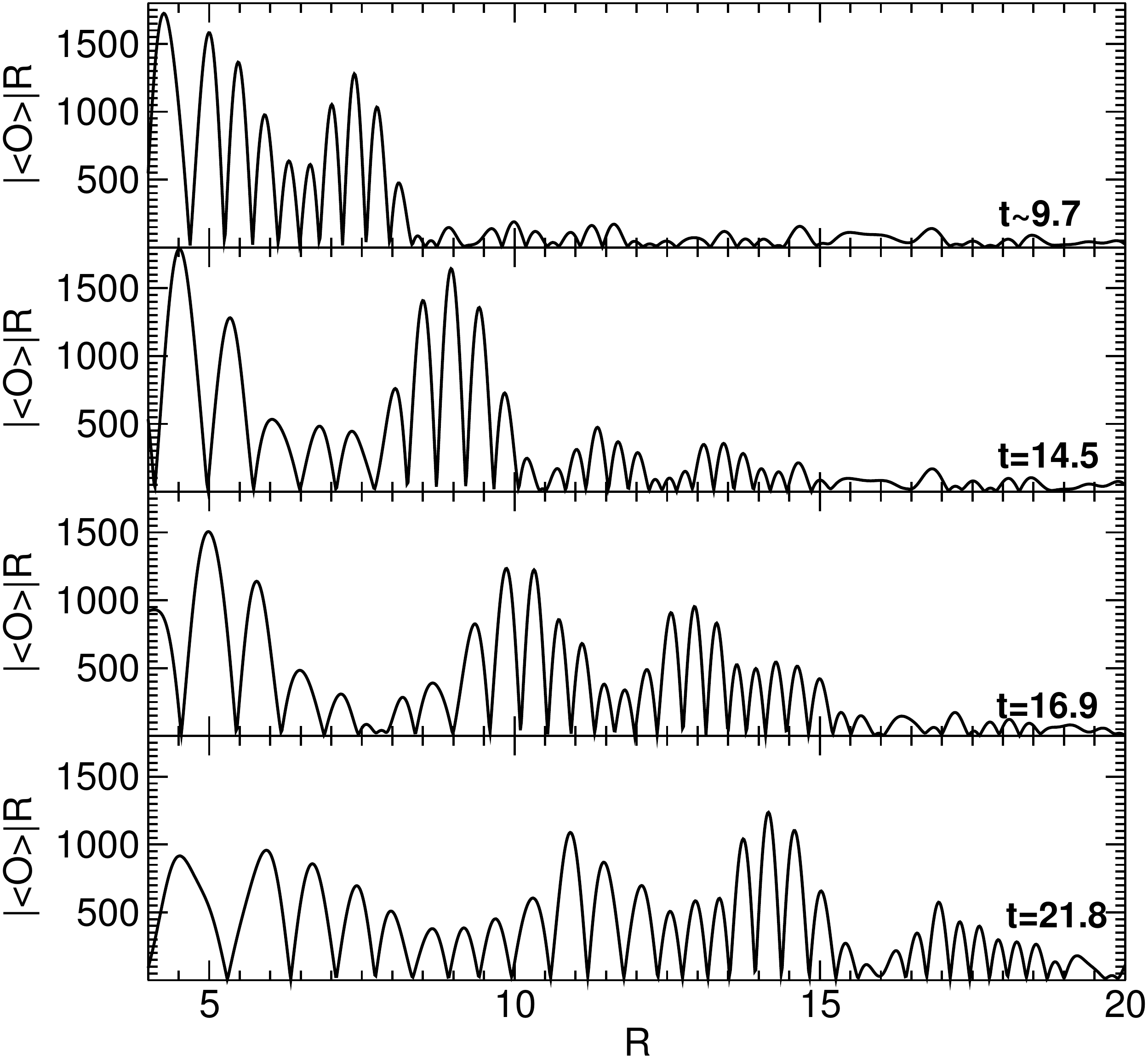}
\caption{The quantity $|\langle{\cal O}(t,R)\rangle|R$, with the static part subtracted as in eq.~\eqn{subtracted},  as a function of 
$R$ for $\omega_{\rm cutoff}\sim13$ and $N>3$. Each panel displays a 
different time snapshot.
\label{fig:scalar_time}}
\end{center}
\end{figure*}

\begin{figure*}[htb]
\begin{center}
\begin{tabular}{cc}
\includegraphics[width=7.4cm,angle=0,clip=true]{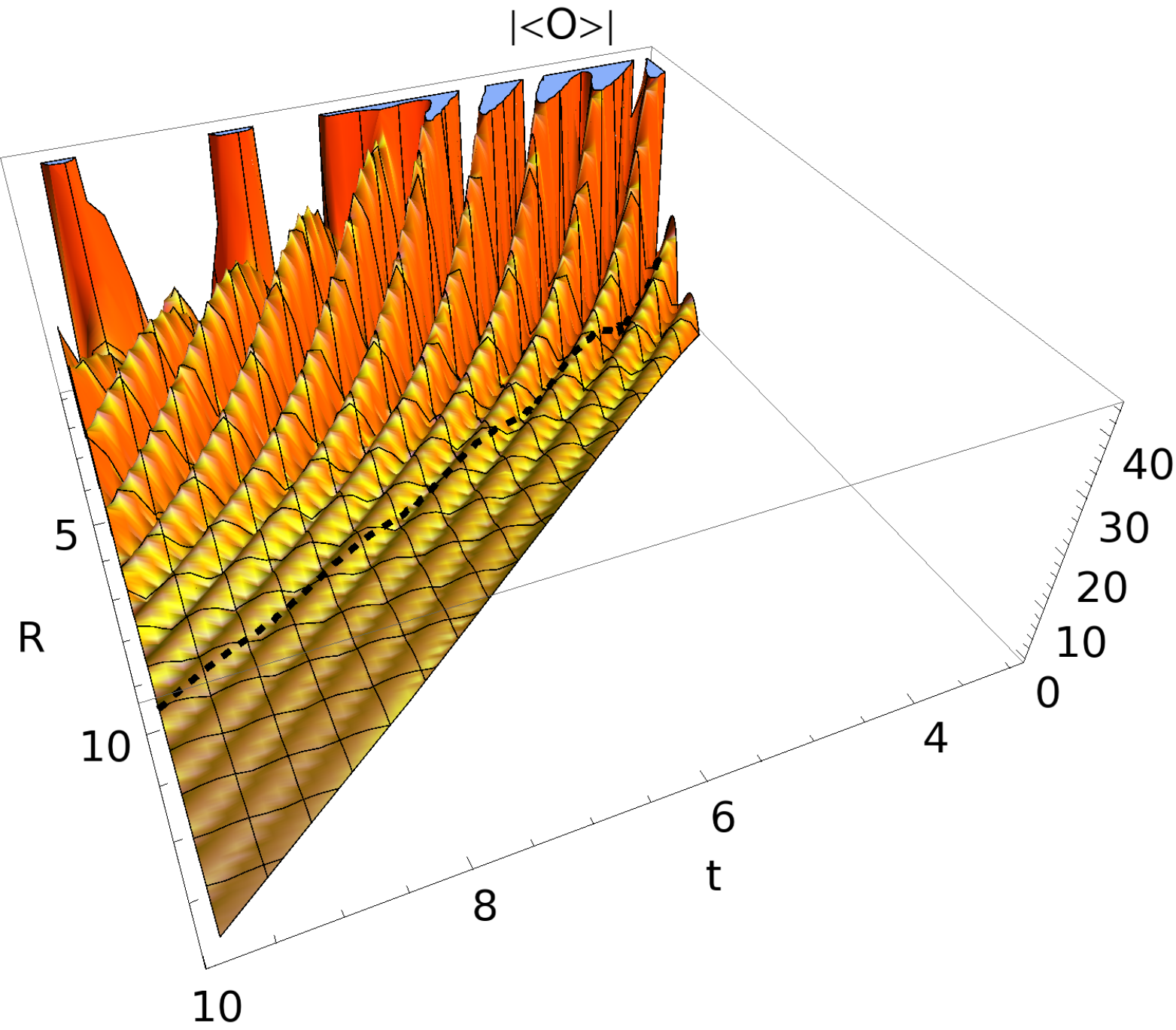}&
\includegraphics[width=7.4cm,angle=0,clip=true]{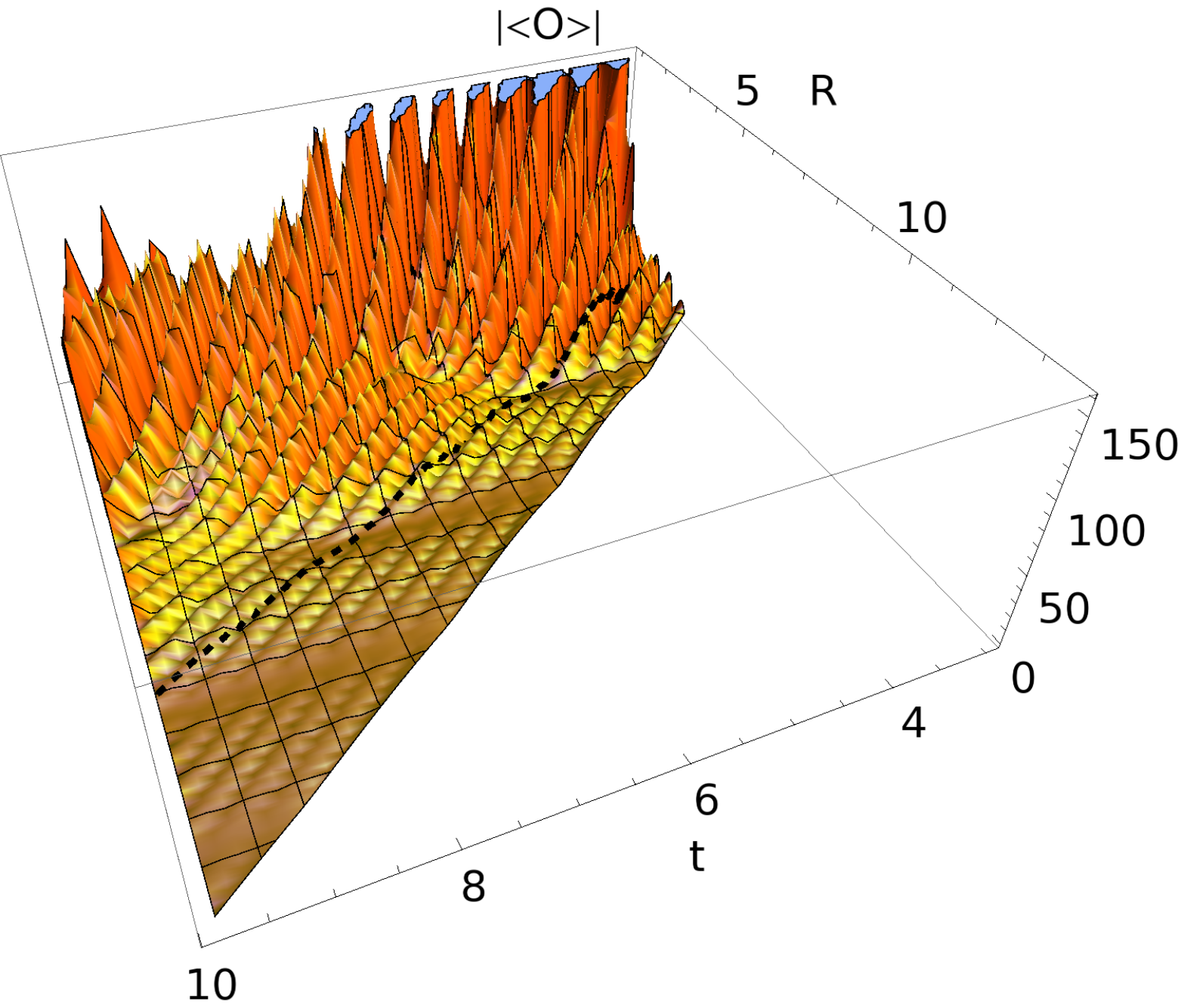}
\end{tabular}
\caption{The time-domain quantity $|\langle{\cal O}(t,R)\rangle|$ as a function of $t$ and $R$ and for a cutoff frequency $\omega_{\rm cutoff} \sim 5.4$ and $\omega_{\rm cutoff} \sim 8$.
The black dashed curve superimposed on each plot represents the time of arrival of the first signals that can reach the radius $R$ after the instantaneous collision. The locus of this line can be easily computed using Eq.~\eqref{tnull}. We get $t_{\rm arrival}(R) = \sqrt{(1.2535 L^2/r_0)^2 + R^2}$.
\label{fig:scalar_3D_time}}
\end{center}
\end{figure*}

With the frequency-domain quantities under control, we can Fourier-transform back to the time-domain and discuss the time evolution of the scalar operator.
Numerically, this is achieved by evaluating the inverse-Fourier amplitude of a generic function $\tilde{\psi}(\omega)$:
\be
 \psi(t_j)=\frac{1}{N}\sum_{k=0}^{N-1}\tilde\psi(\omega_k) 
e^{i\omega_k t_j}\,,\label{FFT}
\ee
where $\tilde\psi$ is evaluated at a fixed spatial position, $\omega_k=k \Delta \omega$ with $2\Delta \omega=\omega_{\rm max}/(N-1)$ and we assume a frequency domain $[-\omega_{\rm max},\omega_{\rm max}]$ discretized in $N$ equidistant points. The resolution in time is given by $2\pi/\omega_{\rm max}$, so that the larger the frequency domain the more refined is the resolution of the time evolution.

In Fig.~\ref{fig:scalar_time} we show several snapshots of $\langle O\rangle$ in the time domain as a function of $R$. Several pulses (corresponding to modes with different masses) propagate with different velocities. As expected from causality, the waveform must vanish when $R> t$. This is consistent with our results to within our numerical accuracy. 
Note that the details of the waveform depend on the frequency cutoff, but the qualitative behavior is generic. The larger the frequency cutoff, the larger the number of massive modes that can be excited and the waveform displays some beating effects. The full time-domain dependence of $|\langle{\cal O}(t,R)\rangle|$ is shown in Fig.~\ref{fig:scalar_3D_time} for different cutoff frequencies. 

\begin{figure*}[htb]
\begin{center}
\includegraphics[width=9.5cm,angle=0,clip=true]{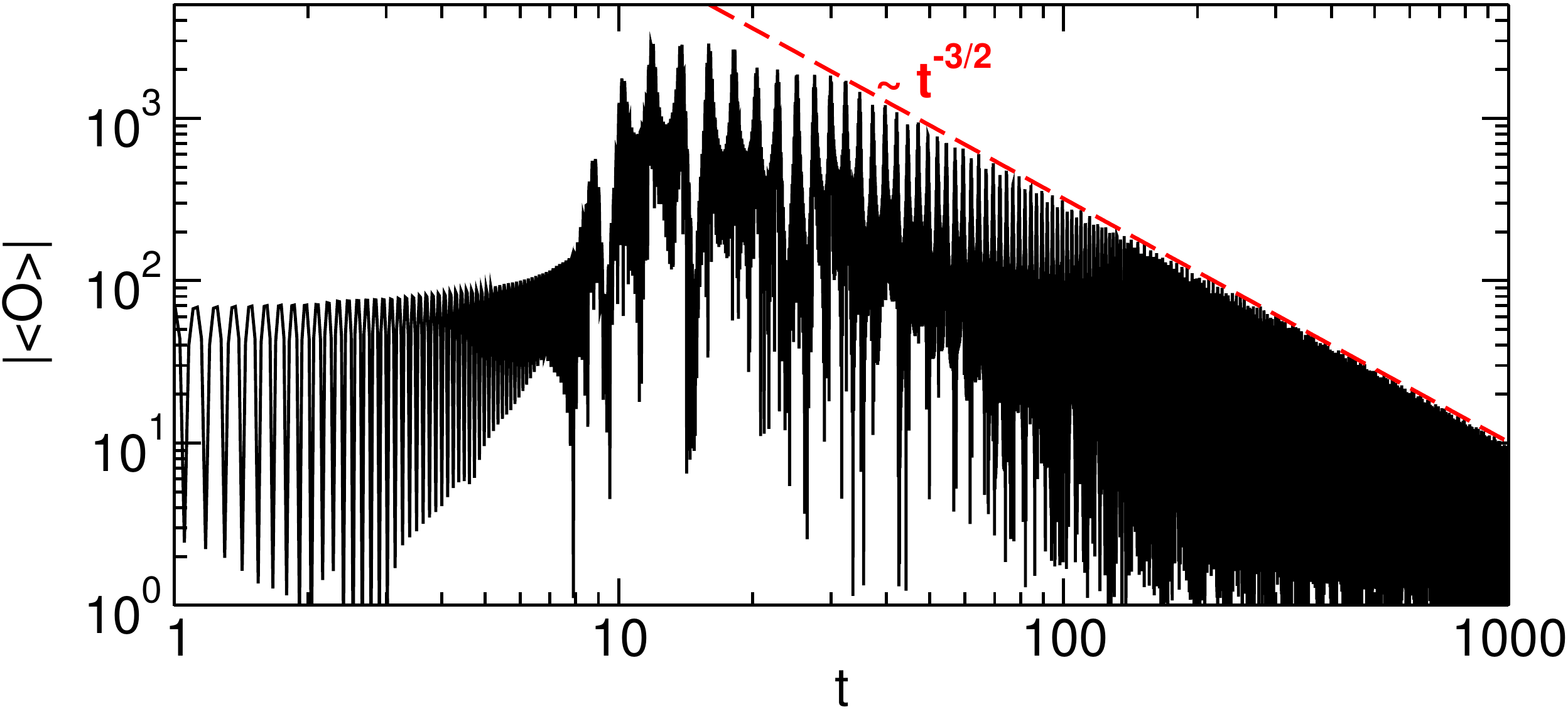}
\caption{The time-domanin quantity $|\langle{\cal O}(t,R)\rangle|$ as 
a function of $t$ for $R=8$, $\omega_{\rm cutoff}\sim25.6$ and $N>8$.
\label{fig:timeslope}}
\end{center}
\end{figure*}

Finally, there is a nontrivial dispersion relation, and different frequencies propagate at different speeds.  In other words, Huygens' principle is not satisfied and there is propagation inside the entire light cone~\cite{feshbach,Cardoso:2003jf}.  This peculiarity gives rise to a wake behind the main pulse, which dies off at late times as~\cite{feshbach}
\be
\langle{\cal O}\rangle - \langle{\cal O}_\mt{static}\rangle
\sim \sum_n\frac{\sin{(\tw{n}\,t\,r_0/L^2)}}{t^{3/2}}\,,\quad t\to \infty\,,
\label{subtracted}
\ee
where $\langle{\cal O}_\mt{static}\rangle$ is the static, Yukawa-like potential to which  $\langle{\cal O}\rangle$ asymptotes at late times.
The $t^{-3/2}$ fall-off can be proven analytically from the properties of the retarded Green's function of a massive scalar field in four dimensions. This takes the form
\be
G(t,x; t', x') = \theta(t-t') \left[ 
\delta(\sigma) - \theta(\sigma) \, 
\frac{m J_1 \left( m \sqrt{2 \sigma} \right)}{\sqrt{2 \sigma}} \right]   \,,
\ee
where 
\be
\sigma = \frac{1}{2} \left[ (t-t')^2 - (x-x')^2 \right] \,.
\ee
The delta function only contributes on the light-cone, whereas the Bessel function contributes inside the light-cone. Because of this, the field generated by a particle at a point $p$ is the integral of the Green's function along the world line of the particle from the remote past to the latest time $t_\mt{ret} (t)$ from which the particle could causally affect $p$. For a particle that has been sitting at $x'=0$ forever, we can write the resulting field at time $t$ schematically as 
\be
\langle{\cal O}_\mt{static}\rangle (t) = \int_{-\infty}^{t_\mt{ret} (t)}G (t,t') dt' \,.
\ee
This of course yields the static Yukawa potential. In contrast, the first term on the left-hand side of \eqn{subtracted} is only sourced from $t'=0$, so the difference in that equation is
\be
\langle{\cal O}\rangle (t) - \langle{\cal O}_\mt{static}\rangle(t) =
-\int_{-\infty}^{0} G (t,t') dt' \,.
\ee
We thus see that for $t\to \infty$ this difference is generated at times that are in the far past of the point of observation. Consequently, they are controlled by the fall-off at large $\sigma$ of the Bessel function, which is 
\be
\frac{m J_1 \left( m \sqrt{2 \sigma} \right)}{\sqrt{2 \sigma}} \simeq
 \frac{\sqrt{m}}{\left( 2\sigma \right)^{3/4}} \sim \frac{1}{t^{3/2}}\,.
\ee
Figure~\ref{fig:timeslope} shows that even when the signal includes several modes, the tail of the waveform in time is precisely $t^{-3/2}$ as predicted by the formula above.  
%

\subsubsection{High-energy collisions, final state with no scalar charge}
\label{sec:scalar_collisions_2}

In this case, $\epsilon_\mt{BH}=0$, and subleading terms have to be taken into account.  At large distance $r$ and for $b\sim r_0$ we have
\be
\Psi(\omega,k_1,k_{\rho})\sim\frac{S(\omega,k_1)}{5r^5a_{\tilde \omega}}\,,
\ee
where we defined $k_{\rho}^2=k_2^2+k_3^2$, and
\be
S(\omega,k_1)\equiv -\frac{8\pi Q L^5 i\omega}{r_0^4E_t(\omega^2-v^2k_1^2)}\,,
\ee
where $E_t = \gamma r_0/L$.
Finally, introducing cylindrical coordinates\footnote{Here, $\rho=\sqrt{x_2^2+x_3^2}$ should not be confused with the dimensionless holographic coordinate $r/r_0$ used in Sec.~\ref{sec:scalar_modes}.} we find, for large $r$,
\be
\Phi(\omega,x_1,\rho,r)\sim-\frac{8i\pi Q L^5\omega}{(2\pi)^2 r_0^4 E_t 5r^5}\int_{-\infty}^{+\infty} dk_1 \frac{e^{ik_1x_1}}{\omega^2-v^2k_1^2}\int_0^{+\infty}dk_{\rho}\; k_{\rho}J_0(k_{\rho} \rho)\frac{1}{a_{\tilde \omega}}\,,\label{intcylcoord}
\ee
with $J_0(x)$ a Bessel function of the first kind~\cite{Abramowitz:1970as}.  Using Eq.~\eqref{anpoles}, close to the poles we have 
\be
a_{\tilde\omega}\sim- \frac{f_n (k_1-k_{1,n})k_{1,n}L^4}{\tw{n}r_0^2} \,,
\ee
where 
\be
k_{1,n}^2=\omega^2-k_\rho^2 - 
\frac{r_0^2\,\tw{n}^2}{L^4} \,.  
\ee
By first integrating over $k_1$ we get
\be
\Phi(\omega,x_1,\rho,r)=\frac{8i\pi Q L^5}{(2\pi)^2 r_0^4 E_t 5\omega r^5}\, 2\pi i \, \int_0^{+\infty}dk_{\rho}\; k_{\rho}J_0(k_{\rho} \rho)\Psi(\omega,x_1,k_\rho)\,,\label{intkrho}
\ee
where we have defined
\be
 \Psi(\omega,x_1,k_\rho)=\frac{i\omega}{v\tilde{a}_{\tilde\omega}}\sin\frac{\omega}{v}x_1+
   \frac{r_0^2}{L^4}\sum_n \frac{\omega^2\,\tw{n}e^{ik_{1,n}x_1}}{f_n k_{1,n}\left[(1-v^2)\omega^2+v^2(k_\rho^2+r_0^2\,\tw{n}^2/L^4)\right]}\,,\label{Psi}
\ee
and $\tilde{a}_{\tilde\omega}$ is computed for $k_1^2=\omega^2/v^2$.  Therefore $\Psi(\omega,x_1,k_\rho)$ has poles at 
\be
k_\rho^2=-\frac{1-v^2}{v^2}\, \omega^2-\frac{r_0^2}{L^4}\, \tw{n}^2
\ee
coming from the first term of the equation above, and at 
\be
k_\rho^2=\omega^2-r_0^2\,\tw{n}^2/L^4 \sac 
k_\rho^2=\frac{v^2-1}{v^2}\omega^2-\frac{r_0^2\,\tw{n}^2}{L^4} \,, 
\ee
which come from the second term in Eq.~\eqref{Psi}.  In the ultrarelativistic limit, $v\to1$, both terms have poles at $k_\rho=\pm i r_0 \tw{n}/L^2$.  From Eq.~\eqref{intkrho}, we obtain
\be
\langle{\cal O}(\omega,x_1,\rho)\rangle=-\frac{2 Q L^5}{r_0^4 E_t \omega}\, I_2(\omega,x_1,\rho)\,,\label{intkrho2}
\ee
where 
\be
 I_2(\omega,x_1,\rho)\equiv \int_{-\infty}^{+\infty}dk_{\rho}\, k_{\rho}\, H_0^{(1)}(k_{\rho} \rho)\, \Psi(\omega,x_1,k_\rho)\,, \label{I2collisionscalar2}
\ee
and $H_n^{(1)}$ is the Hankel function of the first kind. 

\begin{figure}[tb]
\begin{center}
\begin{tabular}{lr}
\includegraphics[width=7.4cm,angle=0,clip=true]{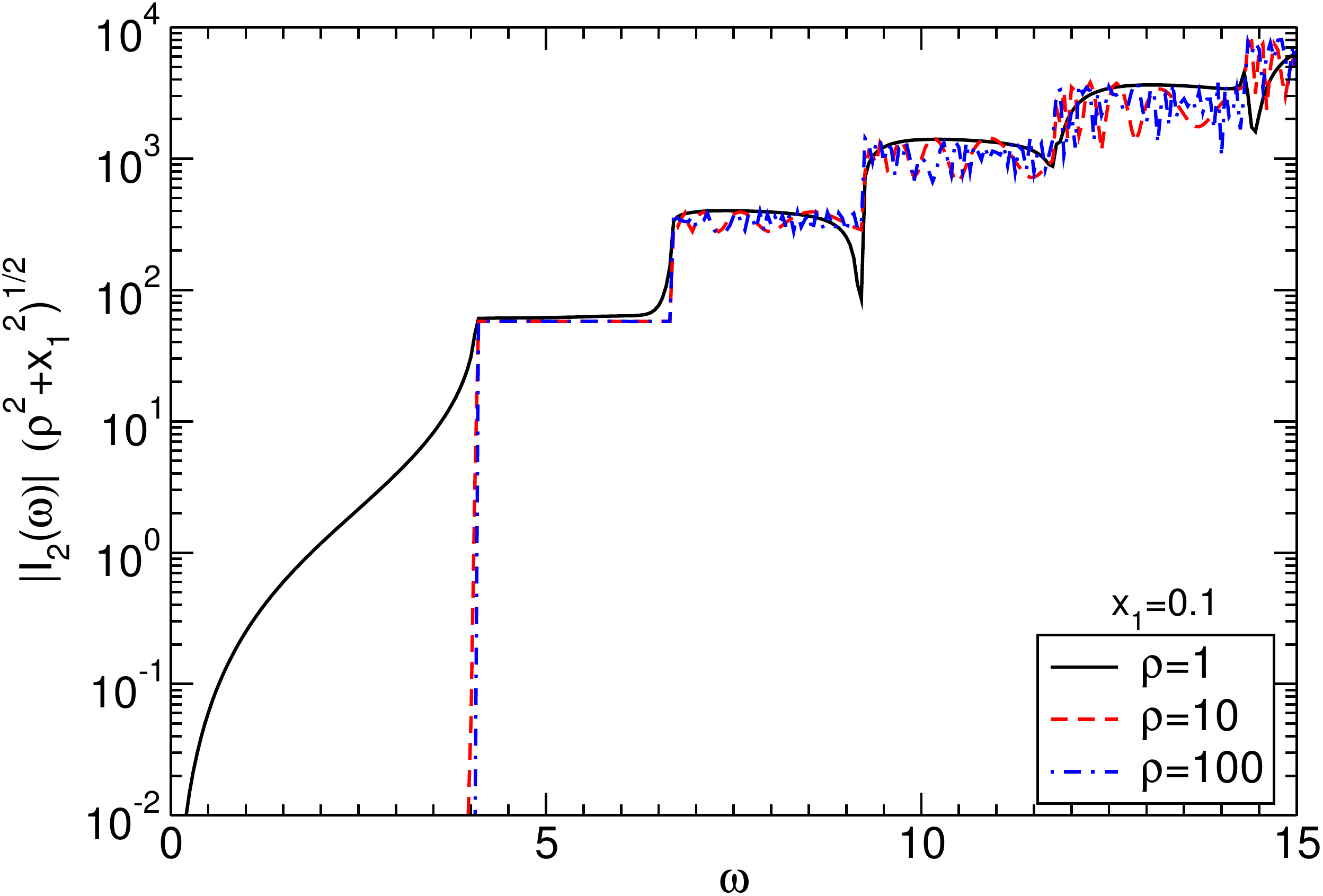}&
\includegraphics[width=7.4cm,angle=0,clip=true]{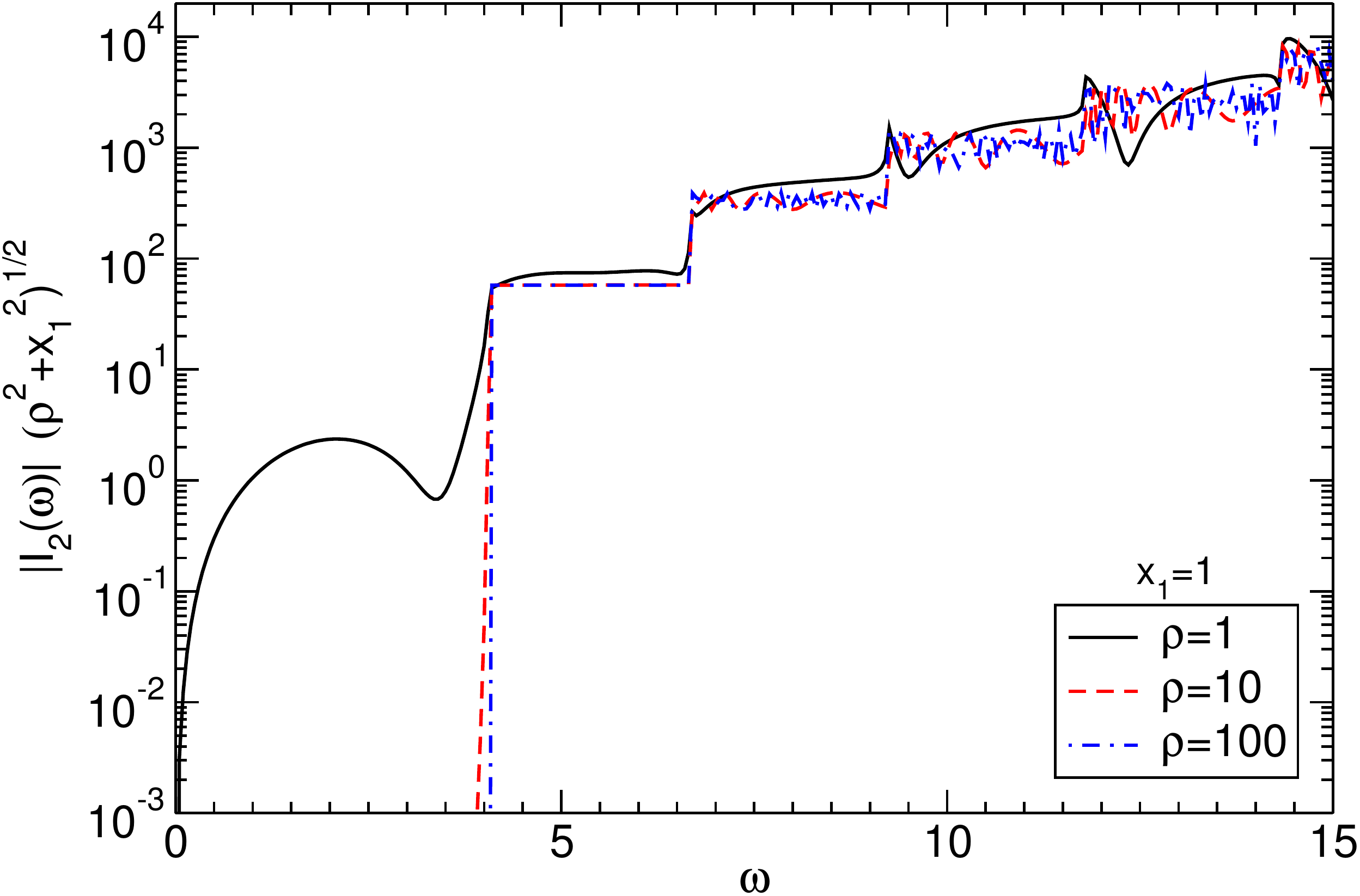}
\end{tabular}
\caption{\small The quantity $\omega\sqrt{\rho^2+x_1^2}\langle{\cal O}(\omega,x_1,\rho)\rangle$ (modulo a coefficient proportional to $Q/E_t$, cf. Eq.~\eqref{intkrho2}) in the case of black hole formation (no charge conservation) and for the case where the cutoff frequency encloses five resonant modes. The operator is not spherically symmetric and it depends on $x_1$ and $\rho$. Left panel: $x_1=0.1$, Right panel: $x_1=1$. Similarly to the spherically symmetric case, the spectrum is exponentially suppressed at large values of $\rho$ when $\omega$ is smaller than the fundamental frequency, $\tw{1}\sim4.062$. When $\omega>\tw{1}r_0/L^2$, the spectrum shows an approximate scaling as $(\rho^2+x_1^2)^{-1/2}$.
\label{fig:spectrum_scalar2}}
\end{center}
\end{figure}

Let us solve the integral above in the ultrarelativistic limit, $v\to1$.  We split it into two contributions, $I_2=I_2^{(a)}+I_2^{(b)}$, accordingly to the two terms in Eq.~\eqref{Psi}.  The first term has poles at 
$k_\rho=\pm i r_0 \tw{j}/L^2$ and, by using the residue theorem, the integral is equal to the residue of the pole in the upper plane:
\be
 I_2^{(a)}=2\pi\omega\sin\omega x_1\frac{r_0^2}{L^4} \sum_n \frac{\tw{n}}{f_n} H_0^{(1)}\left(i\frac{r_0}{L^2}\tw{n}\rho\right)\,,\label{I2a}
\ee
where we have used the fact that 
\be
a_{\tilde\omega}\sim - 
\frac{f_n k_{\rho,n}(k_\rho-k_{\rho,n})L^4}{\tw{n} r_0^2} \,.  
\ee
The contribution $I_2^{(b)}$ is more involved, because the second term in Eq.~\eqref{Psi} has both two poles at $k_\rho=\pm i\frac{r_0^2}{L^4}\tw{j}$ and two branch points at $k_\rho^2=\omega^2-\frac{r_0^2}{L^4}\tw{n}^2$.  When $\omega^2<\frac{r_0^2}{L^4}\tw{n}^2$ the branch points are on the imaginary axis, so that the function to be integrated is regular on the real axis.  On the other hand, when  $\omega^2>\frac{r_0^2}{L^4}\tw{n}^2$ the branch points are on the real axis. In this case we can still integrate numerically,\footnote{In principle, the integral~\eqref{I2collisionscalar2} can be evaluated fully analytically using contour techniques in the complex $k_\rho$ plane, but particular attention must be paid to properly include the branch cut contribution.} but the branching points must be suitably excluded from the integration domain.  The quantity $\omega\sqrt{\rho^2+x_1^2}\langle{\cal O}(\omega,x_1,\rho)\rangle$ obtained by integrating numerically and summing the two contributions is shown in Fig.~\ref{fig:spectrum_scalar2} (modulo a coefficient proportional to $Q/E_t$).

We find the same qualitative features observed in the spherically symmetric case.  When the frequency is smaller than the fundamental mode, $\omega<4.062 r_0/L^2$, the spectrum is exponentially suppressed at large $\rho$.  As $\omega$ increases, several mass barriers can be overcome and single contributions may show an oscillatory behavior.  Finally, when $\omega>4.062 r_0/L^2$ the spectrum shows an approximate decay as $1/\rho$ at large distance.

\begin{figure*}[htb]
\begin{center}
\begin{tabular}{cc}
\includegraphics[width=7.4cm,angle=0,clip=true]{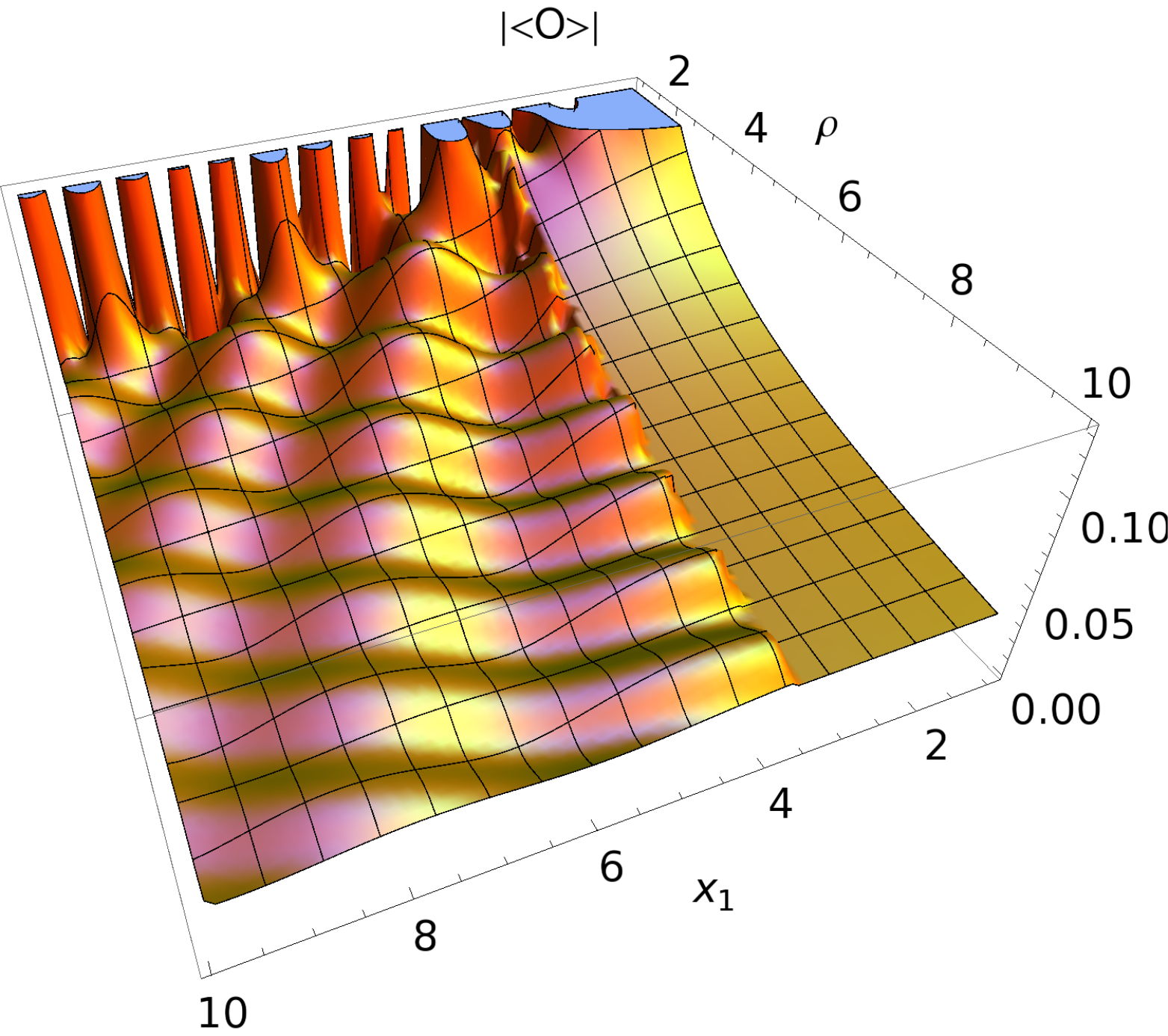}&
\includegraphics[width=7.4cm,angle=0,clip=true]{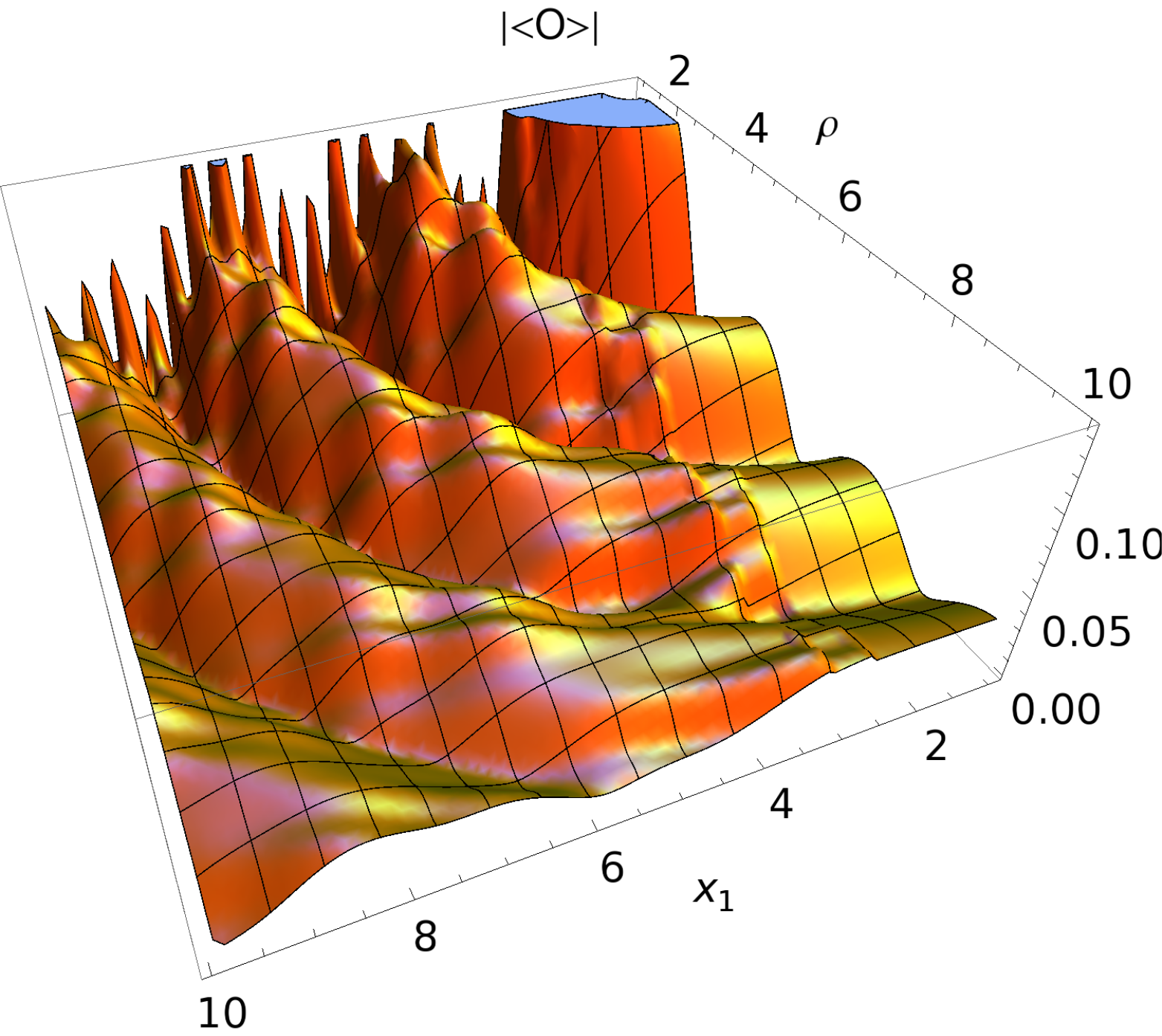}
\end{tabular}
\caption{The frequency-domain quantity $|\langle{\cal O}(\omega,x_1,\rho)\rangle|$ as a function of $x_1$ and $\rho$ for $\omega_{\rm cutoff}\sim 5.4$ (left panel) and for $\omega_{\rm cutoff}\sim 8$ (right panel).
\label{fig:scalar_cyl_3D}}
\end{center}
\end{figure*}

Finally, the full dependence of $|\langle{\cal O}(\omega,x_1,\rho)\rangle|$ in the frequency domain is shown in Fig.~\ref{fig:scalar_cyl_3D} as a function of $x_1$ and $\rho$ for $\omega_{\rm cutoff}\sim 5.4$ (left panel) and for $\omega_{\rm cutoff}\sim 8$ (right panel).

\subsection{Sources extended in the holographic direction}
\label{sec:vacuum_scalar}

In the previous sections, we constructed the field of point-like scalar charges, showed it decays exponentially fast in the gauge theory directions and collided them along one of the flat directions.  What are the effects of finite-size on the previous results?  To investigate this, we now construct a general class of solutions which are {\em not} localized in the holographic direction, and which still display Yukawa-type asymptotics.  
The formalism below is quite general and can handle any finite sized object in the holographic direction, reducing the problem to a Minkowski space evolution of massive fields.

Up to now we have kept factors of $r_0$ and $L$ in most of the equations.  In this section and in the remainder of the paper we will set $r_0=L=1$ to reduce cluttering of the equations.  This can always be accomplished by appropriately rescaling the holographic coordinate, together with the fields.  However, we shall explicitly reinstate such factors in the main results.

Consider then the Klein-Gordon equation with the general source
\be
-4\pi T=-\frac{4\pi}{r^2}\delta(x_1)\delta(x_2)\delta(x_3)\rho(r)\,,
\ee
where for the moment $\rho(r)$ is an arbitrary regular function.  If we look for spherically symmetric solutions by introducing a radial coordinate $R=\sqrt{x_1^2+x_2^2+x_3^2}$ we get the following equation
\be
\frac{1}{R^2}\frac{\partial}{\partial R}\left(R^2\frac{\partial}{\partial R}\Phi\right)
+\frac{1}{r^2}\frac{\partial}{\partial r}\left(r^4F\frac{\partial}{\partial r}\Phi\right)=-\frac{\rho(r)}{R^2}\delta(R)\,.
\ee
Separable solutions to this problem exist, and we can study them by using the following decomposition:
\beq
\rho(r)&=&\sum_n a_n\Psi_n(r)\,,\\
\Phi(r,R)&=&\sum_n \Psi_n(r)Z_n(R)\,,
\eeq
where $\Psi_n$ are the vacuum eigenfunctions studied in Section~\ref{sec:scalar_modes}, with eigenvalue $\tw{n}^2=-k_n^2$.  We then get the following ODE for $Z_n(R)$,
\be
\frac{1}{R^2}\frac{d}{d R}\left(R^2\frac{d}{d R}Z_n\right)-k_n^2Z_n=-\frac{a_n}{R^2}\delta(R)\,.
\ee
This is nothing but the equation for the field of a point-like particle coupled to a (massive) Klein-Gordon field in \emph{flat} spacetime, whose solutions have the classical Yukawa-like form:
\be
 Z_n(R)=-a_n \frac{e^{-k_nR}}{R}\,.
\ee
Thus, a generic distribution in the holographic direction can be understood as the sum of the field generated by point-like particles in a Minkowski background and carrying a massive interaction.  In the generic case, the field is a superposition of such solutions and the coefficients $a_n$ are evaluated as an overlap of different eigenfunctions with weight $\sim r^2$,
\be
a_n= \int_1^{\infty}dr\,r^2\rho(r)\Psi_n (r) \,.
\ee

Suppose as a first example that $a_n=\delta_{n\,n_0}$, so that the stress-tensor profile in the holographic direction coincides with one of the normal modes of the field.  We then get
\be
\Phi=-\frac{e^{-k_{n_0}R}}{R}\psi_{n_0}\,,
\ee
and at large distances we find 
\be
\Phi \sim \eta_{n_0}\frac{e^{-k_{n_0}R}}{R\, r^5} \,, 
\ee
where $\eta_{n_0}$ is a normalization constant, which for the fundamental mode is $\eta_0\approx 3.4$.

Let us recover as a final example the point particle results within this approach.  In this case, $\rho=\delta(r-1)$ and we find for the fundamental mode contribution (an infinite tower of modes contribute to the point particle field)
\be
\Phi (R,r) =- \Psi_0(1) \frac{e^{-k_0R}}{R} \, \Psi_{0} (r)\,.
\ee
This can be evaluated to be, at large holographic distances,
\be
\Phi=-\frac{12.212}{r^5}\frac{e^{-k_0 R}}{R}\,,
\ee
which agrees to within $0.01\%$ with a fit of our numerical results for the point particle calculation (which yields $\eta_0\approx 12.213$, to be compared with the analytical approximate results, $\eta_0=c_0/(10\pi^2)\approx 11.516$, cf. Eqs.~\eqref{pot_scalar_inf}--\eqref{c_i}). 
We also get $c_1/c_0\sim -5.4$, in rough agreement with Eq.~\eqref{c_i}.\footnote{The discrepancy is most likely due to numerical inaccuracy as it is very challenging to extract exponentially suppressed contributions.}  It is interesting to note that at large distances the field generated by the same total charge $Q$ does depend on the charge distribution.  In fact, the field generated by a point particle is roughly 3 times stronger than that created by a smooth $\rho(r)$ distribution identical to the fundamental mode.

Collisions of these non-pointlike configurations can also be studied with well-known methods.  Using the same notation as above, our equation now becomes a flat-space massive field equation:
\be
Z_n(k,\omega)=\frac{4\pi a_n}{\omega^2-k^2-\tw{n}^2}\left(\frac{2\omega}{iE_t(\omega^2-v^2k_1^2)}-\frac{2\epsilon_\mt{BH}}{i\omega}\right)\,.
\ee
For $\epsilon_\mt{BH}=1$ and in the $E_t\to \infty$ limit, the solution in the space domain reads
\be
Z_n(R,\omega)=\frac{2a_n}{\,\omega} \frac{e^{ik_n R}}{R}\,.
\ee
Generically, the collision of extended particles is quantitatively different but qualitatively identical to the collision of point particles.  The output can be quantitatively the same by correcting only the static profile.  For instance, the collision of a point particle with a spectrum cut at the fundamental mode results in a spectrum 3 times larger than the collision of a smooth extended distribution along $r$, with $\rho(r)=\psi_0$ and with the same total charge.

\section{Gravitational interactions}
\label{sec:grav}

We now turn to the main interest of this paper: the study of gravitational perturbations resulting from the head-on collision of two point particles in the AdS-soliton background.  We tackle this problem step by step, as we did for the scalar toy model, first addressing the normal modes of the spacetime, then investigating the gravitational field created by a static particle and finally considering the collision process. From our results we will be able to infer the behavior of the stress-energy tensor of the dual gauge theory.
 
Recall we are now setting $r_0=L=1$ for simplicity.  We will explicitly display such factors only in the main results.

\subsection{Stability and normal modes of the gravitational waveguide}
\label{sec:scalar_grav}

Let us start by studying the gravitational normal modes of the AdS-soliton.  Gravitational perturbations have more degrees of freedom, but we will be mainly interested in perturbations that keep some of the symmetries of the background intact.  In particular, our goal here is not to perform a full perturbative decomposition of the gravitational field, so we now focus on the type of perturbations which are more directly relevant for the physics we wish to understand.  In Appendix~\ref{app:vector_gravitational} we show the existence of a special type of vector-like gravitational perturbations, which are {\it not} excited by colliding objects head-on.  These vector-type perturbations, are in principle excited in other, more generic situations; we show in the appendix that their spectrum shares the same main features as the ones we discuss below.

We focus on a subset of gravitational perturbations, appropriate for the symmetries we want to consider.  In $(t,r,x_1,x_2,x_3,y)$ coordinates the metric reads
\be
ds^2=ds_{\rm soliton}^2+\epsilon h_{\mu\nu}dx^\mu dx^\nu\,,
\label{metricpert}
\ee
where the perturbation quantities are defined as
\bea
h_{\mu \nu}= \left[
 \begin{array}{cccccc}
 r^2h_{tt}(r) & h_{tr}(r) & r^2 h_{tx}(r) & 0 & 0 & 0\\
 h_{tr}(r) & \frac{h_{rr}(r)}{F(r)} &h_{rx}(r) & 0 & 0 & 0 \\ 
 r^2 h_{tx}(r) & h_{rx}(r) &r^2h_{xx}(r) & 0 & 0 & 0\\ 
 0 & 0 & 0 & r^2 h_{\perp}(r) & 0 & 0\\ 
  0 & 0 & 0 & 0 & r^2h_{\perp}(r) & 0\\ 
 0 & 0 & 0 & 0 & 0 & F(r) h_{yy}(r)
\end{array}\right]  e^{-i\omega t +ik_i x_i}.
\label{ansatz_staticp4v1b}
\eea
Here, an integral over $\omega$ and $k_i$ is implicit, as well as a summation over $i=1,2,3$.  We have singled out the coordinate $x_1$ to be aligned with the collision axis and for notational convenience we are setting $x_1\equiv x$.  The transverse directions $x_2$ and $x_3$ are on an equal footing and will be denoted indistinctly by $x_\perp$. In the configurations that we consider the stress-energy tensor of the particles does not have any components $T_{tx_\perp}$. Thus, when working in transverse gauge, the components $h_{tx_\perp}$ obey homogeneous equations with trivial boundary conditions and must vanish. From the viewpoint of the gauge theory this may seem counterintuitive, since we expect that the collision generates radiation with momentum in the $x_\perp$ directions. We will see that this apparent puzzle is actually resolved by the fact that the change of coordinates from \eqn{metricpert} (which are a natural gauge choice from the bulk viewpoint) to Fefferman-Graham coordinates (appropriate for boundary observables) generates the expected components ${\cal T}_{t x_2}, {\cal T}_{t x_3}$ of the boundary stress tensor.

Inserting the ansatz~\eqref{ansatz_staticp4v1b} into the Einstein equations we get four algebraic equations for $h_{tx}$, $h_{yy}$, $h_{\perp}$  and $h_{tr}$:
\bea
h_{tx}&=&\frac{k_1 (h_{\perp}-h_{xx})}{\omega}\,, \label{htX}\\ [1mm]
h_{yy}&=& -h_{rr}+h_{tt}-h_{xx}  \,, \label{hyy}\\ [2mm]
h_{\perp}&=&\frac{i \omega  \Big[ (rF'+2F) h_{tr}+rF h_{tr}' \Big] + r \left(-\omega ^2 h_{tt}+k_1^2 h_{xx}\right)}{r \left(k_1^2+\omega ^2\right)}\,,\label{hxx}\\ [1.5mm]
h_{tr}&=&\frac{ir}{2\omega \left(k_1^2+\omega^2+\tilde{\omega}^2\right) F} \Big\{ 2 \left(k_1^2+\omega ^2\right) \left(rF'+2F\right) h_{rr} \nn \\[1mm]
&&+  \left(k_1^2+\omega ^2\right) \left(rF'-2F\right) (h_{xx}-h_{tt}) + 2rF \Big[ (k_1^2+\omega^2) h_{rr}'+\omega ^2  h_{tt}'- k_1^2  h_{xx}'\Big] 
\Big\} \,. \,\,\,\,\,\,\,\,\,\,
\label{htr}
\eea
In addition, the function $h_{rx}$ satisfies a homogeneous second order equation.  We completely fix the gauge by requiring $h_{rx}=0$.  Although the denominator of $h_{tr}$ above vanishes as $r\to 1$, it is easy to show that in the same limit the numerator vanishes with the same power of $(r-1)$, cf.~Table~\ref{tab:asymp} in Appendix~\ref{app:PPgrav} for detail.\footnote{Note however that the metric perturbation~\eqref{ansatz_staticp4v1b} considered in this section is more general than the one considered in Appendix~\ref{app:PPgrav}, which restricts to the static case.}  Therefore, $h_{tr}$ is regular at $r\sim 1$ if the other functions are also regular.  Finally, we get three dynamical equations for $h_{xx}$, $h_{rr}$ and $h_{tt}$.
The latter can be simplified by introducing two new variables $z_-(r)$ and $z_+(r)$ such that
\be
 h_{tt}=-\frac{z_+ + z_-}{2}\,,\qquad  h_{xx}=\frac{z_+ - z_-}{2}\,. \label{zpm}
\ee
The perturbation equations in these new variables read
\bea
 z_-''&=& -\frac{\tilde{\omega}^2 z_-+r \left(rF'+4F\right) z_-'}{r^2 F}  \,,\label{eqzm}\\ [2mm]
 z_+''&=&\frac{4F h_{rr}+\left(rF'+13F-15r^2 - \tilde{\omega}^2\right) z_+ - r \left(rF'+4F\right)z_+'}{r^2 F}\,,\label{eqzp}\\ [2mm]
 h_{rr}''&=&\frac{1}{2 r^2 F^2} \Big[ \left( 5rF(F'-3r)+r^2F'^2+F^2\right) z_+ \nn \\[1.5mm]
&& +2 \left(r^2 F'^2 - 10F^2 - \tilde{\omega}^2 F \right) h_{rr} - 2rF \left(rF'+4F\right) h_{rr}' \Big]\,. \label{eqhrr}
\eea
Here, we used the identity $r^2F''=2(3F-rF')$ to avoid the explicit appearance of the second derivative of the metric function $F(r)$.  

It is worth noting that our ansatz~\eqref{ansatz_staticp4v1b} includes, but it is not limited to, gravitational scalar modes. The latter are defined as those that, in their rest frame, transform as scalars under the little group $SO(3)$. This is equivalent to the condition $h_{xx}=h_{\perp}$ at $k^2=0$ in the ansatz~\eqref{ansatz_staticp4v1b}. Using this condition and the perturbation equations above, it is easy to show that Eq.~\eqref{eqzm} is identically satisfied. Therefore, the gravitational scalar sector is entirely described by the system~\eqref{eqzp}--\eqref{eqhrr}. On the other hand, Eq.~\eqref{eqzm} is decoupled from the other two and it describes a subsector of the gravitational vector perturbations. Notice that Eq.~\eqref{eqzm} is equivalent to the scalar equation~\eqref{waveeqscalarfields}, so that this subset of gravitatonal vector modes coincides with the scalar spectrum, as shown in Table~\ref{tab:resonance}. In Table~\ref{tab:resonance} we refer to these modes as ``vector II'' family, to distinguish them from the ``vector I'' family presented in Appendix~\ref{app:vector_gravitational} which, however, is not excited in the collision discussed in Sec.~\ref{collision_grav}.

Furthermore, Eqs.~\eqref{eqzm}--\eqref{eqhrr} depend only on the combination $\tilde{\omega}^2$, which was also the case in Section~\ref{sec:scalar}. 

Close to the bottom $r=1$, we impose regularity of the perturbation functions:
\be
 h_{rr}\sim \sum_{j=0}^\infty A_b^{(j)}(r-1)^{j}\,,\qquad
 z_+\sim \sum_{j=0}^\infty B_b^{(j)}(r-1)^{j}\,,\qquad
 z_-\sim \sum_{j=0}^\infty C_b^{(j)}(r-1)^{j}\,,
 \label{asymptoticgrav_scalar_r0}
\ee
where the coefficients $A_b^{(j)}$ and $B_b^{(j)}$ can all be written in terms of just two independent parameters, namely $A_b^{(1)}$ and $B_b^{(0)}$.
The equation for $z_-$ is decoupled and the coefficients $C_b^{(j)}$ can all be expressed in terms of $C_b^{(0)}$.  

Close to infinity we get
\be
h_{rr}\sim\sum_{j=0}^\infty A_\infty^{(j)}r^{-j}\,, \qquad
z_+\sim \sum_{j=0}^\infty B_\infty^{(j)}r^{-j}\,, \qquad
z_-\sim \sum_{j=0}^\infty C_\infty^{(j)}r^{-j}\,,
\label{asymptoticgrav_scalar_infinity}
\ee
where the expansion coefficients $A_\infty^{(j)}$ and $B_\infty^{(j)}$ can be written in terms of four independent parameters and the coefficients $C_\infty^{(j)}$ can be written in terms of two independent parameters.  We guarantee that all metric perturbations decay at infinity by fixing 
\be
 A_\infty^{(2)}=B_\infty^{(0)}=C_\infty^{(0)}=0\,. \label{BCgravINF}
\ee
By imposing the conditions above, the asymptotic behavior of the metric functions reads 
\bea
 h_{ab}&=&\frac{A_{ab}}{r^3}+\frac{B_{ab}}{r^5}+{\cal O}(r^{-7})\,,\qquad (a,b)\neq(t,r)\\
 h_{tr}&=&\frac{A_{tr}}{r^4}+\frac{B_{tr}}{r^6}+{\cal O}(r^{-8})\,,
\eea
and the explicit form of the constants $A_{ij}$ and $B_{ij}$ is given in Appendix~\ref{appsec:asymptotics}.  There we show that the large-distance behavior only depends on the three parameters
\be
A_\infty^{(3)}\equiv A_{rr} \sac B_\infty^{(5)}\equiv B_{tt}+B_{xx} \sac 
C_\infty^{(5)}\equiv B_{tt}-B_{xx} \,.
\ee 

We have integrated Eqs.~\eqref{eqzp}--\eqref{eqhrr} imposing the expansion~\eqref{asymptoticgrav_scalar_r0} at $r=1$ and requiring Eqs.~\eqref{BCgravINF} at infinity.  With these boundary conditions, all metric components are guaranteed to be regular at $r=1$ and to vanish as $r\to\infty$.  We have searched for the eigenvalues using two different methods, one of which we now describe.  An alternative method based on Frobenius expansions is outlined in Appendix~\ref{app:Frobenius}.

The most efficient method is a standard technique to deal with matrix-valued eigenvalue problems (cf. Ref.~\cite{Pani:2013pma} for a review).  First, we perform two integrations starting from $r=1$ with $(A_b^{(1)},B_b^{(0)})=(1,0)$ and $(A_b^{(1)},B_b^{(0)})=(0,1)$.  By extracting the functions $h_{rr}$ and $z_+$ at infinity we construct the matrix
\bea
 \label{matrix_coupled}
\mathbf{S}(\tilde\omega,k)=
\begin{pmatrix}
A_\infty^{(2),I} & A_\infty^{(2),II}\\ 
B_\infty^{(0),I} & B_\infty^{(0),II}\\ 
\end{pmatrix}\,,
\eea
where the superscripts $(I,II)$ denote the two choices of $(A_b^{(0)},B_b^{(0)})$, respectively.  The latter also correspond to two sets of solutions, $(h_{rr}^{I},z_+^{I})$ and $(h_{rr}^{II},z_+^{II})$.  Finally, the eigenfrequency $\tilde{\omega}$ is obtained by searching for the roots of $\det{\mathbf{S}}$.  We find a discrete  set of modes, which are listed in Table~\ref{tab:resonance}.  Curiously, the modes of the system~\eqref{eqzp}--\eqref{eqhrr} include, but they are not limited to, the modes of Eq.~\eqref{eqzm}.  We stress that the latter coincide with the scalar modes previously discussed.

\begin{figure*}[tb]
\begin{center}
\begin{tabular}{lr}
\includegraphics[width=7.4cm,angle=0,clip=true]{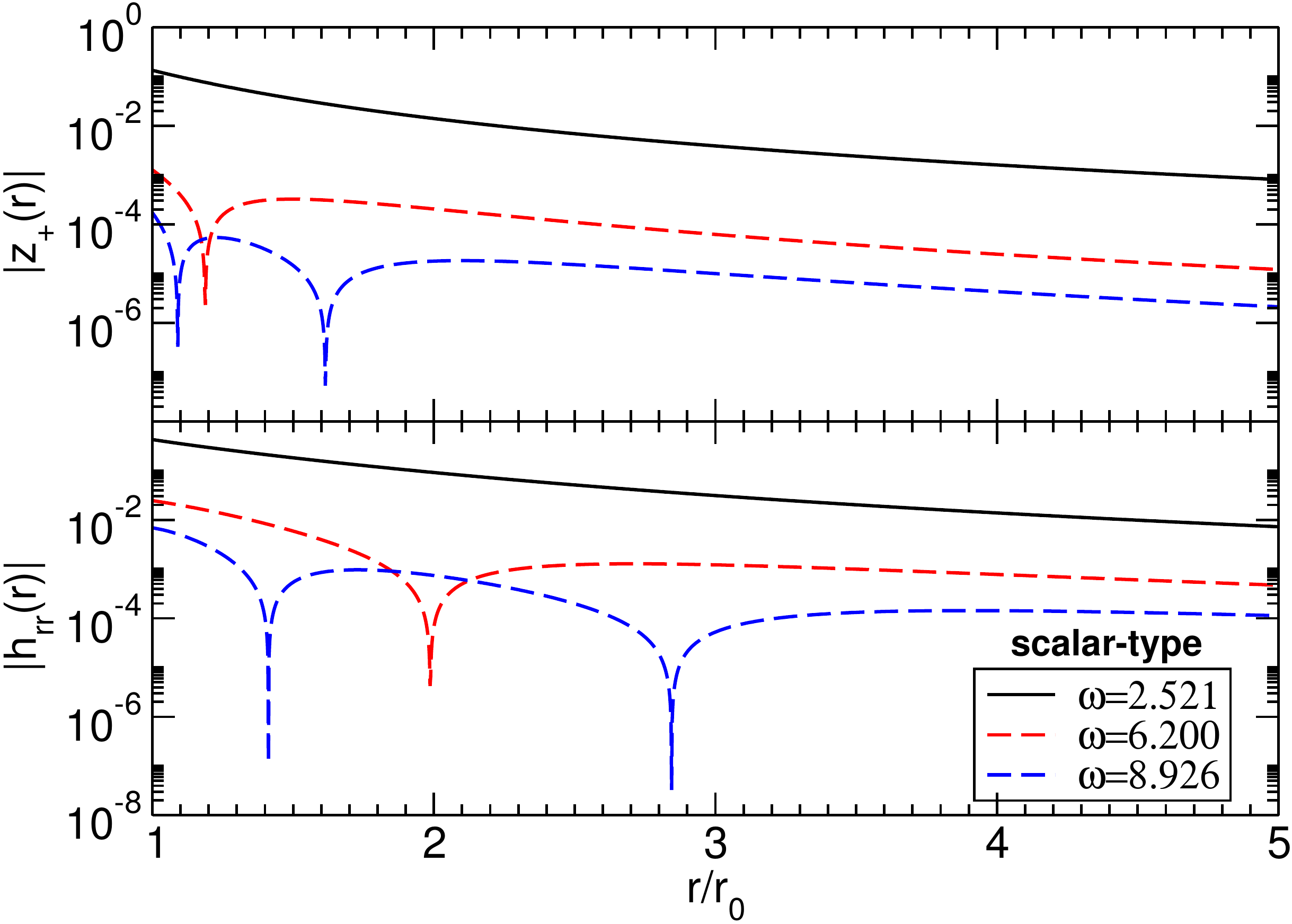} &
\includegraphics[width=7.4cm,angle=0,clip=true]{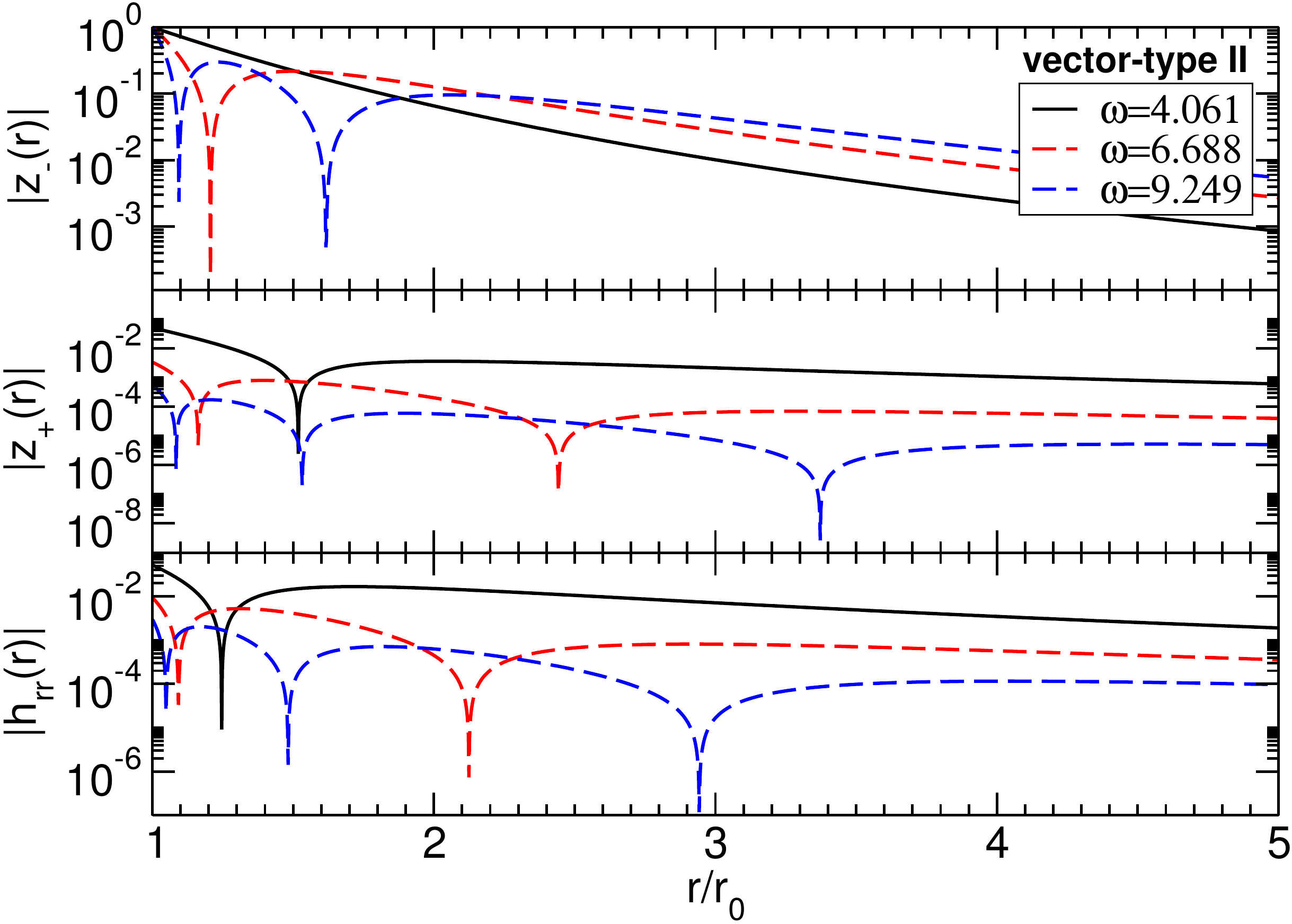}
\end{tabular}
\caption{Eigenfunctions corresponding to the fundamental mode and the first few overtones of scalar-type gravitational perturbations (left panels) and of vector-type II gravitational perturbations (right panels) as listed in Table~\ref{tab:resonance}. \label{fig:eigenfunctions}}
\end{center}
\end{figure*}

In Fig.~\ref{fig:eigenfunctions} we show the eigenfunctions corresponding to the first three gravitational scalar-type and first three vector-II type modes in Table~\ref{tab:resonance}.  Within this direct integration method, the eigenfunctions are defined as
\bea
 h_{rr}(r)&=& \alpha_1 h_{rr}^{I}(r) -\alpha_2 h_{rr}^{II}(r) \,,\\
 z_+(r)&=& \beta_1 z_+^{I}(r)-\beta_2 z_+^{II}(r)\,,
\eea
where $\alpha_1=B_\infty^{(0),I}$, $\alpha_2=B_\infty^{(0),II}$, $\beta_1=A_\infty^{(2),I}$, $\beta_2=A_\infty^{(2),II}$ are constants obtained after the determinant of the matrix $\mathbf{S}$ in Eq.~\eqref{matrix_coupled} has been minimized.  By construction, since $\det\mathbf{S}=0$, the functions above are automatically eigenfunctions. In Fig.~\ref{fig:eigenfunctions}, the classification of modes into two different families is manifest.  Finally, we show in Appendix~\ref{app:zero_mode} that $\omega=k=0$ is {\it not} a regular solution of the problem, and therefore that no zero modes exist in this background. In the gauge theory this means that the glueball spectrum is gapped, as expected. 

\subsection{A static point particle in the AdS-soliton background}
\label{sec:staticPP}

In this section, we investigate the gravitational field generated by a static point-like source located at the tip $r=r_0=1$. In $(t,r,x_1,x_2,x_3,y)$ coordinates the metric reads as in Eqs.~\eqref{metricpert}--\eqref{ansatz_staticp4v1b} but with all off-diagonal terms set to zero, in addition to $\omega=0$ and $h_\perp=h_{xx}$.
Inserting this ansatz into the Einstein equations we get two coupled, second-order differential equations for $h_{xx}$ and $h_{rr}$:
\bea
h_{rr}''&=& \frac{-r^2\tilde{T}_{tt}(r)/2+\left(k^2 r^2-30 r^4\right) h_{rr}-\left(6 r^5-1\right) \left(2 h_{rr}'-h_{xx}'\right)}{r(r^5-1)} \,, \label{e1}\\[2mm]
h_{xx}''&=&\frac{-r^2\tilde{T}_{tt}(r)/2+10 r^4 h_{rr}+k^2 r^2 h_{xx}+2 \left(r^5-1\right) h_{rr}'+\left(-8 r^5+3\right) h_{xx}'}{r(r^5-1)}\,,\label{e2}
\eea
where $\tilde{T}_{tt}(r)$ is the Fourier transform (defined as in Eq.~\eqref{antifourier}) of the only nonvanishing component of $T_{\mu\nu}$, $\tilde{T}_{tt}(r)=\mu\,\delta(r-1)$.  Note that the Einstein equations allow a source term of this form, which is consistent with the fact that in this spacetime a point-particle can be static only at the tip $r=1$.  The remaining perturbation functions $h_{yy}$ and $h_{tt}$ are algebraically related to $h_{xx}$, $h_{rr}$, and their derivatives:
\bea
 h_{yy}&=&-h_{rr}+h_{tt}-h_{xx}\,, \label{algebraic}\\[2mm]
 h_{tt}&=&\frac{2 \left(4 r^5+1\right) h_{rr}+5 h_{xx}+2 r \left(r^5-1\right) \left(h_{rr}'-h_{xx}'\right)}{5}\,,\label{algebraic2}
\eea
and they can be directly computed once the system~\eqref{e1}--\eqref{e2} is solved.  In Appendix~\ref{app:PPgrav} we explicitly solve Eqs.~\eqref{e1}--\eqref{e2} using Green's function techniques.  Here, we just report the final results, where we have reinserted factors of $r_0$ and $L$.

In the Fourier domain, we obtain
\bea
  h_{rr}(k_i,r)&=&\frac{\mu L^4}{10 r_0^3 \left(a_k d_k-b_k c_k\right)}\left[ b_k h_{rr}^{(\infty,1)}(r)-a_k h_{rr}^{(\infty,2)}(r)\right]\,,\label{solhrrF}\\
 h_{xx}(k_i,r)&=&\frac{\mu L^4}{10 r_0^3 \left(a_k d_k-b_k c_k\right)}\left[ b_k h_{xx}^{(\infty,1)}(r)-a_k h_{xx}^{(\infty,2)}(r)\right]\,.\label{solhxxF}
\eea
The series expansions in Table~\ref{tab:asymp} then determine the large (holographic) distance behavior:\footnote{The $1/r^3$ fall-off of $h_{xx}$ may seem surprising, but this is simply due to the fact that $r$ does not coincide with the Fefferman-Graham coordinate $\bar r$ near the boundary. In terms of the latter coordinate the fall-off is instead $1/\bar r^5$, as expected (see Eq~\eqn{surprise}).}
\be
h_{rr}(k_i,r) \sim \frac{\mu L^4 \, b_k }{10 r_0^3 \left(a_k d_k-b_k c_k\right)}\left(\frac{r_0}{r}\right)^3\,,\qquad
h_{xx}(k_i,r) \sim -\frac{h_{rr}(k_i,r)}{3}\,.
\ee
The dimensionless functions $a_k(k^2)$, $b_k(k^2)$, $c_k(k^2)$ and $d_k(k^2)$ are related to the behavior of the gravitational perturbations at $r\sim r_0$.  They can be constructed by a numerical integration of the homogeneous system, cf.~Appendix~\ref{app:PPgrav} for details.
Finally, in the space domain at leading order in $r$ we obtain
\be
 h_{rr}(x_i,r)=-3h_{xx}(x_i,r)=\frac{1}{(2\pi)^3}\int_{-\infty}^{+\infty} d^3k\, e^{i k_i x_i}  h_{rr}(k_i,r)= \frac{\mu L^4}{80\pi^3 r_0^3 r^3} \, {\cal I}(x_i)\,,
\ee
so that all the information about the metric perturbations is encoded in the following integral 
\begin{equation}
 {\cal I}(x_i)\equiv\int_{-\infty}^{+\infty} d^3k \frac{b_k e^{i k_i x_i}}{a_k d_k-b_k c_k}\,.\label{intIgrav}
\end{equation}
The functions $b_k$ and $D_k=a_k d_k-b_k c_k$ have behaviors qualitatively similar to that of the function $a_{\tilde\omega}$ in the negative $x-$axis shown in Fig.~\ref{fig:a_VS_wt2}.  In order to evaluate the integral~\eqref{intIgrav}, we proceed as in the scalar case:
\be
{\cal I}(x_1,x_2,x_3)={\cal I}(R)=\frac{2\pi}{iR} \int_0^{+\infty}dk\, k\, b_k(k)\frac{e^{ikR}-e^{-ikR}}{D_k(k)}\,.
\label{integral_grav}
\ee

\begin{figure*}[tb]
\begin{center}
\begin{tabular}{lr}
\includegraphics[width=7.4cm,angle=0,clip=true]{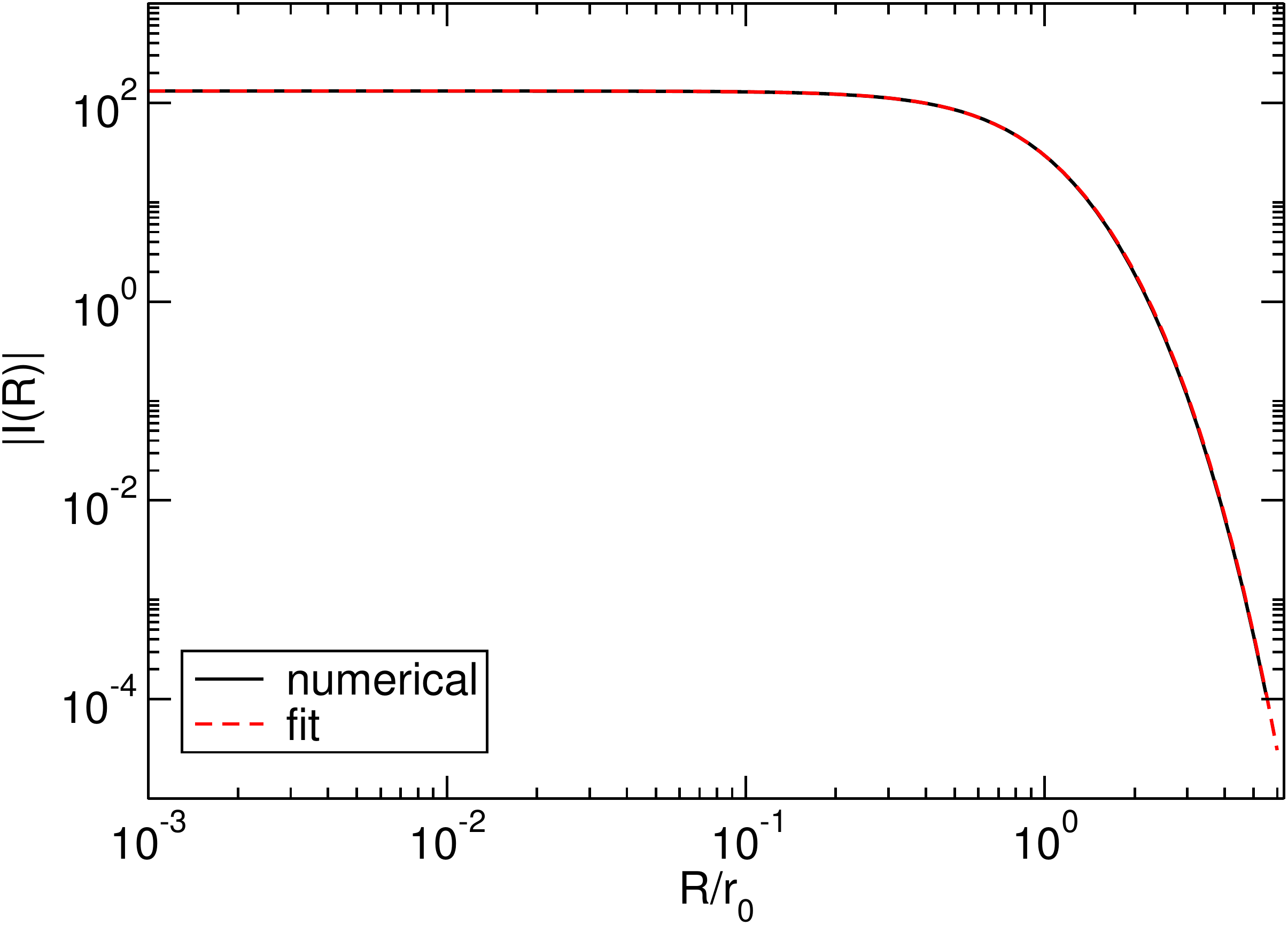} &
\includegraphics[width=7.4cm,angle=0,clip=true]{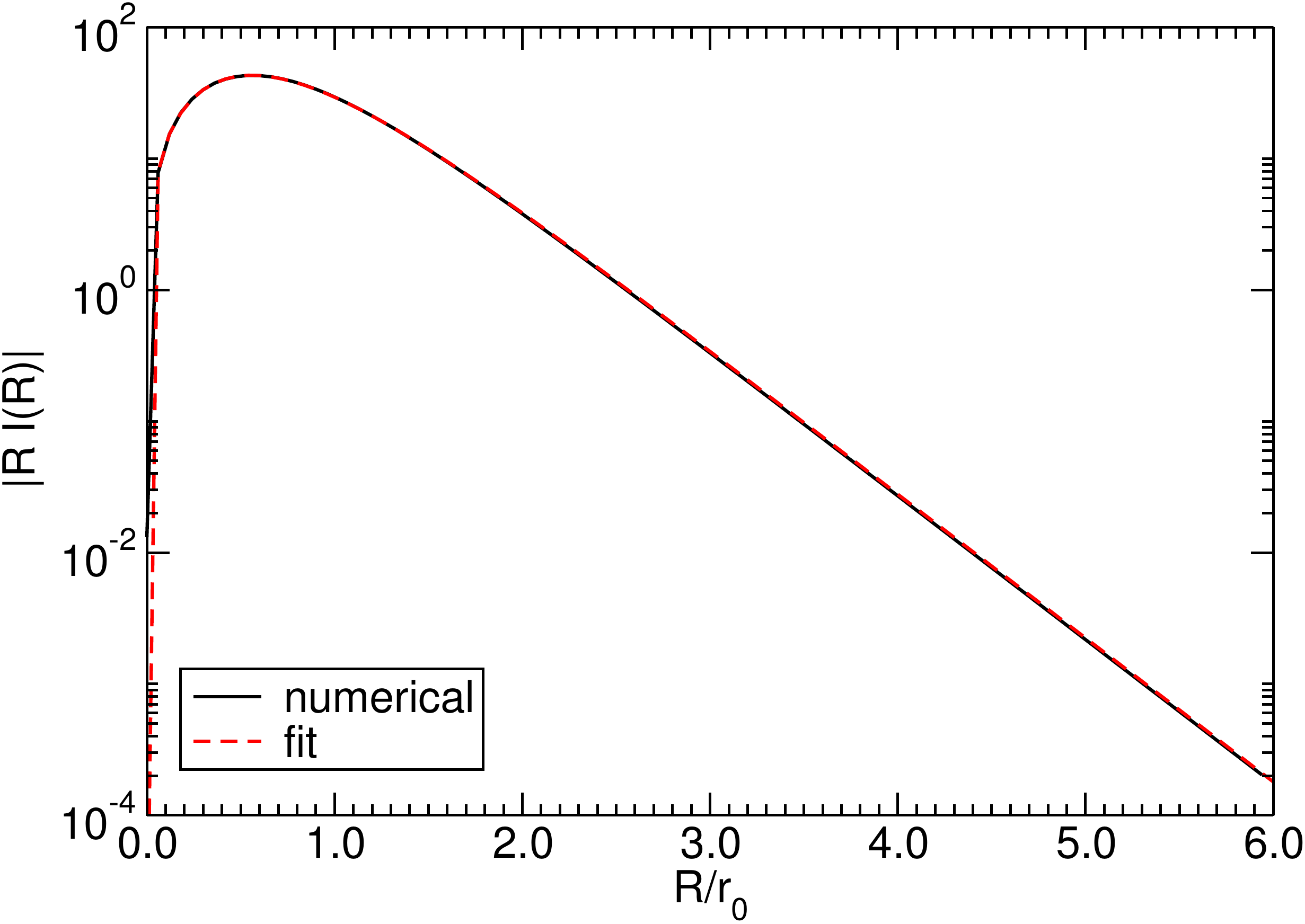}
\end{tabular}
\caption{Left: the integral ${\cal I}(R)$~\eqref{intIgrav} as a function of $R$ compared with a fit of the superposition of normal mode solutions~\eqref{scalar_superposition_normal} with $N=3$ and using the gravitational scalar-type modes in Table~\ref{tab:resonance}.  We consider the fundamental mode and the first three overtones including both families listed in Table~\ref{tab:resonance}. Right: same for the quantity $R\, {\cal I}(R)$.
\label{fig:int_grav}}
\end{center}
\end{figure*}

The integral above is shown in Fig.~\ref{fig:int_grav}, where we again compare the numerical results with a superposition of normal-mode solutions analogous to Eq.~\eqref{scalar_superposition_normal}, but where now $\mu_n$ are only the first four gravitational modes listed in Table~\ref{tab:resonance} (we considered the fundamental mode plus three overtones including both scalar-type and vector-II type modes).

\subsection{High-energy collision of particles}
\label{collision_grav}

In this section, we compute the linear gravitational emission during the head-on collision of two point-particles with mass $m$ boosted with speed $v$, each following a straight geodesic along the $x_1$ direction in the hyperplane defined by $r=r_0$.  We consider Einstein's equations with the stress-energy tensor 
\bea
T^{\mu\nu} &=& \frac{mL^3r_0}{r^4\gamma} \delta(x_2)\delta(x_3)\delta(r-r_0)
\Theta(-t)\left[ u_{(1)}^\mu u_{(1)}^\nu \delta(x_1-vt)
+u_{(2)}^\mu u_{(2)}^\nu \delta(x_1+vt) \right] \nn \\ [2mm]
 && +\frac{\Theta(t) ML^4}{r^4} u_{(3)}^\mu u_{(3)}^\nu \delta(x_1)\delta(x_2)\delta(x_3)\delta(r-r_0)\,,\label{ZFLStressGrav}
\eea
where $\gamma=1/\sqrt{1-v^2}$ and, at first order, the mass of the final (static) object is equal to the total energy in the initial system, $M=2m\frac{L}{r_0}\gamma$.  Accordingly, the spacetime velocity vectors of the three particles are
\be
u_{(1)}^\mu=\frac{L}{r_0}\gamma(-1,0,v,0,0,0)\,, \quad\;\;\;
u_{(2)}^\mu=\frac{L}{r_0}\gamma(-1,0,-v,0,0,0)\,, \quad\;\;\;
u_{(3)}^\mu=(-1,0,0,0,0,0)\,.
\ee
Recall we are using coordinates $x^\mu=(t,r,x_1,x_2,x_3,y)$ but, to avoid cluttering subsequent formulas, we shall adopt the notation $x_1\equiv x$ as in Section~\ref{sec:scalar_grav}.  In the final results we revert to the original notation.

The Fourier transform, defined as in Eq.~\eqref{antifourier}, of the stress-energy tensor (with covariant indices) may be expressed as follows:
\be
 \tilde{T}_{\mu\nu}(\omega,k_i,r) = {-\frac{2imL\gamma}{r_0}\delta(r-r_0)}\left[\frac{1}{\omega^2-v^2k_1^2}
 			\left(\begin{array}{c|c}
                                                                     \mathbf{P}_i	&  \mathbf{0}	\\ \hline
                                                                     \mathbf{0}	&  \mathbf{0}
                                                                    \end{array}\right)
-\frac{1}{\omega}\left(\begin{array}{c|c}
                                                                     \mathbf{P}_f	&\mathbf{0}	\\ \hline
                                                                     \mathbf{0}	&\mathbf{0}	
                                                                    \end{array}\right)\right]\,,
\ee
where
\be
 \mathbf{P}_i = \left(\begin{array}{ccc}
                                                                     \omega	&0	& -v^2 k_1	\\
                                                                     0		&0	& 0			\\
                                                                     -v^2 k_1&0	& v^2\omega
                                                                     \end{array}\right)\,, \qquad
 \mathbf{P}_f= \left(\begin{array}{ccc}
                                                                     1	&0	&0	\\
                                                                     0	&0	&0	\\
                                                                     0	&0	&0
                                                                     \end{array}\right)\,, \qquad
 \mathbf{0}= \left(\begin{array}{ccc}
                                                                     0	&0	&0	\\
                                                                     0	&0	&0	\\
                                                                     0	&0	&0
                                                                     \end{array}\right)\,.
\ee
Note two qualitative differences with respect to the scalar case: (i) because kinetic energy gravitates, the stress-energy tensor depends on $v$ also in the ultrarelativistic limit $v\to 1$ through the $\gamma$ term; (ii) $\tilde{T}_{\mu\nu}$ explicitly depends on $k_1$ even in the ultrarelativistic limit, so that the source term is not spherically symmetric in the $k$-space.  This is analogous to the case of black hole formation with no scalar charge, $\epsilon_\mt{BH}=0$, discussed in Section~\ref{sec:scalar_collisions_2}.  Note also that the nonvanishing components of $\tilde{T}_{\mu\nu}$ are related to each other by $\tilde T_{xx}=\omega^2 \tilde T_{tt}/k_1^2$ and $\tilde T_{tx}=-k_1 \tilde T_{xx}/\omega$. By using these relations, in the following we shall write the perturbation equations in terms of $\tilde T_{tt}$ only.

A consistent ansatz for the metric perturbation of two point particles boosted along the $x_1$ direction is given by Eq.~\eqref{ansatz_staticp4v1b}.  The metric perturbations $h_{tx}$, $h_{yy}$, $h_\perp$ and $h_{tr}$ read as in Eqs.~\eqref{htX}--\eqref{htr}, whereas the dynamical variables $h_{xx}$, $h_{tt}$ and $h_{rr}$ satisfy an inhomogeneous system of equations.
The latter takes a simpler form after introducing the functions $z_\pm$ as defined in Eq.~\eqref{zpm}. 
We obtain one decoupled inhomogeneous equation for $z_-$,
\be
 z_-''=-\frac{r \left(rF'+4F\right) z_-' + \tilde{\omega}^2 z_- - 2 \left(1+\frac{\omega^2}{k_1^2}\right) {{\tilde T}_{tt}}}{r^2 F}\,, \label{sourcezm}
\ee
and a system of two coupled inhomogeneous equations for $z_+$ and $h_{rr}$,
\bea
z_+''&=&\frac{4F h_{rr}+\left(rF'+13F-15r^2 - \tilde{\omega}^2\right) z_+ - r \left(rF'+4F\right)z_+' 
  + \left(1-\frac{\omega^2}{k_1^2}\right) {\tilde T}_{tt} }{r^2 F}\,, \,\,\,\,\,\,\,\,\,\label{sourcezp}\\ [2mm]
 h_{rr}''&=&\frac{1}{2 r^2 F^2} \big[ \left( 5rF(F'-3r)+r^2F'^2+F^2\right) z_+ \nn \\ [1.5mm]
&& +2 \left(r^2 F'^2 - 10F^2 - \tilde{\omega}^2 F \right) h_{rr} - 2rF \left(rF'+4F\right) h_{rr}' \big]\,. \label{sourcehrr}
\eea
Note that no source term appears in Eq.~\eqref{sourcehrr}, which reads as in the vacuum case.

Our final goal is to compute the expectation value of the holographic stress-energy tensor in the dual theory.  This is computed explicitly in Appendix~\ref{appsec:holoTmunu} 
and the entire computation is detailed in Appendix~\ref{app:collisions}.  Here we only give the final result. In Fourier space and in the $v\to1$ limit, we get
\bea
 {\cal T}_{tt}(\omega,k_i)&=& \xi\left[\frac{30 \left(3 k_1^2-k^2\right) \left(k_1^2+\omega ^2\right)}{\omega  \left(\omega ^2-k_1^2\right) \left(2 \omega ^2-k^2+k_1^2\right) a_{\tilde\omega }} \right. \nn \\ [1mm]
   && \left. +\frac{15 \left(k^2+k_1^2\right) \alpha _{\tilde\omega }-L^4 \left(k^4+k_1^2 \omega ^2-3 k^2 \left(k_1^2+\omega ^2\right)\right) \beta _{\tilde\omega }}{\omega  \left(2 \omega ^2-k^2+k_1^2\right) \Delta _{\tilde\omega }}\right]\,, \label{holoTtt}\\ [2mm]
 {\cal T}_{x_1x_1}(\omega,k_i)&=&\xi\left[\frac{30 \left(k_1^2+\omega ^2\right)}{\omega\left(\omega^2-k_1^2\right) a_{\tilde\omega }}+\frac{15 \alpha _{\tilde\omega }+L^4 \left(\omega ^2-k^2+2 k_1^2\right) \beta _{\tilde\omega }}{\omega  \Delta _{\tilde\omega }}\right]\,, \label{holoTXX}\\ [2mm]
 {\cal T}_{tx_1}(\omega,k_i)&=&\xi\left[\frac{30 k_1 \left(k_1^2+\omega ^2\right)}{\omega ^2 \left(k_1^2-\omega ^2\right) a_{\tilde\omega }}-\frac{k_1 \left(15 \alpha _{\tilde\omega }+L^4 \left(k^2+\omega ^2\right) \beta _{\tilde\omega }\right)}{\omega ^2 \Delta _{\tilde\omega }}\right]\,, \label{holoTtX}\\
 [2mm]
 {\cal T}_{yy}(\omega,k_i)&=& \xi\left[\frac{4 L^4 (\omega^2-k^2) \beta _{\tilde\omega }-30 \alpha _{\tilde\omega }}{\omega  \Delta _{\tilde\omega }}\right]\,, \label{holoTyy}\\ [2mm]
 {\cal T}_{tx_2}(\omega,k_i)&=&\xi\left[\frac{30 k_2 \left(k_1^2+\omega ^2\right)}{\omega ^2 \left(2 \omega ^2-k^2+k_1^2\right) a_{\tilde\omega }} \right. , \nn \\ [2mm]
   && \left. - \frac{k_2 \left(15 \left(k_1^2+\omega ^2\right) \alpha _{\tilde\omega }+L^4 \left(k^2 (k_1^2-\omega^2 )+\omega ^2 \left(k_1^2+3 \omega ^2\right)\right) \beta _{\tilde\omega }\right)}{\omega ^2 \left(2 \omega ^2-k^2+k_1^2\right) \Delta _{\tilde\omega }}\right]\,, \,\,\,\,\,\,\,\,\,\,\,\,\,
   \label{holoTtx2}\\ [2mm]
 {\cal T}_{x_2x_2}(\omega,k_i)&=& \xi\left[-\frac{30 \left(k_1^2+\omega ^2\right)}{\omega  \left(2 \omega ^2-k^2+k_1^2\right) a_{\tilde\omega }} + \frac{15 \left(k_1^2+\omega ^2\right) \alpha _{\tilde\omega }}{\omega  \left(2 \omega ^2-k^2+k_1^2\right) \Delta_{\tilde\omega }} \right.\nn\\[1.5mm]
 && \hspace{-3.5em} \left.-\frac{L^4 \left(2 k_1^2 \left(k_1^2+k_3^2\right)+\left(3 k_1^2+4 k_3^2\right) \omega ^2-3 \omega ^4+k^2 \left(\omega ^2-3 k_1^2-2 k_3^2\right)\right) \beta _{\tilde\omega }}{\omega  \left(2 \omega ^2-k^2+k_1^2\right) \Delta _{\tilde\omega }}\right].\label{holoTx2x2}
\eea
where 
\be
\xi=\frac{im\gamma}{240 \pi G L r_0^4} \sac 
\Delta_{\tilde \omega}=\alpha_{\tilde \omega}\delta_{\tilde \omega}-\beta_{\tilde \omega}\gamma_{\tilde \omega} \,,
\ee
and $\alpha_{\tilde \omega}, \beta{\tilde \omega}, \gamma{\tilde \omega}$ and $\delta_{\tilde \omega}$ are related to the behavior near $r_0$ of the metric functions (see Table \ref{tab:asymp_collision}). 
The other five nonvanishing components, ${\cal T}_{x_3x_3}$, ${\cal T}_{tx_3}$, ${\cal T}_{x_1x_2}$, ${\cal T}_{x_1x_3}$ and ${\cal T}_{x_2x_3}$, can be obtained by using symmetry arguments and the tracelessness and divergence-free conditions
\be
 {\cal T}\equiv \eta^{mn}{\cal T}_{mn}=0\,,\qquad \partial_m {\cal T}^{mn}=0\,,
\ee
where $\eta_{mn}$ stands for the Minkowski metric in the five-dimensional space covered by coordinates $x^m=(t,x_1,x_2,x_3,y)$.  As a check on our calculations, we have computed all components of ${\cal T}_{mn}$ and checked that the conditions above are satisfied by virtue of the Einstein equations.

A relevant quantity is the energy flux across a sphere of radius $R$,
\be
 {\cal F}(\omega)=R^2\int_0^{2\pi}d\varphi \int_{-1}^{1} d\cos\theta \, {\cal T}_{tR}(\omega,R,\cos\theta,\varphi)\,,
\ee
where we have introduced spherical coordinates $(R=\sqrt{x_1^2+x_2^2+x_3^2},\theta,\varphi)$ and ${\cal T}_{tR}$ is the $R-t$ component of the holographic stress-energy tensor.  By performing a change of coordinates, the latter reads
\be
 {\cal T}_{tR}=\frac{x_1 {\cal T}_{tx_1}+x_2 {\cal T}_{tx_2}+x_3 {\cal T}_{tx_3}}{R}\,,
 \ee
where, by symmetry, ${\cal T}_{tx_3}$ can be obtained from Eq.~\eqref{holoTtx2} by replacing $k_2 \leftrightarrow k_3$.  Even though ${\cal T}_{tx_2}$ and ${\cal T}_{tx_3}$ have no particular symmetry, one can show that ${\cal T}_{tR}$ in Fourier space reads
\be
{\cal T}_{tR}(\omega,k_i)=\frac{1}{(2\pi)^3} \int d^3x_i\int d^3k_i'\, e^{-i(k_i-k'_i)x_i}\, \widehat{\cal T}_{tR}(\omega,x_i,k'_i) \,,
\label{TtR_F}
\ee
with
\be
\widehat{\cal T}_{tR}(\omega,x_i,k'_i) = \frac{k'_1 x_1 f_1(\omega^2,{k'_1}^2,{k'_\rho}^2)+(k'_2 x_2+k'_3 x_3) f_2(\omega^2,{k'_1}^2,{k'_\rho}^2)}{R}\,,
\ee
where the functions 
\be
f_1(\omega^2,k_1^2,k_\rho^2)=\frac{{\cal T}_{tx_1}}{k_1} \sac f_2(\omega^2,k_1^2,k_\rho^2)=\frac{{\cal T}_{tx_2}}{k_2} 
\ee
 have cylindrical symmetry in $k$-space, and recall $k_\rho^2\equiv k_2^2+k_3^2$.

\subsection{Numerical results for the stress-energy tensor of the dual theory}

The results~\eqref{holoTtt}--\eqref{holoTx2x2} were obtained in Fourier space.  By inverse-Fourier transforming from $k$-space, we can obtain the spatial dependence of the operators.  Note that Eqs.~\eqref{holoTtt}--\eqref{holoTtX} and Eq.~\eqref{TtR_F} explicitly depend on $k^2$ and $k_1^2$ only and the inverse Fourier transform can be performed in cylindrical coordinates as shown in the scalar case (cf. Eq.~\eqref{intcylcoord}). The term~\eqref{holoTyy} only depends on $k^2$ and its transform can be evaluated as in the spherically symmetric scalar case (cf. Eq.~\eqref{intsphersymm}). Finally, the last two terms~\eqref{holoTtx2} and~\eqref{holoTx2x2} are less symmetric because they explicitly depend on $(k_1,k_2,k_3)$.

After a tedious but straightforward manipulation, we obtain
\begin{flalign}
& {\cal T}_{tt}(\omega,x_i)=-\frac{i \xi  r_0^2}{4L^4} \left\{\sum_n^{\rm II}\frac{120 \tw{n} H_0^{(1)}(i \rho  \tw{n}) \sin(x_1\omega)}{f_n}\right.\nn\\
& \left.-\int_{-\infty}^\infty dk_\rho \frac{k_\rho H_0^{(1)}(k_\rho \rho )}{\omega \pi} \left[-\sum_n^{\rm II} \frac{30 \tw{n}  e^{i k_{1,n} x_1} \left(-2 k_{1,n}^2+k_\rho^2\right) \left(k_{1,n}^2+\omega ^2\right)}{k_{1,n}  \left(k_\rho^2-2 \omega ^2\right) \left(k_\rho^2+\tw{n}^2\right) f_n}\right.\right.\nn\\
& \left.\left.-\sum_n^{\rm I,II} \tw{n} e^{i k_{1,n} x_1}\frac{15 \left(2 k_{1,n}^2+k_\rho^2\right) \alpha_n+L^4 \left(2 k_{1,n}^4-k_\rho^4+3 k_\rho^2 \omega ^2+k_{1,n}^2 \left(k_\rho^2+2 \omega ^2\right)\right) \beta_n}{k_{1,n}  \left(2 \omega ^2-k_\rho^2 \right) g_n}\right]\right\}\,,   \label{holoTttF}
\end{flalign}
\begin{flalign}
& {\cal T}_{x_1x_1}(\omega,x_i)=-\frac{i \xi  r_0^2}{4L^4} \left\{\sum_n^{\rm II} \frac{120  \tw{n} H_0^{(1)}(i \rho  \tw{n}) \sin(x_1\omega)}{f_n}-\int_{-\infty}^\infty dk_\rho  \frac{k_\rho H_0^{(1)}(k_\rho \rho )}{\pi\omega}\right.\nn\\
& \left. \times \left[ -\sum_n^{\rm II} \frac{30 \tw{n} e^{i k_{1,n} x_1} \left(k_{1,n}^2+\omega ^2\right)}{k_{1,n} \left(k_\rho^2+\tw{n}^2\right) f_n} - \sum_n^{\rm I,II}\tw{n}  e^{i k_{1,n} x_1}  \frac{15 \alpha_n+L^4 \left(k_{1,n}^2-k_\rho^2+\omega ^2\right) \beta_n}{k_{1,n} g_n} \right]\right\}\,,
\end{flalign}
\begin{flalign}
&{\cal T}_{tx_1}(\omega,x_i)=-\frac{i \xi  r_0^2}{4L^4} \left\{\sum_n^{\rm II} \frac{120 i  \tw{n} H_0^{(1)}(i \rho  \tw{n})\cos(x_1\omega) }{f_n}-\int_{-\infty}^\infty dk_\rho \frac{k_\rho H_0^{(1)}(k_\rho \rho ) }{\pi\omega^2}\right.\nn\\
& \left.\times \left[\sum_n^{\rm II} \frac{30 \tw{n}  e^{i k_{1,n} x_1} \left(k_{1,n}^2+\omega ^2\right)}{\left(k_\rho^2+\tw{n}^2\right) f_n} + \sum_n^{\rm I,II}\tw{n} e^{i k_{1,n} x_1} \frac{15 \alpha_n+L^4 \left(k_{1,n}^2+k_\rho^2+\omega ^2\right) \beta_n}{g_n}\right] \right\}\,,
\end{flalign}
\begin{flalign}
& {\cal T}_{yy}(\omega,x_i)= \frac{\xi}{\pi}\frac{r_0^2}{L^4}\sum_n^{\rm I,II}\frac{\tw{n} e^{i k_n R}}{g_n\omega R}\left[15\alpha_n-2 L^4\tw{n}^2 \beta_n\right]\,,  \label{TyyF}\\  \nn\\
& {\cal T}_{tx_2}(\omega,x_i)= \frac{i \xi  r_0^2}{2L^4\pi ^2 \omega ^2 }  \int_{-\infty}^\infty dk_2\int_{-\infty}^\infty  dk_3 e^{i(k_2 x_2+k_3 x_3)}\left[\sum_n^{\rm II} \frac{15 e^{i k_{1,n} x_1} k_2 \left(k_{1,n}^2+\omega ^2\right) \tw{n}}{k_{1,n} \left(k_2^2+k_3^2-2 \omega ^2\right) f_n}\right.\nn\\
 & \left.-\sum_n^{\rm I,II} \frac{e^{i k_{1,n} x_1} k_2 \tw{n} \left(15 \left(k_{1,n}^2+\omega ^2\right) \alpha_n+L^4 \left(k_{1,n}^4+k_{1,n}^2 \left(k_2^2+k_3^2\right)+\omega ^2 \left(-k_2^2-k_3^2+3 \omega ^2\right)\right) \beta_n\right)}{2 k_{1,n} \left(k_2^2+k_3^2-2 \omega ^2\right) g_n}\right] \,,
\end{flalign}
\begin{flalign}
& {\cal T}_{x_2x_2}(\omega,x_i)= -\frac{i \xi  r_0^2}{2L^4 \pi ^2 \omega}  \int_{-\infty}^\infty dk_2\int_{-\infty}^\infty  dk_3 e^{i(k_2 x_2+k_3 x_3)} \nn\\
& \left[\sum_n^{\rm II} \frac{15 e^{i k_{1,n} x_1} \left(k_{1,n}^2+\omega ^2\right) \tw{n}}{k_{1,n}   \left(k_2^2+k_3^2-2 \omega ^2\right) f_n}-\sum_n^{\rm I,II} \frac{15 \left(k_{1,n}^2+\omega ^2\right) \alpha_n e^{i k_{1,n} x_1} \tw{n}}{2 k_{1,n}   \left(k_2^2+k_3^2-2 \omega^2\right) g_n}\right.\nn\\
& \left.+\frac{L^4 \left(k_{1,n}^4+2 k_3^4-5 k_3^2 \omega ^2+3 \omega ^4+k_{1,n}^2 \left(3 k_2^2+3 k_3^2-4 \omega ^2\right)+k_2^2 \left(2 k_3^2-\omega ^2\right)\right) \beta_n e^{i k_{1,n} x_1} \tw{n}}{2 k_{1,n}  \left(k_2^2+k_3^2-2 \omega ^2\right) g_n}\right]\,.  \label{holoTx2x2F}
\end{flalign}
In the expressions above we omitted the dependence $e^{-i\omega t}$, and $\alpha_n$, $\beta_n$ are shorthand notations for the functions $\alpha_{\tilde\omega}$, $\beta_{\tilde\omega}$ evaluated at $\tilde\omega=\tw{n}$.  The symbol $\sum_n^{II}$ refers to a sum over the infinite ($n=0,1,\ldots$) modes of the vector-II type family in Table~\ref{tab:resonance} (i.e. to the roots of $a_{\tilde\omega}$), whereas $\sum_n^{I,II}$ refers to the sum over both scalar-type and vector-II type families (i.e.~to the roots of $\Delta_{\tilde\omega}$).  Finally, $f_n$ is defined as in Eq.~\eqref{anpoles} and $g_n$ is defined through the behavior of $\Delta_{\tilde\omega}$ close to its poles, ${\Delta_{\tilde\omega}}\sim g_n(\tilde\omega-\tw{n})$.  Our numerical data are well fitted by $g_1=0.015$ and
\be
  g_n=\frac{2\pi}{{\cal P}}14.53(-1)^{n+1}\tw{n}^{-5.98},\quad n>1\,,
\ee
where $\tw{n}$ are the scalar-type gravitational modes for the two families listed in Table~\ref{tab:resonance}. 

\begin{figure*}[tb]
\begin{center}
\includegraphics[width=8.5cm,angle=0,clip=true]{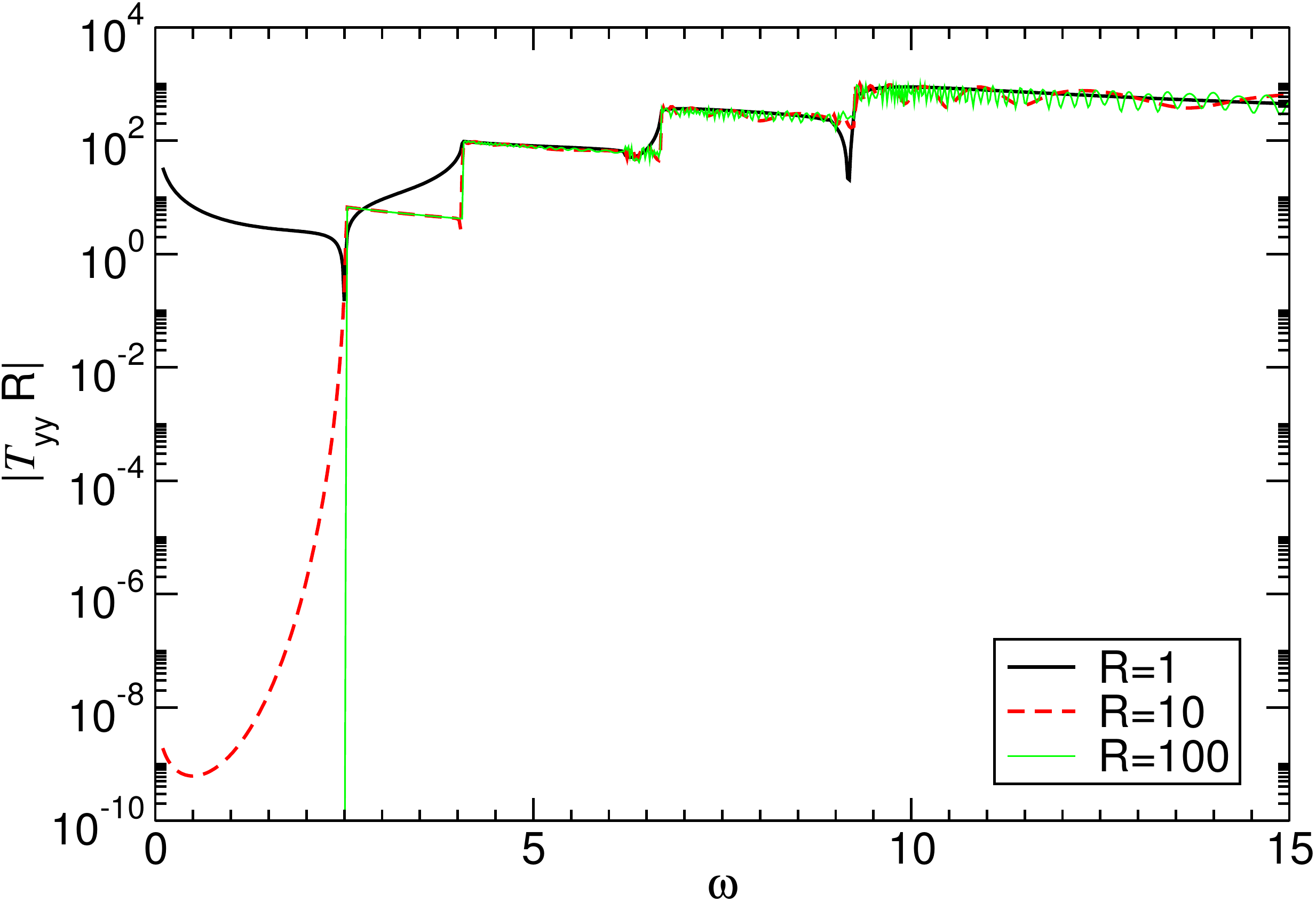}
\caption{The operator $|{\cal T}_{yy}(\omega,R)|R$ (modulo a coefficient proportional to $\xi$, cf. Eq.~\eqref{TyyF}) for $N=5$. The spectrum is qualitatively similar to the spherically symmetric emission in the scalar case, cf. Fig.~\ref{fig:spectrum}.
\label{fig:Tyy}}
\end{center}
\end{figure*}

\begin{figure*}[tb]
\begin{center}
\begin{tabular}{cc}
\includegraphics[width=7.4cm,angle=0,clip=true]{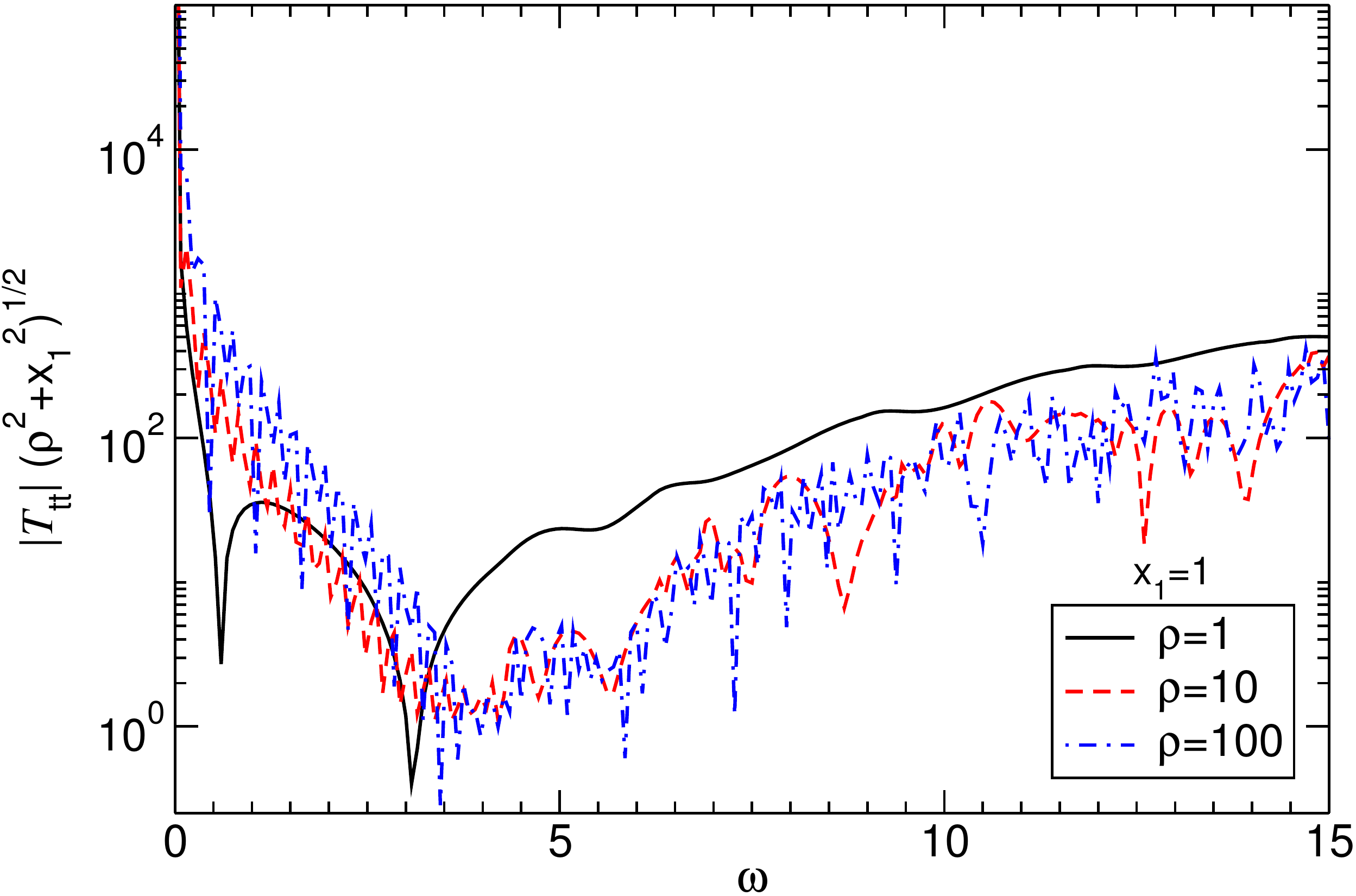}&
\includegraphics[width=7.4cm,angle=0,clip=true]{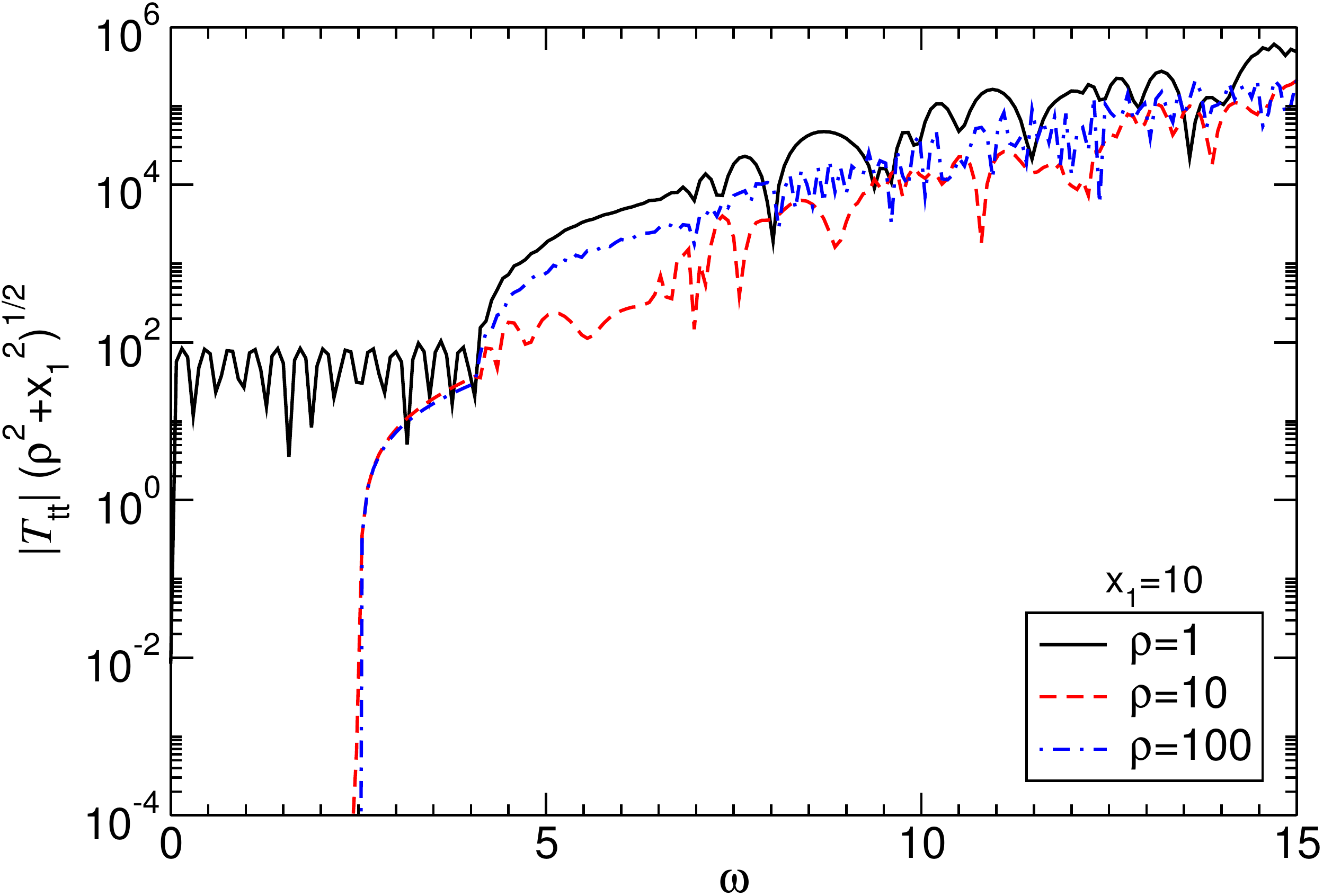}\\
\includegraphics[width=7.4cm,angle=0,clip=true]{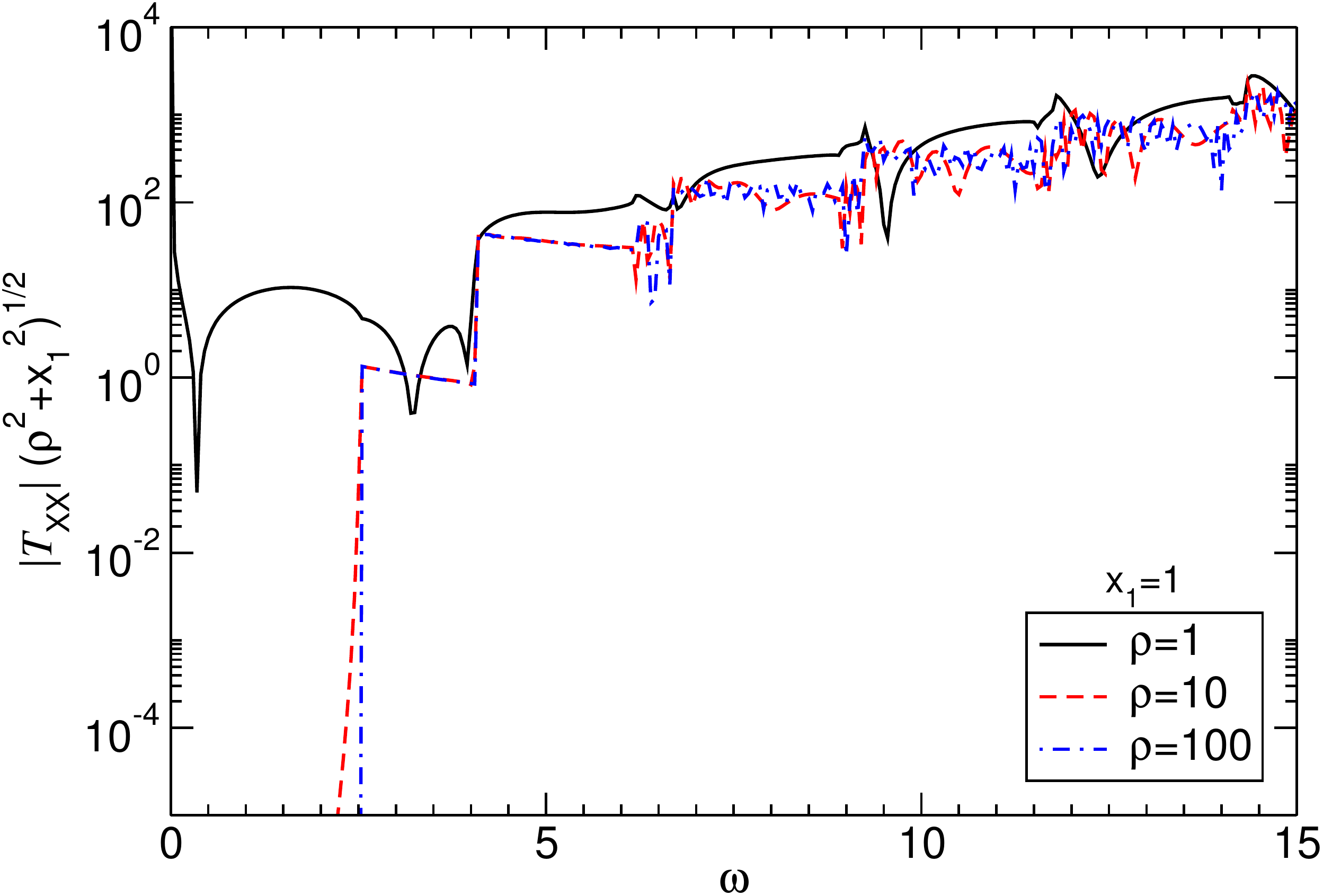}&
\includegraphics[width=7.4cm,angle=0,clip=true]{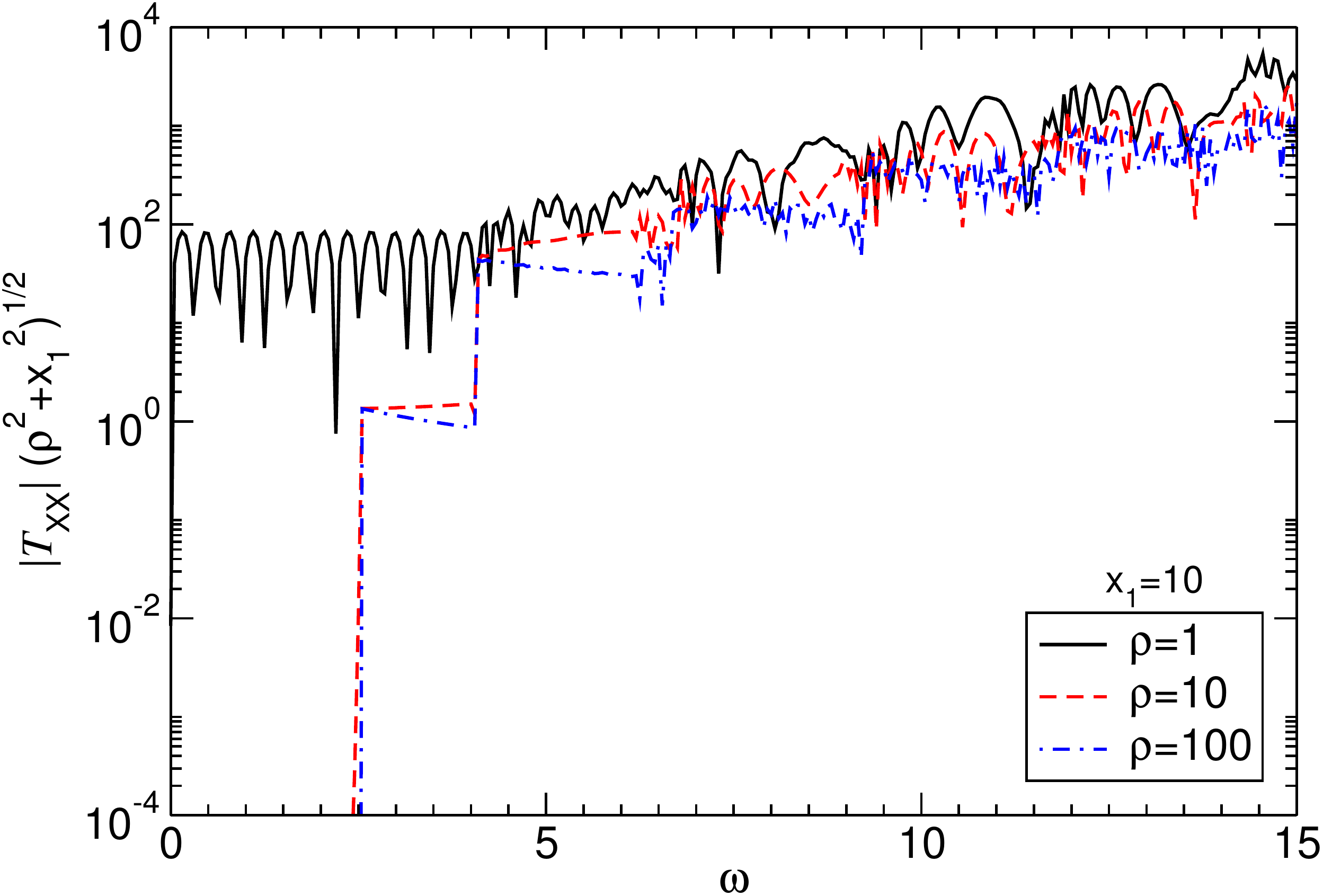}\\
\includegraphics[width=7.4cm,angle=0,clip=true]{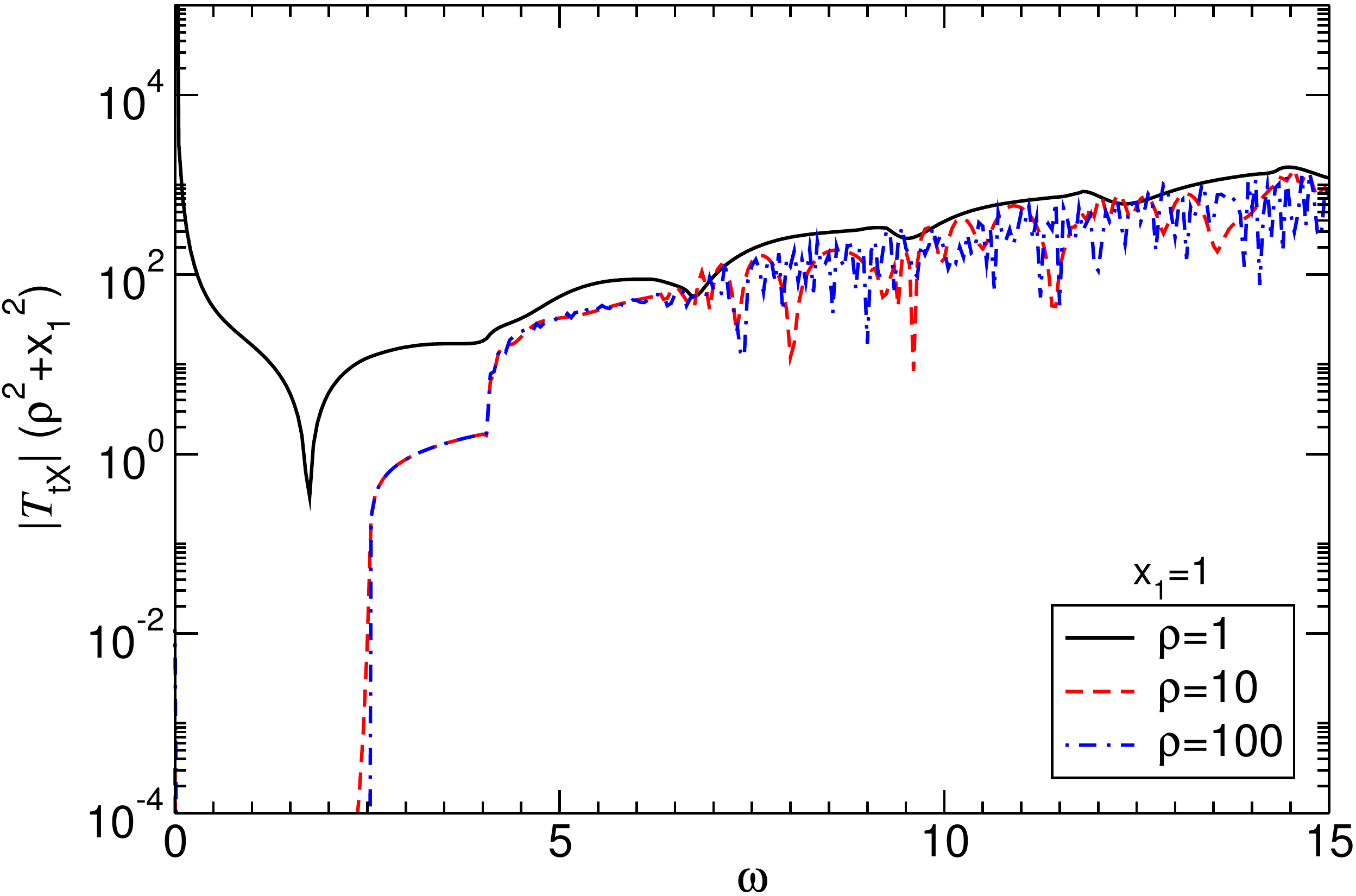}&
\includegraphics[width=7.4cm,angle=0,clip=true]{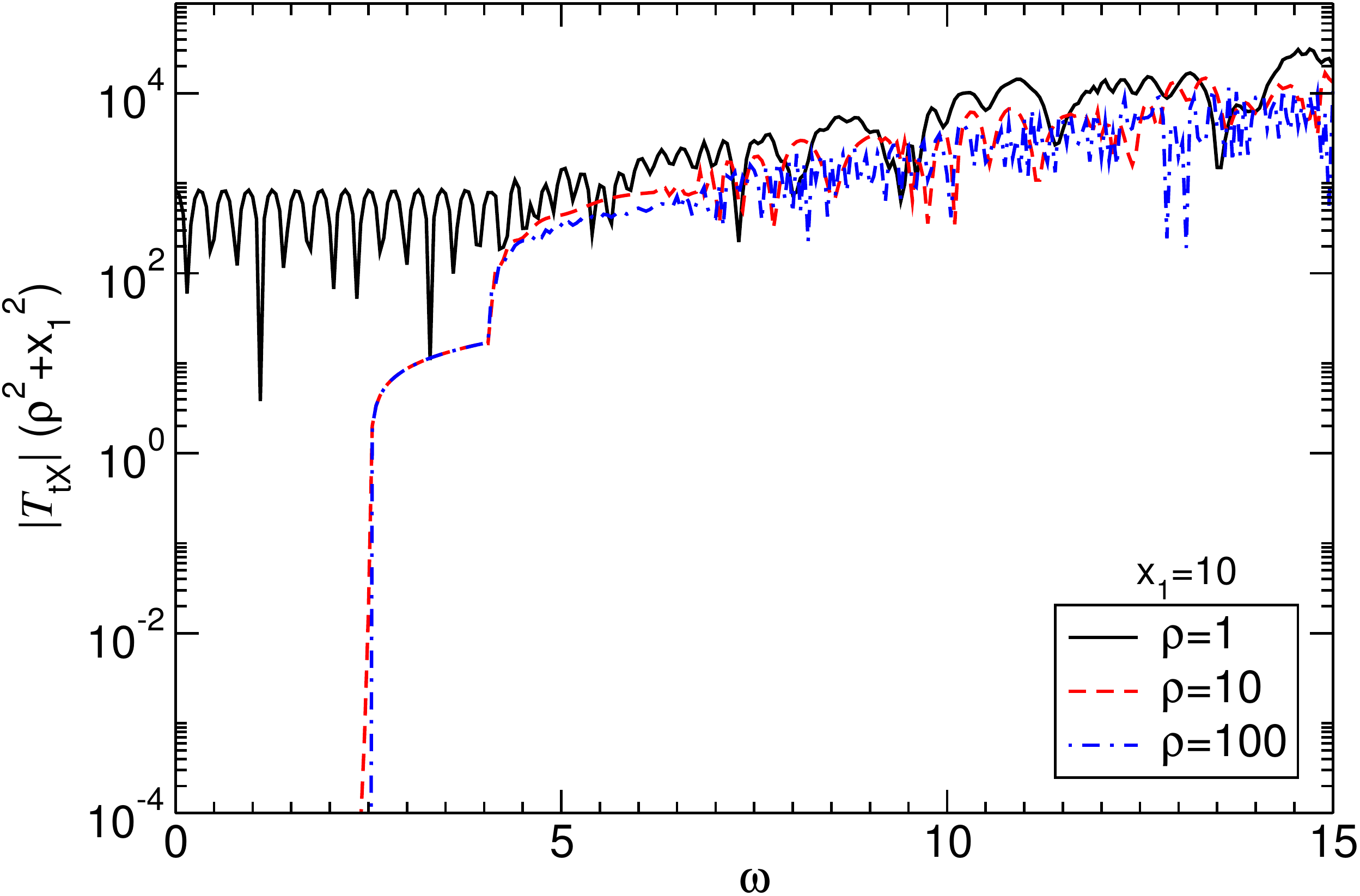}
\end{tabular}
\caption{The operators $|{\cal T}_{tt}(\omega,x_1,\rho)|\sqrt{x_1^2+\rho^2}$, $|{\cal T}_{xx}(\omega,x_1,\rho)|\sqrt{x_1^2+\rho^2}$ and $|{\cal T}_{tx}(\omega,x_1,\rho)|(x_1^2+\rho^2)$ (modulo a coefficient proportional to $\xi$) for $N=11$. Left panels: $x_1=1$, Right panels: $x_1=10$. The spectrum is qualitatively similar to the cylindrically symmetric emission in the scalar case, cf. Fig.~\ref{fig:spectrum_scalar2}.
\label{fig:Ttt_TXX_T_tX}}
\end{center}
\end{figure*}

As in the scalar case, some of the integrals above have to be performed numerically.  In Fig.~\ref{fig:Tyy} we show the operator $|{\cal T}_{yy}(\omega,R)|$, which is spherically symmetric and qualitatively similar to the scalar case shown in Fig.~\ref{fig:spectrum}.  In Fig.~\ref{fig:Ttt_TXX_T_tX} we show the operators that are cylindrically symmetric.  In this case the spectrum is qualitatively similar to the cylindrically symmetric emission in the scalar case, cf. Fig.~\ref{fig:spectrum_scalar2}. 
%

\begin{figure}[tb]
\begin{center}
\includegraphics[width=8.5cm,angle=0,clip=true]{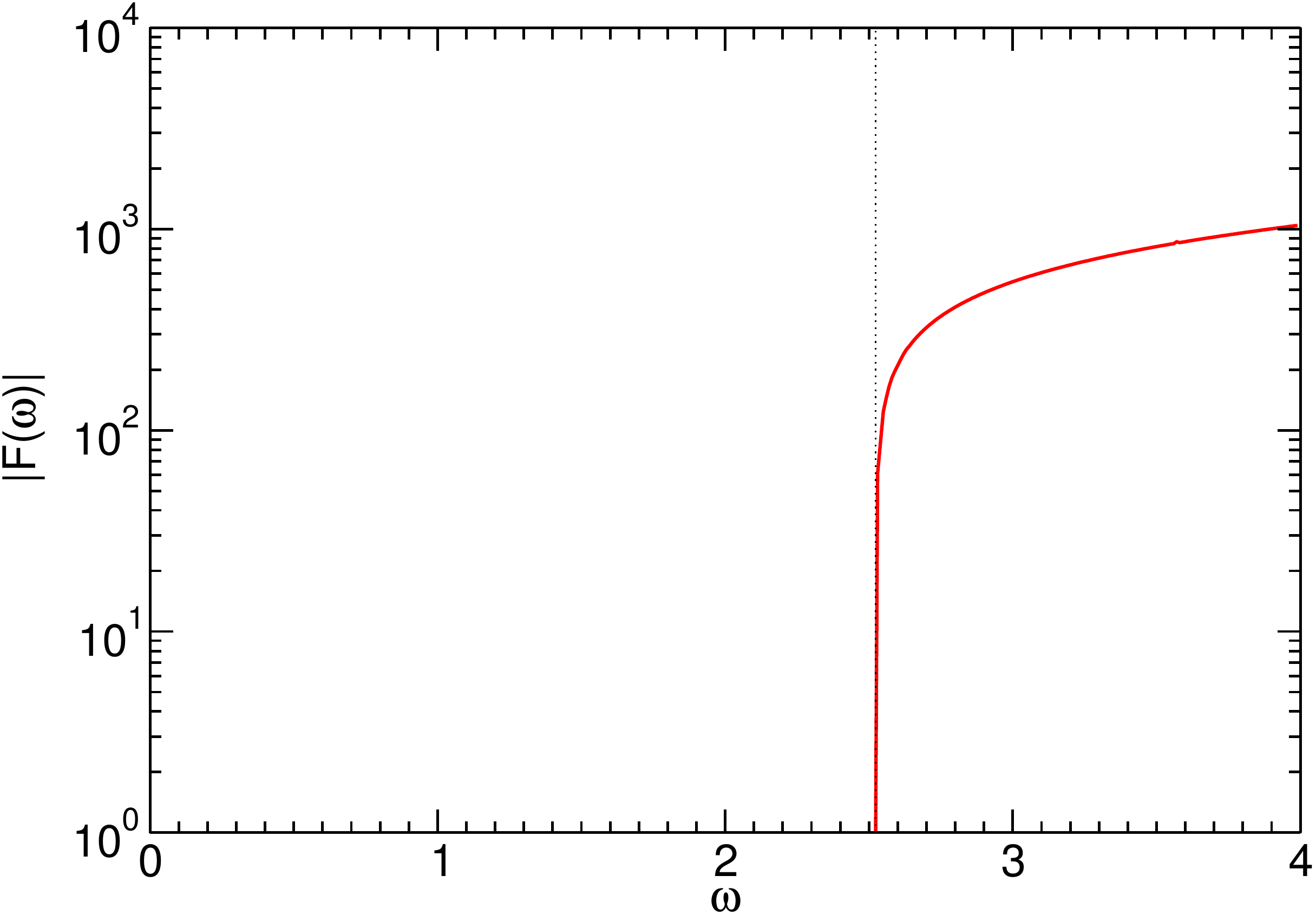}
\caption{Energy flux $|{\cal F}(\omega)|$ (modulo a coefficient proportional to $\xi$) across a sphere at large distance as a function of the frequency. The flux is zero for $\omega$ smaller than the first normal frequency and it increases monotonically for $\omega\gtrsim 2.52$.
\label{fig:flux}}
\end{center}
\end{figure}

Finally, we can compute the energy flux.  Let us first compute the inverse-Fourier transform of ${\cal T}_{tR}$.  From Eq.~\eqref{TtR_F}, we obtain
\bea
 {\cal T}_{tR}(\omega,\rho,x_1)&=&\frac{\pi}{(2\pi)^2R}\int_{-\infty}^{+\infty} dk_1 e^{i k_1 x_1}\int_{-\infty}^{+\infty} dk_\rho k_\rho \left[k_1 x_1 H_0^{(1)}(k_{\rho} \rho) f_1(\omega^2,k_1^2,k_\rho^2) \right. \nn\\
 && \hspace{5cm} \left. + i k_\rho \rho H_1^{(1)}(k_{\rho} \rho)f_2(\omega^2,k_1^2,k_\rho^2)\right]\,,
\eea
where cylindrical coordinates are related to spherical ones via $\rho=R \sin\theta$, $x_1=R\cos\theta$ and there is no explicit dependence on $\varphi$.  Therefore, the energy flux reads
\be
 {\cal F}(\omega)=4\pi R^2\int_0^{1}d\cos\theta \, {\cal T}_{tR}(\omega,R\sin\theta,R\cos\theta)\,,\label{flux}
\ee
where, using the symmetries of the problem, we integrate over half of the $\cos\theta$-space.  The energy flux as a function of the frequency is shown in Fig.~\ref{fig:flux}.
%
%
As expected, the flux is vanishing for frequencies smaller than the first normal mode, $\omega<\tw{1}\sim2.52$.

\section{Discussion and Conclusions}
\label{sec:Conclusion}
Gravity-dominated high-energy collisions are a fascinating topic: the efficiency for gravitational wave emission
is huge, and these processes typically give rise to the largest known luminosities. In addition, fine-tuned
collisions of this kind provide useful tests of Cosmic Censorship. In four-dimensional, asymptotically flat spacetimes the simulations of such events took several decades to perform and turned out to be `mostly linear' in that waveforms are smooth and simple perturbative models capture most of the physics \cite{Berti:2010ce,Cardoso:2002ay,Sperhake:2008ga,Sperhake:2009jz}.

By contrast, full non-linear simulations of collisions in AdS spacetime are in their infancy \cite{Chesler:2010bi,Wu:2011yd,Casalderrey-Solana:2013aba,vanderSchee:2013pia}. 
Validation of {\it any} such simulations requires benchmarking with perturbative results, either for the final ringdown stage or for intermediate
stages of the process. We have explored a simple and compelling model for such collisions. Some of the important physical observables, such as 
total radiated energy and time-dependence of the stress-energy tensor depend quantitatively on the magnitude of the cutoff, which in turn
depends sensitively on the (unknown) final state. Nevertheless, we also obtained what we expect are universal features to be seen in any simulation and experiments:
Yukawa-type potentials for static particles and power-law decay of perturbations at late times, characteristic of massive fields. 
The energy distribution of the particles produced during such events is also cutoff-independent.

As shown in Figs.~\ref{fig:spectrum}, \ref{fig:spectrum_scalar2}, \ref{fig:Tyy} and \ref{fig:Ttt_TXX_T_tX}, the spectrum is exponentially suppressed for frequencies smaller than the fundamental mode of the AdS-soliton, \ie smaller than the mass gap in the gauge theory. Using Eq.~\eqref{memory}, this result implies that the waveform shows no memory effect in this spacetime. This property is most likely related to the fact that the AdS boundaries are timelike and can be reached by the emitted radiation in a finite time.

Furthermore, for larger frequencies the spectrum shows a peculiar upward-stairway structure, which is formed by various plateaux corresponding to the excitation of various normal modes with increasing overtone number. 
From the dual theory perspective, as more massive states becomes available in the confining gauge theory, the energy density $dE/d\omega$ for a given frequency grows monotonically with the energy. For example, for the scalar charge-conserving collisions studied in Section~\ref{sec:chargeconserving}, Eq.~\eqref{Oscalar} predicts a distribution
$dE/d\omega\sim \omega^{2.53}$ at large energies.
For the gravitational emission, our numerical results are less accurate, but the behavior is also consistent with a $w^{\Upsilon}$ dependence, with $2.5\lesssim \Upsilon \lesssim 3$ for $T_{yy}, T_{XX}$ and $T_{tX}$.

These qualitative features are robust because they follow from the fact that the fields in a confining geometry can be decomposed into a discrete set of massive four-dimensional fields propagating in Minkowski spacetime. From the gauge theory viewpoint these are just the different glueball states in the theory. This reduced description in Minkowski spacetime allows us to give a very simple and intuitive picture of the radiation field coming out of the collision, along the same lines as in electromagnetism. Imagine that we regularize the problem by assuming that the two particles come to rest in a small but finite amount of time $\delta t$, so that they start slowing down at $t=0$ and come to a full stop at $t=\delta t$. After this time, the solution has three regions. The  field sufficiently far from the particles, at $R \gg t$, is just the sum of the two boosted fields created by the incoming point particles. In the opposite limit, $R \ll t$, the field is the spherically symmetric solution created by the resulting particle at rest. In between these regions there is a thin shell of thickness $\sim \delta t$ in which the field smoothly connects these two solutions. This `dislocation' is the gravitational wave, and it is not spherically symmetric because it must connect a spherically symmetric solution (at small $R$) to a non-spherically symmetric one (at large $R$). 

Note that the only difference between the present situation and that in electromagnetism is that in our case the effective four-dimensional fields are massive. This is not an essential difference though, since the front wave of a massive field still propagates at the speed of light even if it is followed by slower modes. As in electromagnetism, the regularized picture illustrates the fact that the radiated energy diverges in the limit 
$\delta t \to 0$ and hence explains the need for a cutoff. Indeed, the derivative of the field across the shell of thickness $\delta t$ is of order $v/\delta t$, and hence the total radiated energy scales as $\delta t \times (v /\delta t)^2$.

Although our model has no internal information about the colliding objects, presumably  this is not a serious limitation  if the collision is sufficiently energetic: horizon formation will cloak any multipolar strucutre of the colliding particles
and presumably a point-particle approximation is just as good as any other \cite{East:2012mb,Choptuik:2009ww,Sperhake:2012me}.

A more important limitation is the fact that our linear approximation cannot describe strong-gravity effects, in particular the formation of a black hole and its subsequent relaxation. This means that the part of the gravitational radiation that is accurately captured by our approximation is that near the future lightcone of the collision point. This pulse will be followed by radiation emitted in the relaxation process to the final, equilibrium state. The crudest features of the final state can be accounted for by introducing appropriate cutoffs, as we have explained. However, a more precise determination of the relaxation dynamics to this final state will require a non-linear analysis, which we leave for future work.

\section*{Acknowledgements}

This work was supported by NSF Grant PHY-0900735, by the Intra-European Marie Curie contract aStronGR-2011-298297, by FCT -- Portugal contract 
no.~SFRH/BPD/47332/2008 and by FCT -- Portugal through CERN/FP/123593/2011.

\noindent
Computations were performed on the ``Baltasar Sete-Sois'' cluster at IST, the
cane cluster in Poland through PRACE DECI-7 ``Black hole dynamics in metric
theories of gravity'', on Altamira in Cantabria through BSC grant
AECT-2012-3-0012, on Caesaraugusta in Zaragoza through BSC grants
AECT-2012-2-0014 and AECT-2012-3-0011, XSEDE clusters SDSC Trestles and NICS
Kraken through NSF Grant~No.~PHY-090003.

\noindent
VC acknowledges partial financial support provided under the European Union's FP7 ERC Starting Grant ``The dynamics of black holes: testing the limits of Einstein's theory'' grant agreement no. DyBHo--256667.  This research was supported in part by Perimeter Institute for Theoretical Physics. 
Research at Perimeter Institute is supported by the Government of Canada through 
Industry Canada and by the Province of Ontario through the Ministry of Economic Development 
$\&$ Innovation.

\noindent
DM and RE are supported by grants 2009-SGR-168, MEC FPA2010-20807-C02-01, MEC FPA2010-20807-C02-02 and CPAN CSD2007-00042 Consolider-Ingenio 2010. 

\noindent
DM is also partially supported by  the ERC Starting Grant ``HoloLHC-306605'', and thanks the Kavli Institute for Theoretical Physics at the University of California, Santa Barbara for hospitality during the completion of this work.

\appendix

\section{The function \texorpdfstring{$a_{\tilde\omega}$}{Function aw}\label{app:aomega}}

\begin{figure*}[tb]
\begin{center}
\includegraphics[width=8.5cm,angle=0,clip=true]{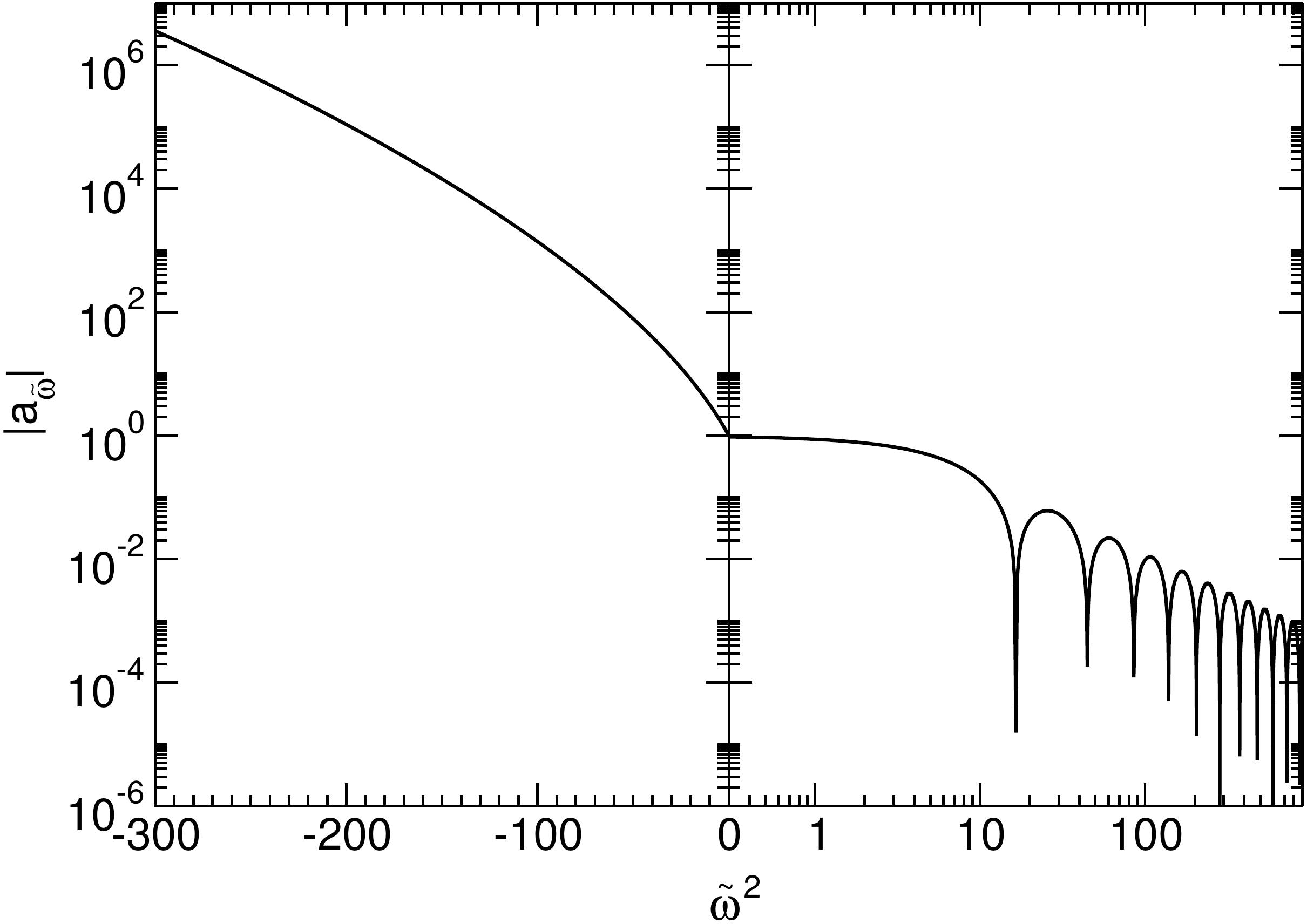}
\caption{The function $a_{\tilde\omega}$ computed numerically for positive and negative values of $\tilde{\omega}^2$. Note that the plot is shown in a log-log scale for $\tilde{\omega}^2>0$ and in a log-linear scale for $\tilde{\omega}^2<0$.
\label{fig:a_VS_wt2}}
\end{center}
\end{figure*}

Because it plays such an important role in our analysis, in this appendix we present the main properties of the function $a_{\tilde \omega}$ introduced in~\eqref{eq:aw}.  The overall behavior is shown in Figure~\ref{fig:a_VS_wt2}. 

\paragraph{Small frequency behavior.}
For $\tilde \omega=0$, the homogeneous equation associated with Eq.~\eqref{eqscalarsource} or more generally with Eq.~\eqref{waveeqscalarfields} can be solved exactly, with the result $\psitwo=\log\frac{r^5}{r^5-r_0^5},\,\psione=1$.  Thus $a_{\tilde\omega}(\tilde\omega=0)=-1$.  A systematic scheme to compute corrections to $a_{\tilde \omega}(\omega^2-k^2)$ for small values of the argument can be found as follows.  The only possible nontrivial expansion is of the form $\Psi=\sum_{n=0}{\tilde \omega}^{2n}\Psi^{(n)}$, in which case one gets
\be
r(r^5-r_0^5)\Psi^{(n)''}+(6r^5-r_0^5)\Psi^{(n)'}=-r_0^2r^2\Psi^{(n-1)}\,.
\ee
The two homogeneous solutions of interest are
\beq
\Psi^{(n)}_{h,2}&=&\log\frac{r^5}{r^5-r_0^2}\,,\\
\Psi^{(n)}_{h,1}&=&1\,.
\eeq
The solution which is regular everywhere is
\be
\Psi^{(n)}=-r_0^2 \left[\Psi^{(n)}_{h,1}(r)\int_r^{\infty} d\rbar\,\frac{\rbar\,\Psi^{(n)}_{h,2}(\rbar)\Psi^{(n-1)}(\rbar)}{(\rbar^5-r_0^5) {\cal W}{^{(n)}}(\rbar)}+\Psi^{(n)}_{h,2}(r)\int_{r_0}^{r} d\rbar\,\frac{\rbar\,\Psi^{(n)}_{h,1}(\rbar)\Psi^{(n-1)}(\rbar)}{(\rbar^5-r_0^5) {\cal W}{^{(n)}}(\rbar)}\right]\,,
\ee
where ${\cal W}^{(n)} \equiv \Psi^{(n)}_{h,1} {\Psi^{(n)}_{h,2}}' - \Psi^{(n)}_{h,2} {\Psi^{(n)}_{h,1}}'$.
With this procedure, what we have accomplished is to maintain the singularity behavior of $\psitwo^{(0)}$ at $r_0$, at the expense of changing the normalization at infinity.  This way, the constant $a_{\tilde \omega}$ needs to be redefined.  In particular, we get that close to the AdS boundary we have
\be
\Psi^{(n)}(r\sim \infty)\simeq -\frac{r_0^7}{r^5}\int_{r_0}^{\infty} d\rbar\,\frac{\rbar\,\Psi^{(n-1)}(\rbar)}{(\rbar^5-r_0^5) {\cal W}{^{(n)}}(\rbar)} = \frac{r_0^2}{5r^5}\int_{r_0}^{\infty} d\rbar\,\rbar^2\Psi^{(n-1)}(\rbar)\,.
\ee

As an example, let's work to second order in $\tilde \omega$.  We find that close to the tip of the soliton, $r=r_0$, the singular solution behaves like $\psitwo=(1+\epsilon\, \tilde\omega^2)\log\frac{r^5}{r^5-r_0^5}$ with
\be
\epsilon=-\frac{1}{5}\int_{1}^{\infty} d\xi\;\xi^2\log\frac{\xi^5}{\xi^5-1}
\approx-0.13228\,.
\ee
Thus, to summarize,  we have 
\be
a_{\tilde\omega}=-1+0.13228\,{\tilde \omega}^2+{\cal O}(\tilde \omega)^4\,.
\ee
The numerical results are in perfect agreement with this prediction:  we find $a_{\tilde \omega}=-1+0.13229\,{\tilde \omega}^2$ from fitting the data to a parabola dependence.  The procedure can be easily extended to higher orders.

\paragraph{Large momentum behavior.}
At large {\it negative values} of ${\tilde \omega}^2$, we find an exponential increase 
\be
a_{\tilde \omega} \sim -7\times 10^{-3}e^{1.2\,|\tilde \omega|}\,,\quad {\tilde \omega}{^2}\to-\infty\,.\label{fit_a_negative}
\ee
The WKB analysis can be used to understand this scaling, giving a large $k$ expansion of the wavefunction~\cite{fedoryuk} 
\be
\Psi(\rho) \sim e^{\alpha(\rho) k}\,,
\ee
with 
\be
\alpha(\rho)=L\int_\rho^{\infty}\frac{d\rho'}{\rho'\sqrt{F(\rho')}}=\frac{2}{3}\left[\rho^{3/2}\sqrt{\rho^5-1}\, {}_2F_1\left(\frac{4}{5},1,\frac{13}{10},\rho^5\right)+\frac{(-1)^{1/5}}{\sqrt{\pi}}\Gamma\left(\frac{1}{5}\right)\Gamma\left(\frac{13}{10}\right) \right]\,.
\ee
Here ${}_2F_1$ is an hypergeometric function, while $F$ was defined in~\eqref{soliton} as a function of $r=r_0\rho$.  Close to the tip, $\rho=1$, we find $\alpha(\rho=1)\simeq 1.25$, in good agreement with a numerical fit of our data.

\paragraph{Large frequency behavior.}
Finally, at large {\it positive values} of ${\tilde \omega}^2$, we find an oscillating power-law decay of the form 
\be
a_{\tilde \omega} \sim 3.9\,{\tilde \omega}^{-2.53}\sin{\left(\frac{2\pi}{5.012} \tilde\omega-5.25\right)}\,,\quad {\tilde \omega}{^2}\to+\infty\,.\label{fit_a_positive}
\ee
Notice that the period of this ringing pattern is roughly twice as large as the spacing of the resonant modes in Table~\ref{tab:resonance}, which is a good consistency check.

\section{Vector-I gravitational perturbations}
\label{app:vector_gravitational}

In the main text we studied gravitational modes which are excited by the axi\-symmetric collision described in this work. These modes were referred to as gravitational scalar and vector-II modes in Table~\ref{tab:resonance}.
However, other types of perturbations are excited more generically.  Here we briefly show the existence of at least one other perturbation mode, which is vector-type (with respect to the $(t,r)-$subspace) and takes the form~\eqref{metricpert} with the perturbation quantities defined as
\bea
h_{\mu \nu}= \left[
 \begin{array}{cccccc}
0 & 0 & 0 & 0 & 0 & h_{ty}(r)\\ 
0 & 0 & 0 & 0 & 0 & h_{ry}(r) \\ 
0 & 0 & 0 & 0 & 0 & 0\\ 
 0 & 0 & 0 & 0 & 0 & 0\\
  0 & 0 & 0 & 0 & 0 & 0\\
h_{ty}(r) & h_{ry}(r) & 0 & 0 & 0 & 0
\end{array}\right]e^{-i\omega t +ik_i x_i}\,,
\label{ansatz_staticp4v1bvector}
\end{eqnarray}
where an integral over $\omega$ and $k_i$ ($i=1,2,3$) is implicit.  Inserting the ansatz above into the Einstein equations we get
\begin{flalign}
E_{xy}:&\;(4r^5+1)h_{ry}+ir^2\omega h_{ty}+r(r^5-1)h_{ry}'=0\,,\\
E_{ry}:&\;-r(r^5-1)\tilde\omega^2 h_{ry}-i\omega(2r^5+3) h_{ty}+i\omega r(r^5-1)h_{ty}'=0\,,\\
E_{ty}:&\;-i\omega r(4r^5+1)h_{ry}+ (k^2r^3+6(r^5-1))h_{ty}-ir(r^5-1)\left(r\omega h_{ry}'-2ih_{ty}'-irh_{ty}''\right)=0\,,
\end{flalign}
where primes stand for derivatives in $r$ and note we are setting $r_0=L=1$.  One can solve $E_{ry}$ for $h_{ry}$ and its derivatives,
\be
h_{ry}=\frac{i\omega}{r(r^5-1)\tilde\omega^2} \left(r(r^5-1)h_{ty}'-(2r^5+3)h_{ty}\right)\,,
\label{hryeq}
\ee
and plug it back in the remaining two Einstein equations.  These are then both equivalent to a single equation for $h_{ty}$,
\be
r^2h_{ty}''+2rh_{ty}'+\left(\tilde\omega^2\frac{r^3}{r^5-1}-6\right)h_{ty}=0\,. \label{eqVectI}
\ee
Regular solutions have to decay like $h_{ty}\sim r^{-3}$ at large holographic direction and as $(r-1)$ when $r \sim 1$, in order to keep $h_{ry}$ finite [the generic dominant behavior is $h_{ty} \sim (r^{2},\,{\rm const.})$ at infinity and $r=1$, respectively].

The resonances of Eq.~\eqref{eqVectI} were searched with standard direct integration, and are presented in Table~\ref{tab:resonance} as ``vector I'' family.  Note that for $\omega=k$ there is an exact solution regular at infinity, $h_{ty}=r^{-3}$.  However, this solution is not compatible with the exact solution for~\eqref{hryeq} regular at $r=1$, $h_{ty}=(r^5-1)/r^3$.  Thus, in particular there are no zero modes.

At large overtone numbers the spacing is again consistent with the geodesic approximation.  In fact, the substitution $r\equiv1/\eta$ leads the equation above to the following WKB-amenable form,
\be
h_{ty}''+{\tilde \omega^2}\left(-\frac{6}{{\tilde \omega^2}\eta^2}+\frac{1}{1-\eta^5}\right)h_{ty}=0\,.
\ee
Following the scalar-field analysis, with the same large ${\tilde \omega}$ behavior, one finds the same asymptotic expression.

\section{Determination of eigenmodes by Frobenius expansions}
\label{app:Frobenius}

In this appendix we describe a method to obtain the eigenmodes of a boundary value problem defined by a system of coupled ODEs.  This procedure is an alternative to the method discussed in Section~\ref{sec:scalar_grav}.

Here we adopt a series solution to the problem, by first defining the wavefunctions $H_{rr}$ and $Z_{+}$  according to
\be
h_{rr}=\eta^2 H_{rr}\,,\qquad z_{+}=\eta^2 Z_{+}\,,
\ee
which are then expanded in a power series around $\eta=1/r=1$:
\be
H_{rr}=\sum_{q=0}^\infty a_q(1-\eta)^q\,,\qquad
Z_{+}=\sum_{q=0}^\infty b_q(1-\eta)^q\,.
\ee
The radius of convergence of this series is at least as large as the distance to the closest singular point, thus it should converge on the entire interval $0<\eta<1$, corresponding to $1<r<\infty$.  The coefficients $a_q,b_q$ can be obtained by direct substitution into the two coupled ODEs.  Because it is linear, we can choose $a_0=1$, and all coefficients will be functions of $\tilde \omega$ and $b_0$.  This expansion satisfies the boundary conditions at the tip $\eta=1$;  to satisfy the boundary conditions at infinity ($\eta=0$) we require that
\be
\sum_{q}a_q=\sum_{q}b_q=0\,.
\ee
This results in two conditions for two quantities, $\tilde \omega$ and $b_0$.  The series has to be truncated at some value; we typically need around 30 terms in the expansion to get an accuracy of $1\%$.  This second method yields values in very good agreement with the direct integration procedure, but computationally it seems more costly.  To get higher overtones, one needs to keep more terms in the series to get a good convergence rate.  This method is an extension of a Frobenius expansion used originally in Refs.~\cite{Horowitz:1999jd,Berti:2009kk,Delsate:2011qp} (see also Ref.~\cite{Pani:2013pma} for a review).

\section{Absence of a zero mode}
\label{app:zero_mode}

Our numerical investigations in Section~\ref{sec:scalar_grav} did not reveal the existence of any zero mode ($\tilde\omega=0$).  Here we explicitly rule out this mode in the static limit, which then dictates generic Yukawa-type decay at large distances $R$.  For $\omega=k=0$, equations~\eqref{e1}--\eqref{algebraic2} simplify to
\bea
 h_{yy}&=&-h_{rr}+h_{tt}-h_{xx}\,, \\
 h_{tt}&=&\frac{2 \left(4 r^5+1\right) h_{rr}+5 h_{xx}+2 r \left(r^5-1\right) \left(h_{rr}'-h_{xx}'\right)}{5}\,,\\
 h_{rr}''&=& -\frac{30 r^4 h_{rr}+\left(6 r^5-1\right) \left(2 h_{rr}'-h_{xx}'\right)}{r(r^5-1)} \,, \\
 h_{xx}''&=&\frac{10 r^4 h_{rr}+2 \left(r^5-1\right) h_{rr}'+\left(-8 r^5+3\right) h_{xx}'}{r(r^5-1)}\,.
\eea
We can solve the third equation for $h_{xx}'$, 
\be
h_{xx}'=\frac{30r^4h_{rr}+2(6r^5-1)h_{rr}'+r(r^5-1)h_{rr}''}{6r^5-1}\,,
\ee
and get a single third-order ODE for $h_{rr}$ from the last equation.  With the behavior~\eqref{asymptoticgrav_scalar_infinity} we find,
\be
h_{tt}=\frac{2rA_\infty^{(4)}}{15}+\left(B_\infty^{(0)}-\frac{2A_\infty^{(5)}}{5}\right)+{\cal O}(r^{-2})\,.
\ee
The solution {\it does} have $B_\infty^{(0)}=A_\infty^{(4)}=0$.  However, $A_\infty^{(5)}\neq 0$, and thus it corresponds to a spacetime with a deformed boundary, because both $h_{tt}$ and $h_{yy}$ asymptote to a constant.  We should exclude these solutions.  Therefore, no zero-mode solution exists in this spacetime.

\section{Green's function analysis for a static point particle in the AdS-soliton background}
\label{app:PPgrav}

In this appendix we discuss a Green's function approach to solve the system~\eqref{e1}--\eqref{e2}.  By setting $\mathbf{Y}\equiv(h_{rr},h_{xx},h_{rr}',h_{xx}')$, Eqs.~\eqref{e1}--\eqref{e2} can be written in the form~\eqref{system} with
\be
  \mathbf{V}= \left(\begin{array}{cccc}
                    0	&	0	&	-1	&	0\\
                    0	&	0	&	0	&	-1\\
                    \frac{-k^2 L^4 r+30 r^3}{r^5-r_0^5}	&	0	&	\frac{2(6r^5-r_0^5)}{r(r^5-r_0^5)}	&	-\frac{6 r^5-r_0^5}{r(r^5-r_0^5)}\\
                    -\frac{10 r^3}{r^5-r_0^5}	&	-\frac{k^2 L^4 r}{r^5-r_0^5}	&	-\frac{2}{r}	&	\frac{8 r^5-3 r_0^5}{r(r^5-r_0^5)}
                   \end{array}
\right)\,, 
\qquad 
 \mathbf{S}=-\frac{L^4 r \mu\delta(r-r_0)}{2 (r^5-r_0^5)} \left(\begin{array}{c}
                    0\\
                    0\\
		    1\\
		    1
                   \end{array}
\right)\,.
\ee
For clarity of notation, we restore all factors of $r_0$ and $L$ in this appendix.

In order to construct the matrix $\mathbf{X}$ (see Section~\ref{sec:Green}), we need four independent solutions of the homogeneous system which satisfy the correct boundary conditions.  Close to $r_0$, the general solution reads
\bea
 h_{rr}^{(r_0)}&\sim& \frac{a_k r_0}{r-r_0}+\left(c_k+\frac{a_k k^2L^4}{5r_0^2}\right)\log\frac{r-r_0}{r_0}+\sum_{i=0}^\infty c_{rr}^{(i)}(r-r_0)^i\,,\label{expINFhrr}\\
 h_{xx}^{(r_0)}&\sim& c_k \log\frac{r-r_0}{r_0}+ \sum_{i=0}^\infty c_{xx}^{(i)}(r-r_0)^i\,.\label{expINFhxx}
\eea
We require the fields to be regular at $r=r_0$, so $a_k=c_k=0$ and the correct asymptotic behavior reads
\be
 h_{rr}^{(r_0)}\sim \sum_{i=0}^\infty c_{rr}^{(i)}(r-r_0)^i\,,\qquad  h_{xx}^{(r_0)}\sim \sum_{i=0}^\infty c_{xx}^{(i)}(r-r_0)^i\,,\label{expr0statgrav}
\ee
where $c_{rr}^{(i)}$ and $c_{xx}^{(i)}$ are constants that can be expressed in terms of two parameters only, $c_{rr}^{(0)}$ and $c_{xx}^{(0)}$, by solving the equation in the $r\to r_0$ limit perturbatively.  Then, two independent solutions can be found by imposing these two parameters to be $(1,0)$ and $(0,1)$.  We shall denote these solutions by $\mathbf{Y}^{(r_0,1)}$ and $\mathbf{Y}^{(r_0,2)}$, respectively.

Likewise, at infinity we have the following general behavior:
\be
 h_{rr}^{(\infty)}\sim \sum_{i=1}^\infty \frac{d_{rr}^{(i)}}{r^i}\,,\qquad h_{xx}^{(\infty)}\sim\sum_{i=0}^\infty \frac{d_{xx}^{(i)}}{r^i}\,,\label{expinfstatgrav}
\ee
where again all the expansion coefficients $d_{rr}^{(i)}$ and $d_{xx}^{(i)}$ can be expressed in terms of four parameters.  We impose that the perturbations decay at infinity.  Using the series expansions above and the solution of $h_{tt}(r)$ in Eq.~\eqref{algebraic2}, we get
\be
 d_{xx}^{(0)}=0\,, \qquad d_{rr}^{(5)} =\frac{k^2L^4 d_{rr}^{(3)}}{6}\,,
\ee
where the second condition comes from requiring $h_{tt}\to0$ at infinity.  After imposing these boundary conditions, the asymptotic behavior of the perturbation functions reads 
\bea
 h_{rr}^{(\infty)}&\sim&d_{rr}^{(3)}\left(\frac{1}{r^3}+\frac{k^2L^4}{6 r^5}+\frac{k^4L^8}{120 r^7}+\frac{5r_0^5}{6 r^8}+\frac{k^6L^{12}}{5040 r^9}\right)+\frac{d_{rr}^{(10)}}{r^{10}} \,,\\
 h_{xx}^{(\infty)}&\sim& d_{rr}^{(3)}\left( -\frac{1}{3 r^3} + \frac{k^2L^4}{30 r^5}+\frac{k^4L^8}{280 r^7}+\frac{k^6L^{12}}{9072 r^9}\right)+\left[\frac{13 k^2L^4r_0^5}{180} d_{rr}^{(3)}-\frac{d_{rr}^{(10)}}{3}\right]\frac{1}{r^{10}}\,,\\
 h_{tt}^{(\infty)}&\sim& \frac{d_{rr}^{(3)}}{3 r^3}+\left[\frac{59 k^2L^4}{90} d_{rr}^{(3)}-\frac{56}{15 r_0^5} d_{rr}^{(10)}\right]\frac{1}{r^{5}}\,,
\eea
Note that the large-distance behavior only depends on two parameters, $d_{rr}^{(3)}$ and $d_{rr}^{(10)}$.  If $d_{rr}^{(3)}\neq0$, both functions $h_{rr}$ and $h_{xx}$ decay as $r^{-3}$, whereas they decay as $r^{-10}$ when $d_{rr}^{(3)}=0$.
Two independent solutions can be found by imposing these parameters to be $(1,0)$ and $(0,1)$, respectively.  We shall denote these solutions by $\mathbf{Y}^{(\infty,1)}$ and $\mathbf{Y}^{(\infty,2)}$, respectively.  Therefore, the matrix of the homogeneous system reads
\be
 \mathbf{X}= \left(\begin{array}{cccc}
                    h_{rr}^{(r_0,1)}	&	h_{rr}^{(r_0,2)}	&	h_{rr}^{(\infty,1)}	&	h_{rr}^{(\infty,2)}\\
                    h_{xx}^{(r_0,1)}	&	h_{xx}^{(r_0,2)}	&	h_{xx}^{(\infty,1)}	&	h_{xx}^{(\infty,2)}\\
                    {h_{rr}^{(r_0,1)}}'	&	{h_{rr}^{(r_0,2)}}'	&	{h_{rr}^{(\infty,1)}}'	&	{h_{rr}^{(\infty,2)}}'\\
                    {h_{xx}^{(r_0,1)}}'	&	{h_{xx}^{(r_0,2)}}'	&	{h_{xx}^{(\infty,1)}}'	&	{h_{xx}^{(\infty,2)}}'
                   \end{array}
\right)\,.
\ee
In Table~\ref{tab:asymp} we show the asymptotic behavior of each field at $r\to r_0$ and at $r\to\infty$.

\begin{table}
\begin{center}
\begin{tabular}{| c | c | c |}
\hline
\hline
			& $r\to r_0$			& $r\to \infty$		\\
\hline
$h_{rr}^{(\infty,1)}$	& $a_k \frac{r_0}{r-r_0}$	& $(r_0/r)^3$		\\
$h_{rr}^{(\infty,2)}$	& $b_k \frac{r_0}{r-r_0}$	& $(r_0/r)^{10}$		\\
\hline
$h_{xx}^{(\infty,1)}$	& $c_k \log(r-r_0)$	& $-(r_0/r)^3/3$		\\
$h_{xx}^{(\infty,2)}$	& $d_k \log(r-r_0)$	& $-(r_0/r)^{10}/3$		\\
\hline\hline
\end{tabular}
\caption{\label{tab:asymp} Schematic asymptotic behavior of the homogeneous solutions of the system~\eqref{e1}--\eqref{e2}. Recall that the superscripts $(1,2)$ in $h_{rr}$ and $h_{xx}$ denote the choice $(1,0)$ or $(0,1)$, respectively, for the couple of independent parameters of the expansion at infinity.  Making similar choices for the independent parameters of the expansion near $r=r_0$ and integrating out to infinity gives the asymptotic behavior of $h_{rr}^{(r_0,i)}$ and $h_{xx}^{(r_0,i)}$. These behaviors are not displayed since in practice we do not need them (they do not contribute when $b\to r_0$).}
\end{center}
\end{table}

Finally, from Eq.~\eqref{gen_sol}, we can write the solutions for $h_{rr}$ and $h_{xx}$, which satisfy the correct boundary conditions in the presence of the source term, as follows:
\bea
 h_{rr}&\equiv&Y_1=\sum_{i=1}^2\left(h_{rr}^{(\infty,i)}(r)I_-^{(i)}(r)+h_{rr}^{(r_0,i)}(r)I_+^{(i)}(r)\right)\,,\label{solhrr}\\
 h_{xx}&\equiv&Y_2=\sum_{i=1}^2\left(h_{xx}^{(\infty,i)}(r)I_-^{(i)}(r)+h_{xx}^{(r_0,i)}(r)I_+^{(i)}(r)\right)\,,\label{solhxx}
\eea
where
\bea
 I_+^{(i)}&=&\frac{\mu L^4}{2}\int_{r}^\infty d\rbar \frac{C_+^{(i)}\rbar}{{\cal W}(\rbar^5-r_0^5)}\delta(\rbar-b)\,,\nn\\
 I_-^{(i)}&=&\frac{\mu L^4}{2}\int_{r_0}^r d\rbar \frac{C_-^{(i)}\rbar}{{\cal W}(\rbar^5-r_0^5)}\delta(\rbar-b)\,,\nn
\eea
with ${\cal W}\equiv \det(\mathbf{X})$.  In writing this we are localizing the particle at $r=b$ but in the end we want to take the limit $b\to r_0$, as in Section~\ref{sec:static_scalar}.  The functions $C_\pm^{(i)}$ depend on the solutions of the homogeneous system.  For completeness, their expressions read
\bea
 C_+^{(1)}&=& h_{rr}^{(\infty,2)} (h_{xx}^{(\infty,1)} (-{h_{rr}^{(r_0,2)}}'+{h_{xx}^{(r_0,2)}}')+h_{xx}^{(r_0,2)} ({h_{rr}^{(\infty,1)}}'-{h_{xx}^{(\infty,1)}}'))\nn\\
 &&+h_{rr}^{(\infty,1)} (h_{xx}^{(\infty,2)} {h_{rr}^{(r_0,2)}}'-h_{xx}^{(r_0,2)} {h_{rr}^{(\infty,2)}}'-h_{xx}^{(\infty,2)} {h_{xx}^{(r_0,2)}}'+h_{xx}^{(r_0,2)} {h_{xx}^{(\infty,2)}}')\nn\\
&&+h_{rr}^{(r_0,2)} (-h_{xx}^{(\infty,2)} {h_{rr}^{(\infty,1)}}'+h_{xx}^{(\infty,1)} {h_{rr}^{(\infty,2)}}'+h_{xx}^{(\infty,2)} {h_{xx}^{(\infty,1)}}'-h_{xx}^{(\infty,1)} {h_{xx}^{(\infty,2)}}')\,,\\
C_+^{(2)}&=& h_{rr}^{(\infty,2)} (h_{xx}^{(\infty,1)} ({h_{rr}^{(r_0,1)}}'-{h_{xx}^{(r_0,1)}}')+h_{xx}^{(r_0,1)} (-{h_{rr}^{(\infty,1)}}'+{h_{xx}^{(\infty,1)}}'))\nn\\
&&+h_{rr}^{(\infty,1)} (-h_{xx}^{(\infty,2)} {h_{rr}^{(r_0,1)}}'+h_{xx}^{(r_0,1)} {h_{rr}^{(\infty,2)}}'+h_{xx}^{(\infty,2)} {h_{xx}^{(r_0,1)}}'-h_{xx}^{(r_0,1)} {h_{xx}^{(\infty,2)}}')\nn\\
&&+h_{rr}^{(r_0,1)} (h_{xx}^{(\infty,2)} {h_{rr}^{(\infty,1)}}'-h_{xx}^{(\infty,1)} {h_{rr}^{(\infty,2)}}'-h_{xx}^{(\infty,2)} {h_{xx}^{(\infty,1)}}'+h_{xx}^{(\infty,1)} {h_{xx}^{(\infty,2)}}')\,,\\
 C_-^{(1)}&=& h_{rr}^{(\infty,2)} (h_{xx}^{(r_0,2)} (-{h_{rr}^{(r_0,1)}}'+{h_{xx}^{(r_0,1)}}')+h_{xx}^{(r_0,1)} ({h_{rr}^{(r_0,2)}}'-{h_{xx}^{(r_0,2)}}'))\nn\\
 &&+h_{rr}^{(r_0,2)} (h_{xx}^{(\infty,2)} {h_{rr}^{(r_0,1)}}'-h_{xx}^{(r_0,1)} {h_{rr}^{(\infty,2)}}'-h_{xx}^{(\infty,2)} {h_{xx}^{(r_0,1)}}'+h_{xx}^{(r_0,1)} {h_{xx}^{(\infty,2)}}')\nn\\
&&+h_{rr}^{(r_0,1)} (-h_{xx}^{(\infty,2)} {h_{rr}^{(r_0,2)}}'+h_{xx}^{(r_0,2)} {h_{rr}^{(\infty,2)}}'+h_{xx}^{(\infty,2)} {h_{xx}^{(r_0,2)}}'-h_{xx}^{(r_0,2)} {h_{xx}^{(\infty,2)}}')\,,\\
 C_-^{(2)}&=& h_{rr}^{(\infty,1)} (h_{xx}^{(r_0,2)} ({h_{rr}^{(r_0,1)}}'-{h_{xx}^{(r_0,1)}}')+h_{xx}^{(r_0,1)} (-{h_{rr}^{(r_0,2)}}'+{h_{xx}^{(r_0,2)}}'))\nn\\
 &&+h_{rr}^{(r_0,2)} (-h_{xx}^{(\infty,1)} {h_{rr}^{(r_0,1)}}'+h_{xx}^{(r_0,1)} {h_{rr}^{(\infty,1)}}'+h_{xx}^{(\infty,1)} {h_{xx}^{(r_0,1)}}'-h_{xx}^{(r_0,1)} {h_{xx}^{(\infty,1)}}')\nn\\
&&+h_{rr}^{(r_0,1)} (h_{xx}^{(\infty,1)} {h_{rr}^{(r_0,2)}}'-h_{xx}^{(r_0,2)} {h_{rr}^{(\infty,1)}}'-h_{xx}^{(\infty,1)} {h_{xx}^{(r_0,2)}}'+h_{xx}^{(r_0,2)} {h_{xx}^{(\infty,1)}}')\,.
\eea

If $r>b\to r_0$, then the solution reads
\bea
 h_{rr}(k_i,r)&=&\frac{\mu L^4}{2}\sum_{i=1}^2 {\cal A}^{(i)} h_{rr}^{(\infty,i)}(r)\,,\label{solhrr2}\\
 h_{xx}(k_i,r)&=&\frac{\mu L^4}{2}\sum_{i=1}^2 {\cal A}^{(i)} h_{xx}^{(\infty,i)}(r)\,,\label{solhxx2}
\eea
where
\bea
 {\cal A}^{(1)}&=&\lim_{b\to r_0} \left.\frac{C_-^{(1)}\rbar}{{\cal W}(\rbar^5-r_0^5)}\right|_{\rbar=b}=\frac{b_k}{5r_0^3 \left(a_k d_k-b_k c_k\right)}\,,\\
 {\cal A}^{(2)}&=&\lim_{b\to r_0} \left.\frac{C_-^{(2)}\rbar}{{\cal W}(\rbar^5-r_0^5)}\right|_{\rbar=b}=-\frac{a_k}{5r_0^3 \left(a_k d_k-b_k c_k\right)}\,,
\eea
and, as in the scalar case with $n_y=0$, this limit is finite.  Finally, we obtain Eqs.~\eqref{solhrrF} and~\eqref{solhxxF} in the main text.

\section{Computation of the holographic stress-energy tensor for high-energy particle collisions in the bulk}
\label{app:collisions}

In this appendix we compute in detail the holographic stress-energy tensor presented in Eqs.~\eqref{holoTtt}--\eqref{holoTx2x2} and we collect some intermediate results that are presented in the main text.  The computation is divided into three steps.  In Section~\eqref{appsec:asymptotics} we analyse the asymptotic behavior of the metric functions defined in Eq.~\eqref{ansatz_staticp4v1b}.  In Section~\ref{appsec:collisiongrav} we solve the inhomogeneous perturbation equations Eqs.~\eqref{eqzm}--\eqref{eqhrr} in the bulk through Green's function techniques.  Finally, these results are used in Section~\ref{appsec:holoTmunu}, where we compute the holographic stress-energy tensor explicitly, via the holographic renormalization scheme~\cite{deHaro:2000xn}.

\subsection{Asymptotic behavior of the metric perturbation}
\label{appsec:asymptotics}

The asymptotic behavior at large holographic distance $r$ of the metric functions defined in Eq.~\eqref{ansatz_staticp4v1b} reads (recall we are choosing a gauge such that $h_{rx}=0$)
\begin{flalign}
&h_{rr}= \frac{A_\infty^{(3)}}{r^3}-\frac{A_\infty^{(3)} r_0^2 \tilde\omega^2}{6 r^5}+{\cal O}(r^{-7})\,,\label{expnonstaticinf1}\\
&h_{xx}=-\frac{A_\infty^{(3)}}{3 r^3}+\frac{B_\infty^{(5)}-C_\infty^{(5)}}{2 r^5}+{\cal O}(r^{-7})\,,\\
&h_{tt}=\frac{A_\infty^{(3)}}{3 r^3}-\frac{B_\infty^{(5)}+C_\infty^{(5)}}{2 r^5}+{\cal O}(r^{-7})\,,\\
&h_{\perp}= -\frac{A_\infty^{(3)}}{3 r^3} + \frac{15 \left((B_\infty^{(5)}+C_\infty^{(5)}) (k^2L^4+r_0^2\tilde\omega^2)+(B_\infty^{(5)}-C_\infty^{(5)}) k_1^2L^4\right) - A_\infty^{(3)} r_0^4 \tilde\omega^4}{\left(30 \left(k^2+k_1^2\right) L^4+60 r_0^2 \tilde\omega^2\right) r^5}+{\cal O}(r^{-7})\,,\\
&h_{yy}= -\frac{A_\infty^{(3)}}{3 r^3}+\frac{ A_\infty^{(3)} r_0^2 \tilde\omega^2-6B_\infty^{(5)}}{6r^5}+{\cal O}(r^{-7})\,,\\
&h_{tx}=\frac{k_1 \left(30 C_\infty^{(5)} k^2 L^4-15 (B_\infty^{(5)}-3 C_\infty^{(5)}) r_0^2 \tilde\omega^2-A_\infty^{(3)} r_0^4 \tilde\omega^4\right)}{30  \omega  \left(\left(k^2+k_1^2\right) L^4+2 r_0^2 \tilde\omega^2\right) }\left[\frac{1}{r^5}-\frac{r_0^2\tilde\omega^2}{14r^7}\right]+{\cal O}(r^{-9})\,,\\
& h_{tr}= \frac{i}{6 L^2 \omega  \left(\left(k^2+k_1^2\right) L^4+2 r_0^2 \tilde\omega^2\right)} \left[\frac{1}{r^4}-\frac{r_0^2\tilde\omega^2}{10r^6}\right] \left[A_\infty^{(3)} r_0^2 \tilde\omega^2 \left((k^2+k_1^2)L^4+r_0^2 \tilde\omega^2\right)\right.  \nn\\
& \hspace{1cm} \left. + 15 \left((B_\infty^{(5)}+C_\infty^{(5)}) (k^2L^4+r_0^2\tilde\omega^2)+(B_\infty^{(5)}-C_\infty^{(5)}) k_1^2L^4\right) \right]  + {\cal O}(r^{-8}) \,,\label{expnonstaticinf5}
\end{flalign}
which straightforwardly give
\bea
 z_+&=& -\frac{2 A_\infty^{(3)}}{3 r^3}+\frac{B_\infty^{(5)}}{r^5} +{\cal O}(r^{-7})\,,\\
 z_-&=& \frac{C_\infty^{(5)}}{r^5}+{\cal O}(r^{-7}) \,.
\eea
Therefore, the coefficients $A_\infty^{(3)}$, $B_\infty^{(5)}$ and $C_\infty^{(5)}$, which shall be crucial for our analysis, are related to the dominant and subdominant terms of $z_+$ and to the dominant term of $z_-$, respectively.

\paragraph{Asymptotic behavior of the solutions of the homogeneous system.}
In order to apply the Green's function technique discussed in Sect.~\ref{appsec:collisiongrav} below, we need the asymptotic behaviors (at $r\sim r_0$ and at infinity) of the solution of the homogeneous field equations, i.e. Eqs.~\eqref{eqzm}--\eqref{eqhrr} without the source terms.  The Green function method requires two independent solutions of the homogeneous system, the first one being regular at $r\sim r_0$ and (for generic values of the frequency $\omega$) irregular at infinity; the second solution is regular at infinity and generically irregular at $r\sim r_0$.  We shall denote these solutions as $X^{(r_0)}$ and $X^{(\infty)}$, respectively, where $X$ collectively denotes any perturbation variable.  By analyzing Eqs.~\eqref{eqzm}--\eqref{eqhrr} at infinity and at $r\sim r_0$, we obtain the behavior of the relevant metric functions given in Table~\ref{tab:asymp_collision}.

\begin{table}
\begin{center}
\begin{tabular}{| c | c | c |}
\hline
\hline
			& $r\to r_0$			& $r\to \infty$		\\
\hline
$h_{rr}^{(\infty,1)}$	& $\alpha_{\tilde\omega}(r-r_0)^{-1}$	& $r^{-3}+{\cal O}(r^{-8})$		\\
$h_{rr}^{(\infty,2)}$	& $\beta_{\tilde\omega}(r-r_0)^{-1}$	& $15 r_0^5 r^{-10}/56+{\cal O}(r^{-12})$		\\
\hline
$z_+^{(\infty,1)}$	& $\gamma_{\tilde\omega}\log(r-r_0)$			& $-2r^{-3}/3+{\cal O}(r^{-7})$		\\
$z_+^{(\infty,2)}$	& $\delta_{\tilde\omega}\log(r-r_0)$			& $r^{-5}+{\cal O}(r^{-7})$		\\
\hline
$z_-^{(\infty)}$	& $a_{\tilde\omega}\log(r-r_0)$			& $r^{-5}+{\cal O}(r^{-7})$		\\
\hline\hline
\end{tabular}
\end{center}
\caption{\label{tab:asymp_collision} Schematic asymptotic behavior of the homogeneous solutions of the system~\eqref{sourcezp}--\eqref{sourcehrr}. Recall that the superscripts $(1,2)$ in $h_{rr}$ and $z_+$ denote the choice $(1,0)$ or $(0,1)$ for the couple of independent parameters of the expansion at infinity, respectively, whereas the superscripts $(\infty)$ and $(r_0)$ denote solutions which are regular at infinity and at $r=r_0$, respectively. As in Table~\ref{tab:asymp}, we do not display the asymptotic behaviors of the solutions $h_{rr}^{(r_0,i)}$, $z_+^{(r_0,i)}$ and $z_-^{(r_0,i)}$ because they do not contribute in the $b\to r_0$ limit.
}
\end{table}

\subsection{Green's function analysis for collisions of particles in the AdS-soliton background}
\label{appsec:collisiongrav}

Let us now solve Eq.~\eqref{sourcezm} and the system~\eqref{sourcezp}--\eqref{sourcehrr} explicitly by using Green's function techniques.  We shall separate the discussion in two parts. In the first part we compute the coefficient $C_\infty^{(5)}$ by solving Eq.~\eqref{sourcezm} and in the second part we compute the coefficients $A_\infty^{(3)}$ and $B_\infty^{(5)}$ by solving the system~\eqref{sourcezp}--\eqref{sourcehrr}.

\paragraph{The coefficient $C_\infty^{(5)}$.}
The solution of Eq.~\eqref{sourcezm} can be written in the form of Eq.~\eqref{system} with $\mathbf{Y}\equiv(z_-,z_-')$ and (restoring factors of $r_0$ and $L$)
\be
 \mathbf{V}= \left(\begin{array}{cc}
                    0 & -1 \\
                    \frac{\tilde{\omega}^2 L^4 r}{r^5-r_0^5} & \frac{6 r^5-r_0^5}{r(r^5- r_0^5)}
                   \end{array}
\right)\,, \quad 
 \mathbf{S}= \left(\begin{array}{c}
                    0\\
                   -\frac{4i L^4 m \gamma v^2 (k_1^2+\omega^2)r}{(r^5-r_0^5)\omega (\omega^2-k_1^2 v^2)}\delta(r-b)
                   \end{array}
\right)
\,, \quad
\mathbf{X}= \left(\begin{array}{cc}
                   z_-^{(r_0)} &z_-^{(\infty)} \\
                   {z_-^{(r_0)}}' &{z_-^{(\infty)}}'
                   \end{array}\right)\,,
\ee
where $z_-^{(r_0)}$ and $z_-^{(\infty)}$ are two independent solutions of the homogeneous system which satisfy the correct boundary conditions at $r_0$ and at infinity, respectively (cf. Table~\ref{tab:asymp_collision}).  Since the homogeneous equation is equivalent to that of the scalar case, these solutions correspond to $\psione$ and $\psitwo$ defined in Section~\ref{sec:scalar}, with the substitution $k^2\to -\tilde{\omega}^2$.  Once again, we are placing the point source at $r=b$ but in the end we will take the limit $b\to r_0$.  The solution of Eq.~\eqref{sourcezm} which satisfies the correct boundary conditions at $r\sim r_0$ and at infinity reads
\begin{flalign}
& z_-(\omega,k_i,r)=-\frac{4 i L^4 m v^2\gamma (\omega^2+k_1^2)}{\omega(\omega^2-v^2 k_1^2)} \nn\\
& \hspace{0.8cm} \times \left[z_-^{(r_0)}(r)\int_r^\infty \frac{\rbar\, z_-^{(\infty)}(\rbar)}{(\rbar^5-r_0^5){\cal W}(\rbar)}\delta(\rbar-b) + z_-^{(\infty)}(r)\int_{r_0}^r \frac{\rbar\, z_-^{(r_0)}(\rbar)}{(\rbar^5-r_0^5){\cal W}(\rbar)}\delta(\rbar-b)\right]\,.
\end{flalign}
Let us evaluate the integral above when $b\to r_0<r$.  Recalling that ${\cal W}(r)\sim a_{\tilde{\omega}}/(r-r_0)$ and $z_-^{(r_0)}\sim1$ at $r\sim r_0$, we get
\be
 z_-(\omega,k_i,r)=-\frac{4 i L^4 m v^2\gamma (\omega^2+k_1^2)}{5r_0^3\omega(\omega^2-v^2 k_1^2)}\frac{ z_-^{(\infty)}(r)}{a_{\tilde{\omega}}}\,.
\ee
From the equation above and the asymptotic behavior of $z_-^{(\infty)}$ shown in Table~\ref{tab:asymp_collision}, we obtain
\be
 C_\infty^{(5)}=-\frac{4 i L^4 m v^2\gamma }{5r_0^3}\frac{(\omega^2+k_1^2)}{a_{\tilde\omega}\omega(\omega^2-v^2 k_1^2)}\,. \label{C5}
\ee
%

\paragraph{The coefficients $A_\infty^{(3)}$ and $B_\infty^{(5)}$.}
In order to solve Eqs.~\eqref{sourcezp} and~\eqref{sourcehrr} we use the same approach as for the static gravitational case, which was also described by a system of two coupled ODEs.  By setting $\mathbf{Y}\equiv(z_+,h_{rr},z_+',h_{rr}')$, Eqs.~\eqref{sourcezp}--\eqref{sourcehrr} can be written in the form~\eqref{system} with (factors of $r_0$ and $L$ restored)
\bea
 \mathbf{V}&=& \left(\begin{array}{cccc}
                    0	&	0	&	-1	&	0\\
                    0	&	0	&	0	&	-1\\
                    \frac{10 r_0^5+L^4 r^3\tilde{\omega}^2}{r^2(r^5-r_0^5)}	&	-\frac{4}{r^2}	&	\frac{6 r^5-r_0^5}{r(r^5-r_0^5)}	&	0\\
                    -\frac{5 r_0^5(6 r^5-r_0^5)}{2r^2(r^5-r_0^5)^2}	&	\frac{6 r^{10}-32 r^5 r_0^5+r_0^{10}+L^4 r^3 \left(r^5-r_0^5\right) \tilde\omega^2}{r^2(r^5-r_0^5)^2}	&	0	&	\frac{6 r^5-r_0^5}{r(r^5-r_0^5)}
                   \end{array}
\right)\,, \\
 \mathbf{S}&=& \frac{2i L^4 m v^2\gamma(\omega^2-k_1^2) r\delta(r-b)}{(r^5-r_0^5)\omega(\omega^2-v^2 k_1^2)} \left(\begin{array}{c}
                    0\\
                    0\\
		 1\\
		 0
                   \end{array}
\right)\,.
\eea
The matrix $\mathbf{X}$ of the solutions of the homogeneous system reads
\be
 \mathbf{X}= \left(\begin{array}{cccc}
		    z_+^{(r_0,1)}	&	z_+^{(r_0,2)}	&	z_+^{(\infty,1)}	&	z_+^{(\infty,2)}\\
                    h_{rr}^{(r_0,1)}	&	h_{rr}^{(r_0,2)}	&	h_{rr}^{(\infty,1)}	&	h_{rr}^{(\infty,2)}\\
                    {z_+^{(r_0,1)}}'	&	{z_+^{(r_0,2)}}'	&	{z_+^{(\infty,1)}}'	&	{z_+^{(\infty,2)}}'\\
                    {h_{rr}^{(r_0,1)}}'	&	{h_{rr}^{(r_0,2)}}'	&	{h_{rr}^{(\infty,1)}}'	&	{h_{rr}^{(\infty,2)}}'                 
                   \end{array}
\right)\,, 
\ee
where the superscripts $(1,2)$ denote the two independent solutions of the homogeneous system, whereas the superscripts $(r_0)$ and $(\infty)$ denote solutions which are regular\footnote{Note that, even if $h_{rr}$ and $z_+$ are regular, we require regularity of \emph{all} metric perturbations which can be algebraically constructed from them using Eqs.~\eqref{htX}--\eqref{htr}.} at $r\sim r_0$ and at infinity, respectively.  The asymptotic behavior of each field at $r\to r_0$ and at $r\to\infty$ is given in Table~\ref{tab:asymp_collision}.

The solutions for $h_{rr}$ and $z_+$, which satisfy the correct boundary conditions in the presence of the source term, can be written as follows
\bea
 z_+&\equiv&Y_1=\sum_{i=1}^2\left(z_+^{(\infty,i)}(r){\cal I}_-^{(i)}(r)+z_+^{(r_0,i)}(r){\cal I}_+^{(i)}(r)\right)\,,\label{zpsource}\\
 h_{rr}&\equiv&Y_2=\sum_{i=1}^2\left(h_{rr}^{(\infty,i)}(r){\cal I}_-^{(i)}(r)+h_{rr}^{(r_0,i)}(r){\cal I}_+^{(i)}(r)\right)\,,\label{hrrsource}
\eea
where
\bea
 {\cal I}_+^{(i)}&=&\frac{i L^4 m v^2\gamma(\omega^2-k_1^2) }{\omega(\omega^2-v^2 k_1^2)}\int_{r}^\infty d\rbar \frac{{\cal C}_+^{(i)}\,\rbar}{(\rbar^5-r_0^5){\cal W}(\rbar)}\delta(\rbar-b)\,,\nn\\
 {\cal I}_-^{(i)}&=&\frac{i L^4 m v^2\gamma(\omega^2-k_1^2) }{\omega(\omega^2-v^2 k_1^2)}\int_{r_0}^r d\rbar \frac{{\cal C}_-^{(i)}\,\rbar}{(\rbar^5-r_0^5){\cal W}(\rbar)}\delta(\rbar-b)\,,\nn
\eea
with $(i=1,2)$ and ${\cal W}\equiv \det(\mathbf{X})$.  The functions ${\cal C}_\pm^{(i)}$, which have lengthy expressions and so we avoided presenting explicitly, are similar to the functions $C_\pm^{(i)}$ defined in Appendix~\ref{app:PPgrav} for the static case.
If $r>b\to r_0$, then the solution reads
\bea
 z_+(\omega,k_i,r)&=&\frac{i L^4 m v^2\gamma(\omega^2-k_1^2) }{\omega(\omega^2-v^2 k_1^2)}\sum_{i=1}^2 {\cal A}^{(i)} z_+^{(\infty,i)}(r)\,,\label{solhrr2COLL}\\
 h_{rr}(\omega,k_i,r)&=&\frac{i L^4 m v^2\gamma(\omega^2-k_1^2) }{\omega(\omega^2-v^2 k_1^2)}\sum_{i=1}^2 {\cal A}^{(i)} h_{rr}^{(\infty,i)}(r)\,,\label{solhxx2COLL}
\eea
where
\bea
 {\cal A}^{(1)}&=&\lim_{b\to r_0} \left.\frac{{\cal C}_-^{(1)}\,\rbar}{{\cal W}(\rbar)(\rbar^5-r_0^5)}\right|_{\rbar=b}=-\frac{2\beta_{\tilde \omega}}{5r_0^3 \Delta_{\tilde \omega}}\,,\\
 {\cal A}^{(2)}&=&\lim_{b\to r_0} \left.\frac{{\cal C}_-^{(2)}\,\rbar}{(\rbar^5-r_0^5){\cal W}(\rbar)}\right|_{\rbar=b}=\frac{2\alpha_{\tilde \omega}}{5r_0^3 \Delta_{\tilde \omega}}\,,
\eea
and $\Delta_{\tilde \omega}=\alpha_{\tilde \omega}\delta_{\tilde \omega}-\beta_{\tilde \omega}\gamma_{\tilde \omega}$.  Therefore, we obtain
\bea
  z_+(\omega,k_i,r)&=&-\frac{2i L^4 m v^2\gamma(\omega^2-k_1^2) }{5 r_0^3 \omega(\omega^2-v^2 k_1^2)}\frac{1}{\Delta_{\tilde \omega}}\left[ {\beta_{\tilde \omega}} z_+^{(\infty,1)}(r)- {\alpha_{\tilde \omega}}z_+^{(\infty,2)}(r)\right]
  \,,\label{solzpColl}\\
 h_{rr}(\omega,k_i,r)&=&-\frac{2i L^4 m v^2\gamma(\omega^2-k_1^2) }{5 r_0^3 \omega(\omega^2-v^2 k_1^2)}\frac{1}{\Delta_{\tilde \omega}}\left[ {\beta_{\tilde \omega}} h_{rr}^{(\infty,1)}(r)- {\alpha_{\tilde \omega}}h_{rr}^{(\infty,2)}(r)\right]
 \,.\label{solhrrCOLL}
\eea
Finally, using the expressions above and the asymptotic behaviors shown in Table~\ref{tab:asymp_collision}, we obtain
\bea
 A_\infty^{(3)}&=& -\frac{2i L^4 m v^2\gamma}{5 r_0^3}\frac{(\omega^2-k_1^2)}{\omega(\omega^2-v^2 k_1^2)}\frac{\beta_{\tilde \omega}}{\Delta_{\tilde \omega}}\,, \label{A3}\\
 B_\infty^{(5)}&=&  \frac{2i L^4 m v^2\gamma}{5 r_0^3}\frac{(\omega^2-k_1^2)}{\omega(\omega^2-v^2 k_1^2)}\frac{\alpha_{\tilde \omega}}{\Delta_{\tilde \omega}}=-\frac{\alpha_{\tilde \omega}}{\beta_{\tilde \omega}}A_\infty^{(3)}\,, \label{B5}
\eea
which, together with Eq.~\eqref{C5}, constitute the main results of this section.  Note that, while $A_\infty^{(3)}$ and $B_\infty^{(5)}$ do not depend on $k_1$ in the ultrarelativistic limit, $C_\infty^{(5)}$ still has explicit dependence on the longitudinal component of the wavevector.

\subsection{Holographic stress-energy tensor}
\label{appsec:holoTmunu}

In this part we shall obtain the stress-energy tensor ${\cal T}_{ab}$ of the holographic boundary theory, as determined by the asymptotic expansion of the bulk gravitational field via holographic renomalization.

The metric we consider is
\beq
ds^2=\frac{r^2}{L^2}\lp \eta_{ab}+h_{ab}\rp dx^a dx^b
+(1+h_{rr})\frac{dr^2}{F(r)}+2 h_{ar} dr dx^a +F(r) (1+h_{yy})dy^2\,,
\label{metric_plus_pert}
\eeq
Here $x^a=(t,x_1,x_2,x_3)$.  The $h_{\mu\nu}$ are linear perturbations, which we require to vanish at infinity, and we only consider cases where 
\beqa
h_{2\mu}&=&0\,,\hspace{0.5cm} \textrm{for } \mu\neq2\,,\\
h_{3\mu}&=&0\,,\hspace{0.5cm} \textrm{for } \mu\neq3\,,\\
h_{22}&=&h_{33}\,.
\eeqa
which reduces to the ansatz defined in Section~\ref{sec:scalar_grav}.

\paragraph{Asymptotics and stress-energy tensor.}
We will obtain the stress-energy tensor by writing the metric near infinity in the Fefferman-Graham gauge
\beq
ds^2=\frac{L^2}{\bar r^2}d\bar r^2+ \frac{\bar r^2}{L^2}\lp \eta_{mn}+\frac1{\bar r^5}\frac{16\pi G L^6}{5} {\cal T}_{mn}+ O(\bar r^{-7})\rp d\bar x^m d\bar x^n\,.
\label{surprise}
\eeq
Here $m,n=t,x_1,x_2,x_3,y$. 

Note that already the unperturbed soliton solution above is not in these coordinates. We write the perturbations asymptotically as
\beq
h_{\mu\nu} =\lp \frac{a_{\mu\nu}}{r^3}+\frac{b_{\mu\nu}}{r^5}+O(r^{-7})\rp e^{i k_a x^a}
\eeq
(with constant $a_{\mu\nu}$, $b_{\mu\nu}$) 
for all perturbations except for
\beq
h_{ar}=i\frac{a_{ar}}{r^4}e^{i k_a x^a} +O(r^{-6})\,.
\eeq
We raise and lower indices of $k^a$ with $\eta_{ab}$, \eg\  $k_a x^a=-\omega t+ k_i x_i$.

In order to bring the metric into Fefferman-Graham form we make the following change of coordinates:
\beqa
r&=&\bar r \lp 1+\frac{r_0^5}{10 \bar r^5}+\lp \frac{a_{rr}}{6\bar r^3}+\frac{b_{rr}}{10\bar r^5}\rp e^{i k_a x^a} + O(\bar r^{-7}) \rp\,,\\
t&=&\bar t+\frac{i L^2}{30\bar r^5}(\omega L^2 a_{rr}-6 a_{tr})e^{i k_a x^a} + O(\bar r^{-7})\,,\\
x_1&=&\bar x_1+\frac{i L^2}{30\bar r^5}(k_1 L^2 a_{rr}+6 a_{1r})e^{i k_a x^a} + O(\bar r^{-7})\,,\\
x_2&=&\bar x_2+\frac{i L^2}{30\bar r^5}k_2 L^2 a_{rr}e^{i k_a x^a} + O(\bar r^{-7})\,,\\
x_3&=&\bar x_3+\frac{i L^2}{30\bar r^5}k_3 L^2 a_{rr}e^{i k_a x^a} + O(\bar r^{-7})\,,\\
y&=&\bar y + O(\bar r^{-2}) \,.
\eeqa
This makes $g_{\bar r\bar r}=L^2/\bar r^2$ and eliminates all other terms $g_{\bar r\mu}$ from the metric. 
A number of conditions on the metric coefficients appear when solving the Einstein equations in the asymptotic region: 
\beq
-a_{tt}=a_{11}=a_{22}=a_{33}=a_{yy}=-a_{rr}/3\,,\qquad a_{t1}=0\,.
\eeq
These conditions have the effect of eliminating terms $\bar r^{-3}$ in the expansion of the metric along holographic directions and with them the metric has the required asymptotic behavior.

With this, we obtain the stress tensor in the form
\beq
16\pi G L^6 {\cal T}_{yy}= -4r_0^5 +e^{i k_a x^a}\lp 
b_{rr}+5 b_{yy}    
\rp 
\eeq
and
\beqa
16\pi G L^6 {\cal T}_{ab}=\eta_{ab}r_0^5+
e^{i k_a x^a}\lp
\eta_{ab}b_{rr}+5 b_{ab}-L^2 (a_{ar}k_b+a_{br}k_a)-L^4k_a k_b \frac{a_{rr}}{3}
\rp\,,
\eeqa
or more explicitly 
\beqa
16\pi G L^6 {\cal T}_{tt}&=& -r_0^5 +e^{i k_a x^a}\lp
-b_{rr}+5 b_{tt}+2L^2\omega a_{tr}-L^4\omega^2 \frac{a_{rr}}{3}
\rp\,, \\
16\pi G L^6 {\cal T}_{11}&=& r_0^5 +e^{i k_a x^a}\lp 
b_{rr}+5 b_{11}-2L^2k_1 a_{1r}-L^4k_1^2 \frac{a_{rr}}{3} 
\rp\,, \\
16\pi G L^6 {\cal T}_{22}&=& r_0^5 +e^{i k_a x^a}\lp 
b_{rr}+5 b_{22}-L^4k_2^2 \frac{a_{rr}}{3}  
\rp\,, \\
16\pi G L^6 {\cal T}_{33}&=& r_0^5 +e^{i k_a x^a}\lp 
b_{rr}+5 b_{22}-L^4k_3^2\frac{a_{rr}}{3}   
\rp\,, \\
16\pi G L^6 {\cal T}_{t1}&=& e^{i k_a x^a}\lp 
5 b_{t1}+L^2(\omega a_{1r}-k_1 a_{tr}) + L^4 k_1\omega \frac{a_{rr}}{3}
\rp\,, \\
16\pi G L^6 {\cal T}_{t2}&=& e^{i k_a x^a}L^2\lp 
-k_2 a_{tr} + L^2 k_2\omega \frac{a_{rr}}{3}
\rp\,, \\
16\pi G L^6 {\cal T}_{t3}&=& e^{i k_a x^a}L^2\lp 
-k_3 a_{tr} + L^2 k_3\omega \frac{a_{rr}}{3}
\rp\,, \\
16\pi G L^6 {\cal T}_{12}&=& e^{i k_a x^a}L^2\lp 
-k_2 a_{1r} - L^2 k_1 k_2 \frac{a_{rr}}{3}
\rp\,, \\
16\pi G L^6 {\cal T}_{13}&=& e^{i k_a x^a}L^2\lp 
-k_3 a_{1r} - L^2 k_1 k_3 \frac{a_{rr}}{3}
\rp\,, \\
16\pi G L^6 {\cal T}_{23}&=& -e^{i k_a x^a}L^4 k_2 k_3 \frac{a_{rr}}{3}\,.
\eeqa

The constraints that the stress-energy tensor be traceless and divergence-free 
\beq\label{Tmnconstraints}
{{\cal T}_m}^m=0,\qquad
\d _m {\cal T}^{mn}=0
\eeq
 have not been imposed yet. They can be expressed as
\begin{flalign}
& b_{tt}=-b_{t1}\frac{k_1}{\omega}-\frac{a_{tr}}{5\omega}L^2\tilde\omega^2-b_{22}\,, \\
& b_{11}=-b_{t1}\frac{\omega}{k_1}-\frac{a_{tr}}{5 k_1}L^2\tilde\omega^2+b_{22}\,, \\
& b_{22}=\frac{1}{5}\lp -b_{rr}+L^2 (-\omega a_{tr}+k_1 a_{1r}) -L^4\tilde\omega^2\frac{a_{rr}}{3}\rp\,, \\
& b_{yy}=\frac{a_{1r}L^2\omega(\tilde\omega^2-2 k_1^2)-a_{tr}L^2 k_1(\tilde\omega^2+2\omega^2)-5 b_{t1}(k_1^2-\omega^2)-\omega k_1(b_{rr} - L^4 a_{rr} \tilde\omega^2)}{5 k_1\omega}\,. \label{constsolv}
\end{flalign}

This leaves five coefficients, \eg\ $a_{rr}$, $b_{rr}$, $b_{tt}$, $a_{1r}$, $b_{11}$ as independent coefficients. Two of them can be eliminated.
If we choose the gauge $h_{1r}=0$, as in Section~\ref{sec:scalar_grav}, then $a_{1r}=0$. Other non-dynamical Einstein equations correspond to constraints due to gauge choices in directions other than the radial one. This is the case with the algebraic equations \eqref{algebraic}
which imply that
\beq
b_{rr}=-\frac{A_\infty^{(3)}\tilde\omega^2}{6}\,.
\eeq
Making contact with Section~\ref{appsec:asymptotics},
\beq
a_{rr}=A_\infty^{(3)}\,, \qquad
b_{tt}=\frac{B_\infty^{(5)}+C_\infty^{(5)}}{2}\,,\qquad 
b_{11}=\frac{B_\infty^{(5)}-C_\infty^{(5)}}{2}\,.
\eeq
One can check that with the values for the coefficients that are given in Eqs.~\eqref{C5}, \eqref{A3} and~\eqref{B5}, the constraints \eqref{Tmnconstraints} are satisfied.

\paragraph{Background subtraction.}
Regularity at $r=r_0$ requires $y\sim y+\Delta y$ with
\beq
\Delta y=\frac{4\pi}{5}\frac{L^2}{r_0}\lp 1+\frac12 [h_{rr}(r_0)-h_{yy}(r_0)] \rp\,. 
\eeq
Since in our case we have $h_{rr}(r_0)=h_{yy}(r_0)$, the $y$ circles in the solution and in the background will match if we take the same parameter $r_0$ for both of them. Then, in order to subtract the background contribution from the stress-energy tensor we simply have to remove the terms $\propto r_0^5$ from ${\cal T}_{mn}$ (this subtraction could equally well be done by adding an appropriate counterterm to the action).

\paragraph{Final result.}

After performing the background subtraction we obtain
\begin{flalign}
 &{\cal T}_{tt}=\frac{1}{96\pi G L^6 } \frac{1}{2 \omega ^2-k^2+k_1^2}e^{ikx-i\omega t} \nn\\
  & \hspace{0.2cm}\times\left[15 B_\infty^{(5)} \left(k^2+k_1^2\right)+\left(k^2-3 k_1^2\right) \left(15 C_\infty^{(5)}+A_\infty^{(3)} k^2 L^4\right)+A_\infty^{(3)} \left(-3 k^2+k_1^2\right) L^4 \omega ^2\right], \\
& {\cal T}_{x_1x_1}=\frac{1}{96\pi G L^6 }    \left(15 B_\infty^{(5)}-15 C_\infty^{(5)}+A_\infty^{(3)} L^4 \left(k^2-2 k_1^2-\omega ^2\right)\right)e^{ikx-i\omega t}\,,\\
& {\cal T}_{yy}= \frac{1}{48\pi G L^6 }   \left(2 A_\infty^{(3)} L^4 \left(\omega ^2-k^2\right)-15 B_\infty^{(5)}\right)e^{ikx-i\omega t}\,,\\
& {\cal T}_{tx_1}= \frac{1}{96\pi G L^6 }\frac{ k_1 \left(-15 B_\infty^{(5)}+15 C_\infty^{(5)}+A_\infty^{(3)} L^4 \left(k^2+\omega ^2\right)\right)}{\omega }e^{ikx-i\omega t}\,,\\
& {\cal T}_{tx_2}= \frac{1}{96\pi G L^6 } \frac{k_2}{ \omega  \left(2 \omega ^2-k^2+k_1^2\right)}e^{ikx-i\omega t} \nn\\
  & \hspace{0.4cm}\times\left[15 C_\infty^{(5)} (k_1^2-\omega^2)-15 B_\infty^{(5)} \left(k_1^2+\omega ^2\right)+A_\infty^{(3)} L^4 \left(k^2 (k_1^2-\omega^2)+\omega ^2 \left(k_1^2+3 \omega ^2\right)\right)\right], \\
& {\cal T}_{x_2x_2}= \frac{1}{96\pi G L^6 } \left[\frac{15 C_\infty^{(5)} \left(\omega ^2-k_1^2\right)+15 B_\infty^{(5)} \left(k_1^2+\omega ^2\right)}{2 \omega ^2-k^2+k_1^2}\right.\nn\\
 & \hspace{0.3cm}\left.+\frac{A_\infty^{(3)} L^4 \left(2 k_1^2 \left(k_1^2+k_3^2\right)+\left(3 k_1^2+4 k_3^2\right) \omega ^2-3 \omega ^4+k^2 \left(\omega ^2-3 k_1^2-2 k_3^2\right)\right)}{2 \omega ^2-k^2+k_1^2}\right]e^{ikx-i\omega t}.
\end{flalign}
The other five non-vanishing components, ${\cal T}_{x_3x_3}$, ${\cal T}_{tx_3}$, ${\cal T}_{x_1x_2}$, ${\cal T}_{x_1x_3}$ and ${\cal T}_{x_2x_3}$, can be obtained by using the tracelessness and divergence-free conditions. As a check on our calculations, we have computed these components explicitly and checked that the tracelessness and divergence-free conditions hold as a consequence of the asymptotic behavior shown in Eqs.~\eqref{expnonstaticinf1}--\eqref{expnonstaticinf5}.
Finally, inserting Eqs.~\eqref{C5},~\eqref{A3} and~\eqref{B5} with $v=1$ in the expressions above, we get Eqs.~\eqref{holoTtt}--\eqref{holoTx2x2}.


\end{document}